\documentclass[amsmath,superscriptaddress,showpacs,aps,prb,twocolumn,longbibliography] {revtex4-2}
\usepackage{bm}
\usepackage{enumerate}
\usepackage{graphicx}
\usepackage[dvips]{epsfig}
\usepackage{epsf}
\usepackage{xcolor}
\usepackage{hyperref}

\makeatletter
\newcommand{\Rmnum}[1]{\expandafter\@slowromancap\romannumeral #1@}

\makeatletter

\begin{document}
\title{Fluctuation-induced odd-frequency spin-triplet pairing in disordered electron liquid}
\author{Vladimir A. Zyuzin}
\affiliation{Department of Physics and Astronomy, Texas A$\&$M University, College Station, Texas 77843-4242, USA}
\affiliation{L.~D.~Landau Institute for Theoretical Physics, Semenova~1-a,~142432, Chernogolovka, Russia}
\author{Alexander M. Finkel'stein}
\affiliation{Department of Physics and Astronomy, Texas A$\&$M University, College Station, Texas 77843-4242, USA}
\affiliation{Department of Condensed Matter Physics, The Weizmann Institute of Science, Rehovot 76100, Israel}
\date{\today}
\begin{abstract}
We consider a two-dimensional disordered conductor in the regime when the superconducting phase is destroyed by the magnetic field.
We analyze a combination of
fluctuations of different origin which results in an effective interaction amplitude suitable for a
spontaneous s-wave odd-frequency pairing instability.
We observe that the end point of the superconductivity is a quantum critical point separating the conventional superconducting phase from a state with the odd-frequency spin-triplet pairing instability. We speculate that this could shed light on a rather mysterious insulating state observed in strongly disordered superconducting films in a broad region of the magnetic fields.

\end{abstract}

\maketitle

\section{Introduction}

Despite decades of studying disordered superconducting films \cite{Finkel'stein1994,GoldmanMarkovic1998,LarkinAnnPhys1999}, the level of understanding of these systems remains far from being satisfactory. In homogeneously disordered films, the degradation of the superconducting temperature $T_{\mathrm{c}}$ with disorder \cite{GraybealBeasley1984} is well described by a theoretical curve obtained in \cite{Finkelstein1987}. Its notable feature is existence of an ending point, which is an example of a quantum critical point induced by disorder. However, in the past few years, a more complex physical picture has emerged in experiments. Namely, a gap has been found \cite{SacepePRL2008} in the scanning tunneling spectroscopy measurements of the density of states of some amorphous films. This gap tends to a finite value in the vicinity of vanishing $T_{\mathrm{c}}$. This observation was attributed to inhomogeneities\cite{Feigel'manPRL2007,Feigel'man2010}, and the observed gap was called a "pseudo-gap", $E_{\mathrm{g}}$. By contrast, when
a point-contact spectroscopy has been applied \cite{ChapelierNatPhys2019}, another gap, $\Delta_{\mathrm{c}}$,  has been unveiled which follows $T_{\mathrm{c}}$ and disappears together with it.
In \cite{ChapelierNatPhys2019}, the superconducting gap $\Delta_c$ has been attributed to the phase-coherent state, while the pseudo-gap $E_g$ to the preformed Cooper pairs. We are skeptical about the possibility of coexistence of two gaps, which originate from the same s-wave spin-singlet pairing, in a strongly disordered material. Besides the existence of the two gaps, a rather mysterious insulating behavior has been observed in many amorphous superconducting films when the superconductivity is destroyed by the applied magnetic field \cite{GantmakherJETPletters2000, ShaharPRL2004, SteinerBoebingerKapitulnikPRL2005, BaturinaPRL2007}.

Here, we consider an odd-frequency triplet pairing as an alternative to the scenario involving preformed $s$-wave pairs.
The possibility of the odd-frequency pairing has been studied since 1974 when Berezinskii \cite{BerezinskiiJETPlett1974} made a breakthrough remark
that the required antisymmetry of the superconducting pairing function
with respect to the electron exchange can be resolved by an odd dependence on its time arguments.
Although the odd-frequency superconductivity is expected under mesoscopic conditions \cite{BergeretVolkovEfetovPRL2001, BergeretVolkovEfetovRMP2005, BuzdinRMP2005, TanakaGolubovPRL2007, TanakaSatoNagaosaJPSJ2008,  KeizerNature2006, AnwarPRB2010, RobinsonWittBlamireScience2010},
so far there was no reliable example of the odd-frequency pairing in bulk materials. For the odd-frequency pairing, an electron interaction should be strongly retarded. In this paper, we consider the electron scattering by impurities (disorder) as a source of a pronounced time dependence of the effective electron interaction.

In the presence of strong scattering caused by the impurities the variants of the off-diagonal long range order are limited to only s-wave orders. As a result,
only two phases are possible: regular (Cooper pairing) s-wave,
spin-singlet (even-frequency) and the s-wave, spin-triplet
(odd-frequency). 
In this paper we identify the processes of the electron interactions which generate the odd-frequency pairing amplitude.
In a search for such processes, we exploit the fact that fluctuations are particularly strong near the boundary of the phase instability. Hence, we mixed fluctuations of different origin such that their combination is suitable for the s-wave odd-frequency spin-triplet pairing. With that in hand,
we have found that in the vicinity of the region where the conventional superconducting phase is destroyed by the magnetic field, the odd-frequency pairing instability must inevitably develop at low enough temperatures. We, however, distinguish this pairing from the odd-frequency superconductivity. The question of the phase coherence, which is needed for true superconductivity, has not been studied enough for the odd-frequency pairing, especially in the presence of the magnetic field.

The rest of the paper is organized as follows.
In the main part of the text (MT) we emphasize the conceptual aspects of the work.
The details of the calculations can be found in four extended Appendices.
To make the paper self-contained, technical aspects of general character are presented in the Appendix \ref{appendixA}.

\section{Preliminaries}

For illustrative purposes, let us start with a four fermion term in the action, which is non-local in time,
\begin{align}
\hat{V}_{\mathrm{int}}= V(t_{1},t_{2})\sum_{\alpha\beta}\bar{\Psi}_{\alpha}(t_{1}) \bar{\Psi}_{\beta}(t_{2}) \Psi_{\beta}(t_{2}) \Psi_{\alpha}(t_{1}),
\end{align}
where $V(t_{1},t_{2})$ is a function of times $t_{1}$ and $t_{2}$, $\alpha$ and $\beta$ denote fermion's spin (generally speaking, $V(t_1,t_2)$ may depend on spins), and $\bar{\Psi}$ and $\Psi$ are the Grassmann fields.
As an example, consider $\alpha \neq \beta$, in which case the interaction can be split in the Cooper channel into a sum,
\begin{align}
\sum_{\alpha\neq\beta}\bar{\Psi}_{\alpha}(t_{1})& \bar{\Psi}_{\beta}(t_{2}) \Psi_{\beta}(t_{2}) \Psi_{\alpha}(t_{1})
\\
=
&
\frac{1}{2}
\left[
\bar{\Psi}_{\uparrow}(t_{1}) \bar{\Psi}_{\downarrow}(t_{2}) - \bar{\Psi}_{\downarrow}(t_{1}) \bar{\Psi}_{\uparrow}(t_{2})
  \right]
\nonumber
\\
&
\times
\left[
\Psi_{\downarrow}(t_{2}) \Psi_{\uparrow}(t_{1}) - \Psi_{\uparrow}(t_{2}) \Psi_{\downarrow}(t_{1})
  \right]
\nonumber
\\
+
&
\frac{1}{2}
\left[
\bar{\Psi}_{\uparrow}(t_{1}) \bar{\Psi}_{\downarrow}(t_{2}) + \bar{\Psi}_{\downarrow}(t_{1}) \bar{\Psi}_{\uparrow}(t_{2})
  \right]
\nonumber
\\
&
\times
\left[
\Psi_{\downarrow}(t_{2}) \Psi_{\uparrow}(t_{1}) + \Psi_{\uparrow}(t_{2}) \Psi_{\downarrow}(t_{1})
  \right];
\nonumber
\end{align}
Here, the first line corresponds to the Cooper pairing in the singlet channel, while the second one corresponds to the pairing in the triplet channel.
It is clear that if $V(t_{1},t_{2}) = \delta(t_{1} - t_{2})$, the term in the triplet channel vanishes due to the anti-commutation relations of the Grassmann field operators.
However,
this term is non-zero for a time-dependent interaction.
Berezinskii
noticed that $V(t_{1},t_{2})$ permits to resolve the required antisymmetry of the superconducting pairing function, $F(1;2)=-F(2;1)$,
by an odd dependence on the time arguments. This is called the odd-frequency superconducting pairing.
Generally, the odd-frequency pairing  opens a possibility to an $s$-wave spin-triplet and also to a $p$-wave spin-singlet superconducting pairings \cite{BalatskyAbrahamsPRB1992,AbrahamsBalatskyScalapinoSchriefferPhysRevB1995}. However, in the presence of a strong disorder,
scattering on impurities smears out the orbital order, and only the $s$-wave spin-triplet may survive \cite{BelitzKirkpatrickPRL1991,BelitzKirkpatrickPRB1999,BergeretVolkovEfetovRMP2005,FominovPRB2015,LinderBalatsky2017}.

In order for the exotic odd-frequency pairing to get a chance to reveal itself, the time dependence of the electron coupling should be well pronounced.
In this paper we rely on the fact that the disorder contributes to the emergence of a strong time dependence of the {\it effective} electron interaction amplitudes.
The point is that matrix elements calculated with the eigenstates obtained in a given realization of the impurities
are strongly energy dependent. Technically, propagation of electrons on large scales in a disordered medium is described by slow diffusive modes, diffusons and Cooperons. These modes make the electron interactions to be effectively time-dependent. The effect is stronger in low dimensions, and we, therefore, concentrate on the electron liquid in disordered films.

Motivated by the experiments on disordered film superconductors, we analyze here a conventional system without any modelling assumptions. The free-electron part of the Hamiltonian, $\hat{H_0} = \frac{{k}^2}{2m} + V_{\mathrm{d}}({\bf r})$, consists of the kinetic part and the Gaussian distributed disorder potential $V_{\mathrm{d}}({\bf r})$, where
$m$ is the fermion's mass, and ${k}$ is the absolute value of momentum. Note that in the presence of a strong disorder further detalization, such as band structure or angular dependencies of the parameters,  is rather meaningless as only zero-harmonic survives. Correspondingly, the electron-electron interactions will be described by only three Fermi liquid amplitudes. The one, denoted as $Z$, is the coupling constant in the charge-density channel, the other one describes interaction in the spin-density channel ($\Gamma_{2}$) and, finally, the interaction amplitude $V_{\mathrm{s}}$ acts in the s-wave spin-singlet Copper channel. All three amplitudes for energies of interest are instantaneous. The amplitude $V_{\mathrm{s}}$ is assumed to be attractive, $V_{\mathrm{s}}<0$. The amplitude responsible for superconducting pairing in the spin-triplet channel, which is of the most interest to us, is absent in the bare Hamiltonian.

To analyze electron modes on large scales, it is preferable to integrate out fast short range single-particle degrees of freedom and describe electrons in terms of the effective action of the diffusive modes. Large scales allow one to consider averaging the action with respect to the disorder. Here, a subtle technical point arises. Averaging is often performed using replicas in the technique of the Matsubara frequencies. Per contra, we prefer to apply the Keldysh contour technique. The advantage of this approach is that one avoids the analytic continuation from the
imaginary frequency axis.
The cost, however, is in the doubling of the field variables in the Keldysh space \cite{Kamenev}.
For each branch of the Keldysh contour, one has to introduce an independent set of the Grassmann numbers.
Thus,
we write the fermion Grassmann numbers $\hat{\Psi}$ in spin ($\uparrow$, $\downarrow$) space S, time-reversal Gor'kov-Nambu ($\psi$, $\bar\psi$) space N, and also in the Keldysh ($+/-$) space K which corresponds to the forward/backward branches of the contour correspondingly. In total, $\hat{\Psi}$ has $2\times2\times2=8$ components in the KSN spaces.
We will use a set of the Pauli matrices $\tau_i$ acting in the Gor'kov-Nambu space, and the set of matrices $\sigma_\alpha$ to describe  real spin degree of freedom. Matrices from the $\tau_i$ and $\sigma_{\alpha}$ sets will not have hats, only matrices in the Keldysh space will be indicated by a hat.

\section{$Q$-matrix description}

In disordered conductors, perturbations of charge and spin relax diffusively at low frequencies and large distances. In a system obeying time-reversal symmetry, the low-energy modes in the Cooper channel also have a diffusive form. These modes, diffusons and Cooperons, fully describe the low-energy dynamics of the disordered electron liquid.
Therefore, it is convenient to describe an ensemble of disordered electrons directly in terms of the diffusive modes. This can be done with the use of the so-called $Q$-matrix technique in the framework of the non-linear sigma model (NL$\sigma$M) that includes the effects of electron-electron (\emph{e-e}) interactions \cite{Finkelstein1983b,Finkelstein1984b,FinkelsteinReview1990} (also see Appendix \ref{appendixA} for details).
The \emph{e-e} interaction causes a re-scattering of various diffusive modes.
In its essence, the NL$\sigma$M is not a model but a minimal microscopic theory, which incorporates all symmetry constraints and conservation laws relevant to the low energy dynamics of electrons in disordered conductors.
\begin{align}
iS_{0} = -\frac{\pi \nu }{8}
 \mathrm{Tr}
\int_{{\bf r}}
\{ D[ \nabla \hat{Q}({\bf r}) ]^2
+ 4i\tau_{3}\epsilon\hat{Q}({\bf r}) \},
\end{align}
where $\nu$ is the density of states, and $D$ is the diffusion coeffiecient.
Matrices $\hat{Q}_{\epsilon\epsilon^{\prime}}({\bf r})$, apart from eight KSN indexes, depend on a spatial coordinate and two frequencies (indicated as indexes), which are Fourier transforms of the two times (unlike the Matsubara variables, the frequencies here are real).
The traces are in frequency (F), K, S, and N spaces, such that $\mathrm{Tr}\equiv \mathrm{Tr}_{\mathrm{FKSN}}$ (see SM for more details).
 Components of the $Q$-matrix represent all possible two-fermion averages with slow time and spatial arguments.
Correspondingly, deviations of the $Q$-matrix from its equilibrium position
describe fluctuations of various quantities such as charge, spin, and Cooper pairs that slowly propagate at large distances. There is a natural reason  to introduce the $Q$-matrix technique for studying the \emph{e-e} interactions. Namely, non-linearity of the $Q$-matrices, imbedded into the constraints imposed on them, automatically produces scattering between the various diffusive modes (a sort of anharmonicity)in the form corresponding to the Goldstone's nature of these modes.

Matrices exploited in the NL$\sigma$M have an "onion-shell" type structure \cite{Wegner1979, EfetovLarkinKhmel'nitskii}:
\begin{align}
\hat{\underline{Q}} = \hat{\cal{U}} \circ \hat{U} \circ \hat{\sigma}_{3} \circ \hat{\bar{U}} \circ \hat{\cal{U}}^{-1}
\equiv  \hat{\cal{U}} \circ \hat{Q} \circ \hat{\cal{U}}^{-1},
\end{align}
where $\circ$ denotes convolution in time.
The Keldysh matrix $\hat{\sigma}_{3}$, which encodes the difference in propagation of the retarded and advanced directions in time, stands in their core. Matrices $\hat{U} = \exp(-\hat{W}/2)$ and $\hat{\bar{U}} = \exp(\hat{W}/2)$, which stand next, parametrize rotations in the Keldysh space about the $\hat{\sigma}_{3}$ matrix, with $\hat{W}$ being the rotation generators.
Fluctuations of the $\hat{W}$-fields considered within the Gaussian approximation describe the manifold of the diffusive modes, which include diffusons and the Cooperons.

So far, no information about the electron state has been introduced.
This is achieved \cite{Kamenev, FeigelmanLarkinSkvortsovPRB2000, SchwieteFinkel'steinPRB2014} by the two flanking matrices $\hat{{\cal U}}$ and  $\hat{\cal{U}}^{-1}$.
\begin{align}
\hat{{\cal U}}_{\epsilon}
= \left[
\begin{array} {cc}
\hat{u}_{\epsilon} & 0 \\
0 & \hat{u}_{\epsilon}^{\mathrm{T}}
 \end{array}
\right]_{\mathrm{N}},
~~~
\hat{u}_{\epsilon} = \left[
\begin{array}{cc}
1 & {\cal F}_{\epsilon} \\
0 & -1
\end{array}
\right]_{\mathrm{K}}.
\end{align}
Their presence is indicated by underscoring the matrix $Q$.
Matrices $\hat{{\cal U}}$ introduce information about the electron state, which is described by the distribution function $\cal F$.
In equilibrium, the distribution function is ${\cal F}_{\epsilon} = \tanh\left( \frac{\epsilon}{2T} \right)$. Indices $\mathrm{N}$ and $\mathrm{K}$ used in the matrices above denote Gor'kov-Nambu and Keldysh spaces of the matrices correspondingly.

\section{ Electron-electron interaction in the Cooper channels}

The \emph{e-e} interactions in combination with the non-linear amplitudes of the diffusive modes, which appear in the expansion of $\hat U$ and $\hat{\bar U}$-matrices beyond the linear order in $\hat{W}$, mix all the diffusive channels (i.e., the charge, spin and the Cooper ones). Therefore, a detailed analysis of the theory requests for the NL$\sigma$M in its complete form, including interactions in the charge-density and spin-density channels ($Z$ and $\Gamma_2$-terms), as well as the interaction in the conventional s-wave spin-singlet Cooper channel ($V_{\mathrm{s}}$-term):
\begin{widetext}
\begin{align}
iS_{\mathrm{ee}} =
&
-  i \frac{\pi^2 \nu^2}{8}
Z
\int_{\epsilon_{1},\epsilon_{1}^\prime,\epsilon_{2},\epsilon_{2}^\prime}
\int_{{\bf r}}
\mathrm{Tr}_{\mathrm{KSN}}[ \hat{\gamma}^{1/2} \tau_{\pm} \underline{\hat{Q}_{\epsilon_{1}\epsilon_{1}^\prime}({\bf r})} ]
\mathrm{Tr}_{\mathrm{KSN}}[ \hat{\gamma}^{2/1} \tau_{\pm} \underline{\hat{Q}_{\epsilon_{2}\epsilon_{2}^\prime}({\bf r})} ]
\delta_{\epsilon_{1} - \epsilon_{1}^\prime, \epsilon_{2}^\prime - \epsilon_{2}}
\label{amplitudeZ}
\\
&
+  i\frac{\pi^2\nu^2}{8}
\Gamma_{2}
\int_{\epsilon_{1},\epsilon_{1}^\prime,\epsilon_{2},\epsilon_{2}^\prime}
\int_{{\bf r}}
\mathrm{Tr}_{\mathrm{KSN}}[ \hat{\gamma}^{1/2} \tau_{\pm} \bm{\sigma} \underline{\hat{Q}_{\epsilon_{1}\epsilon_{1}^\prime}({\bf r})} ]
\mathrm{Tr}_{\mathrm{KSN}}[ \hat{\gamma}^{2/1} \tau_{\pm} \bm{\sigma}  \underline{\hat{Q}_{\epsilon_{2}\epsilon_{2}^\prime}({\bf r})} ]
\delta_{\epsilon_{1} - \epsilon_{1}^\prime, \epsilon_{2}^\prime - \epsilon_{2}}
\label{amplitudeGamma2}
\\
&
+ i\frac{\pi^2\nu}{8}
\int_{\epsilon_{1},\epsilon_{1}^\prime,\epsilon_{2},\epsilon_{2}^\prime}
V_{\mathrm{s}}(\epsilon_{1},\epsilon_{2})
\int_{{\bf r}}
\mathrm{Tr}_{\mathrm{KSN}}[ \hat{\gamma}^{1/2} \tau^{\pm} \underline{\hat{Q}_{\epsilon_{1}\epsilon_{1}^\prime}({\bf r})} ]
\mathrm{Tr}_{\mathrm{KSN}}[ \hat{\gamma}^{2/1} \tau^{\mp}  \underline{\hat{Q}_{\epsilon_{2}\epsilon_{2}^\prime}({\bf r})} ]
\delta_{\epsilon_{1} - \epsilon_{1}^\prime, \epsilon_{2}^\prime - \epsilon_{2}},
\label{eesinglet}
\end{align}
\end{widetext}
here, as it follows from their matrix and frequency structures, each of the \emph{e-e} interaction terms couples different blocks of the $Q$-matrices.
Matrices $\hat{\gamma}^1$ and $\hat{\gamma}^2$ are standard in the Keldysh technique \cite{Kamenev}; the compact notation $\mathrm{Tr}[\hat{\gamma}^{1/2}..]\mathrm{Tr}[\hat{\gamma}^{2/1}..]$ stands for $(\mathrm{Tr}[\hat{\gamma}^{1}..]\mathrm{Tr}[\hat{\gamma}^{2}..]+ \mathrm{Tr}[\hat{\gamma}^{2}..]\mathrm{Tr}[\hat{\gamma}^{1}..])$ summation, and the same for the $\mathrm{Tr}[\tau^{\pm}..]\mathrm{Tr}[\tau^{\mp}..]$. Thus, the expression written in Eq.~\eqref{eesinglet} combines four terms with a different order of matrices. In the term given by Eq.~\eqref{eesinglet}, matrices $\hat{\gamma}$ and $\tau^{\pm} = \frac{1}{2}(\tau_{x}\pm i \tau_{y})$ regulate the structure in the Keldysh and Gor'kov-Nambu spaces, correspondingly.
A combination of $\tau^+$ and $\tau^-$ matrices together with the spin identity-matrices (not shown explicitly) in Eq.~\eqref{eesinglet} restricts the $V_{\mathrm{s}}(\epsilon_{1},\epsilon_{2})$ interaction term to the Cooper pairing in the spin-singlet channel \cite{Finkelstein1984b, FeigelmanLarkinSkvortsovPRB2000}. For technical aspects we refer the reader to the Appendix \ref{appendixA}.

The bare value of $V_{\mathrm{s}}(\epsilon_{1},\epsilon_{2})$
is defined by the electron-phonon interaction at the frequency arguments about the Debye frequency.
However, as is well known, the bare value
gets strongly modified by the re-scattering processes as well as the corrections generated by the Coulomb interaction which suppresses superconductivity in disordered films \cite{Finkelstein1987}.
All such processes developing on the whole interval of energies
down to $\epsilon\agt T_c$ can be properly included into the amplitude
$\Gamma_{\mathrm{s}}(\epsilon_{1},\epsilon_{2})$ which substitutes the initial $V_{\mathrm{s}}(\epsilon_{1},\epsilon_{2})$ in Eq.~\eqref{eesinglet}.
The renormalized amplitude $\Gamma_{\mathrm{s}}(\epsilon_{1},\epsilon_{2})$ will be used below throughout this paper, \emph{cf.} \cite{GammaC} (see Appendix \ref{appendixB} for details).

The interaction in the spin-triplet part of the Cooper channel is
\begin{widetext}
\begin{align}
iS_{\mathrm{triplet}} =
 i\frac{\pi^2\nu}{8}
\int_{\epsilon_{1},\epsilon_{1}^\prime,\epsilon_{2},\epsilon_{2}^\prime}
V_{\mathrm{t}}(\epsilon_{1},\epsilon_{2})
\int_{{\bf r}}
\mathrm{Tr}_{\mathrm{KSN}}[ \hat{\gamma}^{1/2} \tau^{\pm} {\bm \sigma} \underline{\hat{Q}_{\epsilon_{1}\epsilon_{1}^\prime}({\bf r})} ]
\mathrm{Tr}_{\mathrm{KSN}} [ \hat{\gamma}^{2/1} \tau^{\mp} {\bm \sigma} \underline{\hat{Q}_{\epsilon_{2}\epsilon_{2}^\prime}({\bf r})} ]
\delta_{\epsilon_{1} - \epsilon_{1}^\prime, \epsilon_{2}^\prime - \epsilon_{2}}.
\label{eetriplet}
\end{align}
\end{widetext}
Here, $V_{\mathrm{t}}(\epsilon_{1},\epsilon_{2})$ is some interaction amplitude which depends on the frequencies from different trace blocks.
Note that the matrices $\hat{\gamma}$ and $\tau$ in Eqs.~\eqref{eesinglet} and \eqref{eetriplet} stand in the same combinations.
A crucial difference between the singlet and triplet pairing is in the spin structure of the trace blocks.
In the latter case, the spin Pauli matrices ${\bm\sigma}$ select the triplet pairs only.

\begin{figure}[h] \centerline{\includegraphics[clip, width=1  \columnwidth]{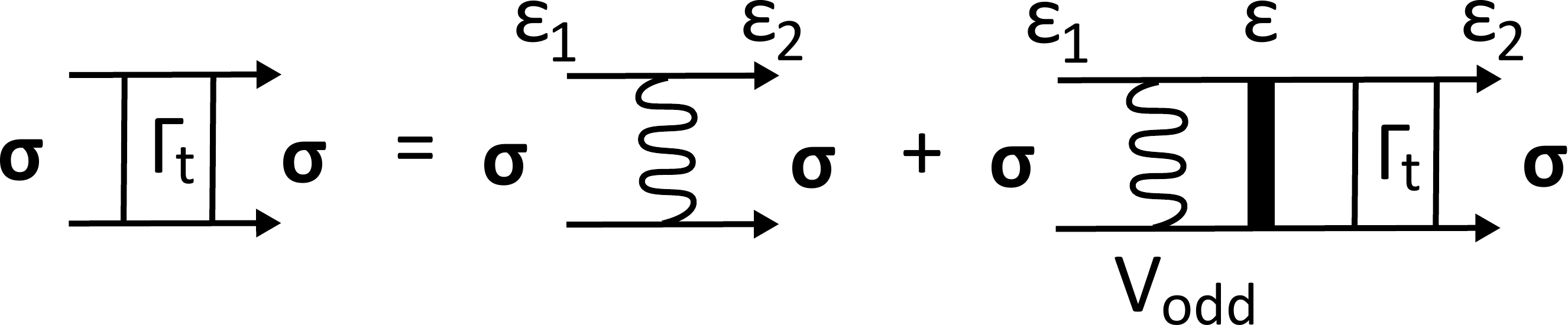}}

\protect\caption{Equation for the effective interaction amplitude in the spin-triplet part of the Cooper channel.
Comments to the odd-frequency interaction amplitude $V_{\mathrm{odd}}$ are in the text.
The black rectangle denotes the Cooperon mode inside each of the sections in the Cooper ladder.
The Pauli matrices $\bf\sigma$ acting in the spin space
select the spin-triplet channel.}

\label{fig:EquationSingletTriplet}

\end{figure}

In contrast to the singlet Cooper channel, in the triplet channel the bare interaction is absent.
However, as we will see in the next section some analogue of the "bare"
interaction can be
generated as a result of mixing different channels.
In order to get an idea about what kind of the \emph{e-e} interaction could be efficient in the triplet channel, we now construct the Cooper ladder
assuming general form of the interaction $V_{\mathrm{t}}(\epsilon_{1},\epsilon_{2})$.
To derive the ladder, we expand the $\hat{Q}$ matrices in Eq.~\eqref{eetriplet} to the first order in $\hat{W}$, and average a series of thus obtained terms within the Gaussian approximation. As a result, we get the following equation for the effective amplitude
in the spin-triplet channel, see Fig.~\ref{fig:EquationSingletTriplet},
\begin{align}
&
\Gamma_{\mathrm{t}}(\epsilon_{1},\epsilon_{2})
=
\frac{1}{2}\left[ V_{\mathrm{t}}(\epsilon_{1},\epsilon_{2})  -   V_{\mathrm{t}}(\epsilon_{1}, -\epsilon_{2}) \right]
\label{equation9}
\\
&
-  \int_{-\infty}^{+\infty} \frac{1}{2}\left[ V_{\mathrm{t}}(\epsilon_{1},\epsilon)  -   V_{\mathrm{t}}(\epsilon_{1}, -\epsilon) \right]\frac{{\cal F}_{\epsilon}}{\epsilon}
\Gamma_{\mathrm{t}}(\epsilon,\epsilon_{2}) \frac{d\epsilon}{2} .
\nonumber
\end{align}
We see that the spin-triplet Cooper ladder describing $\Gamma_{\mathrm{t}}$ is generated by a $\frac{1}{2}\left[ V_{\mathrm{t}}(\epsilon_{1},\epsilon_{2})  -   V_{\mathrm{t}}(\epsilon_{1}, -\epsilon_{2}) \right]$ combination \cite{BelitzKirkpatrickPRL1991,BelitzKirkpatrickPRB1999,FuseyaKohnoMiyakeJPSJ2003, SamokhinMineev2008}, which is odd in its frequency argument $\epsilon_{2}$.
Note that in the case of the singlet pairing, the corresponding combination has a $+$ sign,
i.e. it is even in its frequency arguments. Such a difference is due to the anti-commutation of fermion fields, which in the diagram language corresponds to the exchange of the arguments in the Cooperon mode (black rectangle in Fig.~\ref{fig:EquationSingletTriplet}) connecting two amplitudes in the $\Gamma_{\mathrm{t}}$-ladder.
Moreover, the resulting amplitude in the triplet Cooper ladder
must be symmetric under the exchange of $\epsilon_{1}$ and $\epsilon_{2}$.
Therefore, only the part of the \emph{e-e} amplitude $V_{\mathrm{t}}(\epsilon_{1},\epsilon_{2}) $, which is odd with respect to its both arguments is effective in the triplet channel. The obtained combination, denoted as $V_{\mathrm{odd}}(\epsilon_{1},\epsilon_{2})$ (as it is odd in both arguments), stands in the equation depicted in Fig.~\ref{fig:EquationSingletTriplet}.

We have gotten two very instructive results for the disordered
system:

(i) Not only the odd-frequency pairing is compatible with the triplet pairing, but it is the only possibility:
the triplet part of the Cooper channel automatically selects odd dependence of the Cooper pair on the frequency arguments.

(ii) The amplitude in the triplet part of the Cooper channel can be thought of as the first harmonic in the frequency dependence: only the odd-frequency component passes through the sequence of scattering.

To conclude, the interaction amplitude has to possess a special structure in its frequency arguments in order to not get filtered out for the triplet odd-frequency pairing.

\section{Electron-electron interaction in the odd-frequency Cooper channel}\label{e-einteractioninodd}

We have to identify a process which will be effective for the odd-frequency Cooper channel. Since the bare interaction in the spin-triplet Cooper channel is absent, it has to be generated by the non-linear amplitudes of the NL$\sigma$M which mix the \emph{e-e} interactions acting in different channels. The processes of this kind (a sort of anharmonicity) inevitably lead to a small parameter $\rho=\frac{1}{(2\pi)^2 \nu_{2\mathrm{d}} D}$,
where $\nu_{2\mathrm{d}}$ is the effective two-dimensional density of states.
Amorphous superconductors films used in the experiments are relatively thick.
In these systems the electron excitations are not quantized in the transverse direction.
On the contrary, the diffusive modes (diffusons and Cooperons) are effectively two-dimensional. Hence, the appearance of $\nu_{2\mathrm{d}}$.

We start with the processes \cite{Finkelstein1983b,Finkelstein1984b,FinkelsteinReview1990} which result in transferring of the interaction amplitudes Eqs. (\ref{amplitudeZ})-(\ref{eesinglet}) into the spin-triplet part of the Cooper channel.
We conclude that to the first order in $\Gamma_{2}$ or $Z$ the obtained interaction in the
spin-triplet part of the Cooper channel is rather weak.
The spin-triplet pairing instability in this case may exist, but it is of a threshold type.
Namely, unlike in the BCS theory, where for any attractive interaction there is a transition temperature, the instability in the spin-triplet channel will occur only for effective interactions larger than a threshold value, which is controlled by the parameter $\rho$.
It turned out that the threshold value corresponds to $\rho\sim 1$, i.e., to the amount of disorder
at which the system is close to the Anderson insulating state.
However, the theory we are developing in this paper
aims at physical systems, which are in the metallic regime.
To complete the analysis in the first order with respect to the \emph{e-e} interaction amplitudes, we have checked that it is
impossible to generate an interaction amplitude in the triplet part of
the Cooper channel with the use of the spin-singlet amplitude $\Gamma_{\mathrm{s}}$ only.

In view of this experience, it is natural to anticipate the occurrence of the odd-frequency spin-triplet pairing in the
vicinity of some other instability.
In this case the smallness of $\rho$ could be compensated by the effective amplitude of the \emph{e-e} interaction related to this instability.
Motivated by the experiments, in the analysis of the second order in the \emph{e-e} interaction amplitudes, we have chosen to exploit divergence of the $\Gamma_{\mathrm{s}}$ amplitude when
the system approaches a phase transition into the conventional s-wave spin-singlet superconducting state.
As is well known, in the vicinity of the superconducting transition, the critical fluctuations are non-local in time and space \cite{AslamazovLarkinFTT1968,GalitskiLarkinPRB2001}. 
Therefore, they may generate an effective interaction in the spin-triplet Cooper channel.
We have checked that out of the $\Gamma_{\mathrm{s}}\Gamma_{2}$ and $\Gamma_{\mathrm{s}}Z$ combinations,
only the diagrams with $\Gamma_{2}$ appeared to be relevant for the spin-triplet part of the Cooper channel, while
the $\Gamma_{\mathrm{s}}Z$ combination does not contribute.
This is not surprising, as the spin-density amplitude $\Gamma_{2}$ is the only process that can re-scatter superconducting fluctuations $\Gamma_{\mathrm{s}}$ from the spin-singlet into the spin-triplet part of the Cooper channel.

\begin{figure}[h] \centerline{\includegraphics[clip, width=0.9  \columnwidth]{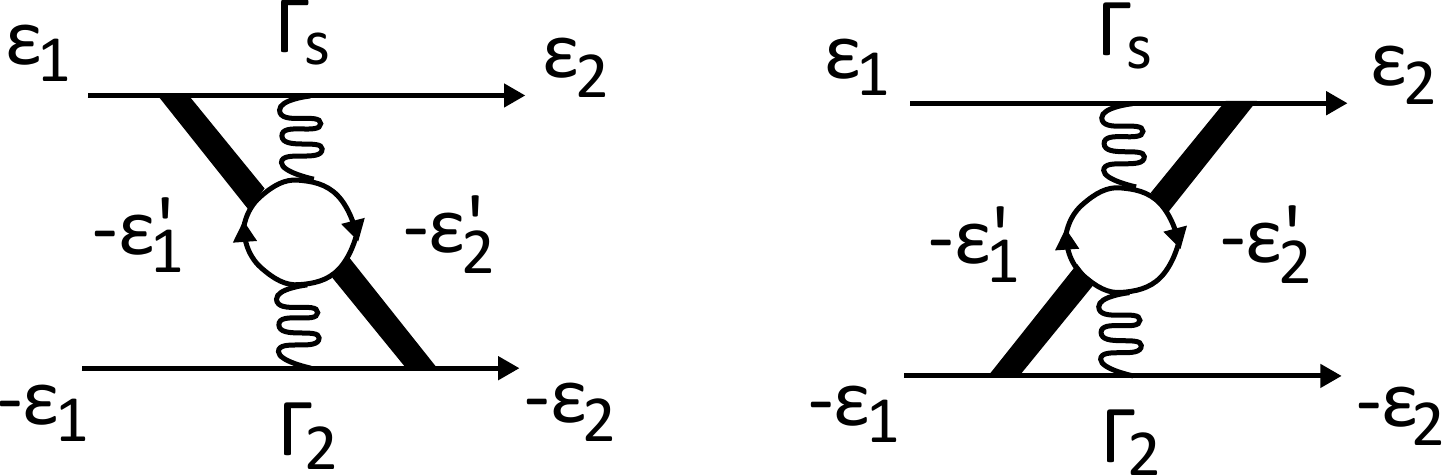}}

\protect\caption{Amplitude efficient for the spin-triplet pairing. $\Gamma_{\mathrm{s}}$ is the amplitude in the spin-singlet Cooper channel, $\Gamma_{2}$ describes the spin-density interaction.  The \emph{e-e} interaction terms are equipped with a Cooperon and a diffuson both depicted by black blocks, but distinguished by the arrow structure.
The Cooperon accompanies the interaction amplitude in the Cooper channel, while the diffuson stands together with the spin-density amplitude.}

\label{fig:diagramCDtext}

\end{figure}

Thus, to the leading order in the \emph{e-e} interactions, the processes that result in the amplitude for the spin-triplet part of the Cooper channel are presented in Fig.~\ref{fig:diagramCDtext} (see Secs. \ref{sectionmixed1} and \ref{sectionmixed2} for
details of the derivation).
Here, one of the \emph{e-e} interaction amplitudes is in the spin-singlet Cooper channel ($\Gamma_{\mathrm{s}}$), while the other one describes the spin-density interaction ($\Gamma_2$). In addition, the \emph{e-e} interaction terms are equipped with a Cooperon and a diffuson. The Cooperon accompanies the amplitude $\Gamma_{\mathrm{s}}$, while the diffuson appears together with the amplitude $\Gamma_2$.
This is in line with the knowledge that in the disordered
systems, fluctuations are enhanced by the soft diffusive modes described by Cooperons and diffusons.

Note that the amplitudes similar to those with the Cooperon and diffuson presented in Fig.~\ref{fig:diagramCDtext} but with two Cooperons or two diffusons are not relevant. This conclusion is also confirmed by our numerical estimates. In other words, these Renormalization Group terms do not contribute as they mostly cancel out in the process of extraction of the odd-frequency dependent terms.

As we demonstrate below, the frequency dependence of the process presented in Fig.~\ref{fig:diagramCDtext} is rather remarkable. We are interested in a case when a sum of incoming frequencies (as well as a sum of outgoing ones) is equal to zero. Next, for fixed incoming frequency $\epsilon_{1}$, we study $V_{\mathrm{odd}}(\epsilon_{1},\epsilon_{2})$ as a function of the outgoing frequency $\epsilon_{2}$ (see Sec. \ref{sectionmixed2} for technical details).
For a  model situation when $\Gamma_{\mathrm{s}}$ is taken to be a constant as a function of frequencies, the dependence of the amplitude $V_{\mathrm{odd}}(\epsilon_{1},\epsilon_{2})$ on the two frequencies $\epsilon_{1}$ and $\epsilon_{2}$
is illustrated in Fig.~\ref{fig:diagramCDfig1}.
There, the obtained frequency dependence reminds that of a $\mathrm{sign}( \epsilon_{2})$ function. In particular, it has a jump at $\epsilon_{2}=0$ and then extends over a large interval of frequencies. The jump at $\epsilon_{2}=0$ persists at finite temperatures.

A note here is in order. The derived effective interaction $V_{\mathrm{odd}}(\epsilon_{1},\epsilon_{2})$ is time-reversal symmetric as it remains invariant under the sign change of both frequencies, $V_{\mathrm{odd}}(\epsilon_{1},\epsilon_{2}) = V_{\mathrm{odd}}(-\epsilon_{1},-\epsilon_{2})$.

Importantly, as it follows from
Fig.~\ref{fig:diagramCDfig1}, the largest of the two frequencies (incoming or outgoing) does not cut off the
interaction amplitude presented in Fig.~\ref{fig:diagramCDtext}. This
picture is very different from the logarithmic integrals commonly encountered in the Renormalization Group approach.
Namely, the integrals determining the odd-frequency amplitude are limited by
the smallest of the two frequencies, or other low-energy cut-offs such as the temperature or the magnetic field.
In contrast to the Renormalization Group, the largest of the two frequencies drops out, and on the ultraviolet side the amplitude $V_{\mathrm{odd}}$ is limited only by the dependence of the \emph{e-e} interactions amplitudes on their frequencies.
This unique property makes the odd-frequency pairing remarkably effective.
Namely, when, as a result of the renormalizations, the amplitude in the singlet channel grows up and becomes of the order one or larger, $\vert \Gamma_{\mathrm{s}} \vert \gtrsim1$, the amplitude in the triplet channel presented in Fig.~\ref{fig:diagramCDtext} exploits this renormalized value $\emph{in the whole interval}$ of its outgoing frequency.
Owing to this very peculiar property, the obtained amplitude $\Gamma_{\mathrm{t}}$ appears to be so efficient.

\begin{widetext}

\begin{figure}[h] \centerline{
\begin{tabular}{ccc}
\includegraphics[width=0.32\textwidth,height=0.15\textheight]{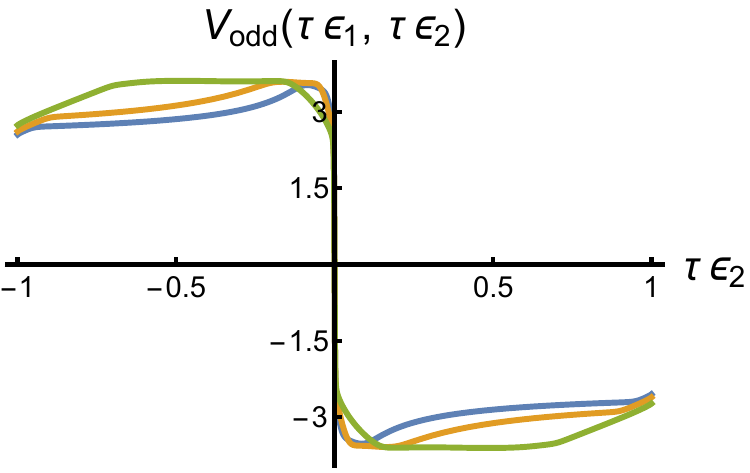}~~&
\includegraphics[width=0.32\textwidth,height=0.15\textheight]{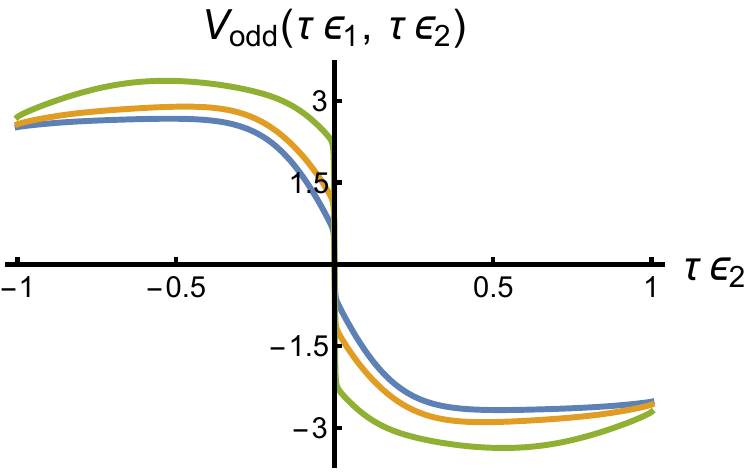}~~&
\includegraphics[width=0.32\textwidth,height=0.15\textheight]{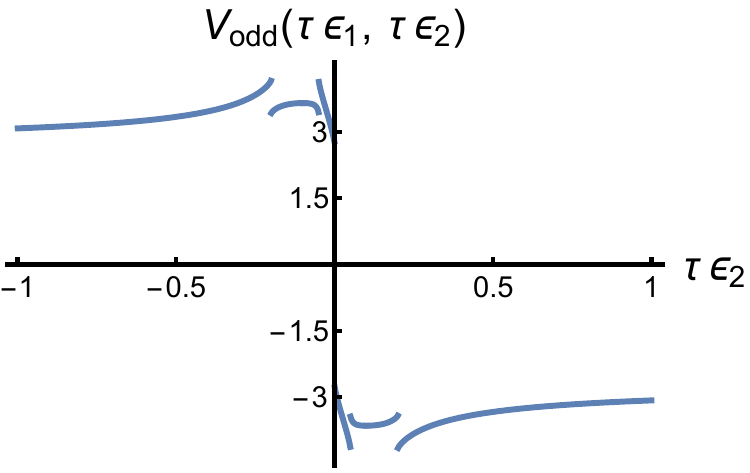}\\
\end{tabular}
}

\protect\caption{Plot of the odd-frequency interaction amplitude $V_{\mathrm{odd}}(\epsilon_{1},\epsilon_{2}) $ in the absence of the magnetic field for: $\epsilon_{1}\tau = 0.05$ (blue), $\epsilon_{1}\tau = 0.1$ (orange), and $\epsilon_{1}\tau = 0.3$ (green); left $T\tau = 0.01$, center $T\tau=0.1$. On the right is a plot of analytical expression of the interaction amplitude for $\epsilon_{1}\tau = 0.1$ at $T=0$, cf. plot on the left. Combination $\rho \vert \Gamma_{\mathrm{s}} \vert \Gamma_2$ is taken to be equal $1$.}

\label{fig:diagramCDfig1}
\end{figure}

\begin{figure}[h] \centerline{
\begin{tabular}{ccc}
\includegraphics[width=0.32\textwidth,height=0.15\textheight]{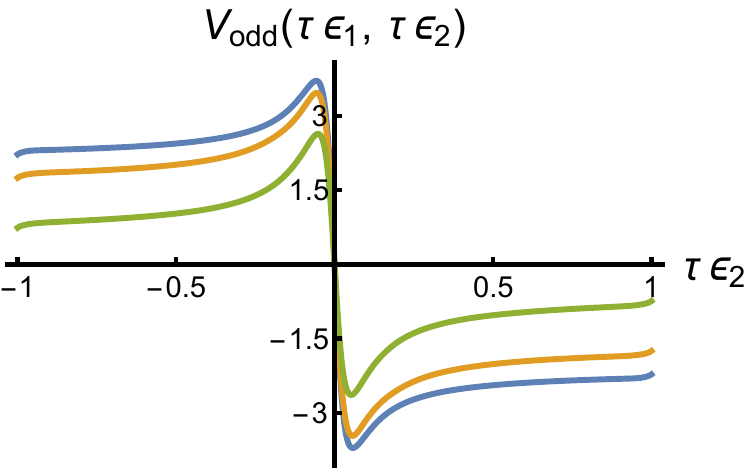}~~&
\includegraphics[width=0.32\textwidth,height=0.15\textheight]{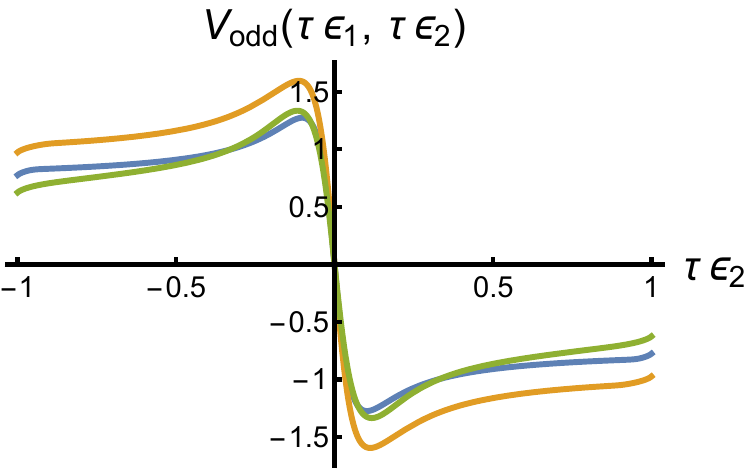}~~&
\includegraphics[width=0.32\textwidth,height=0.15\textheight]{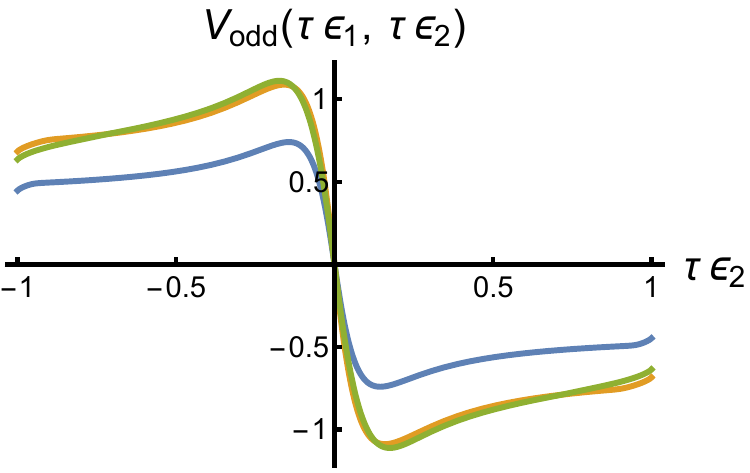}\\
\end{tabular}
}

\protect\caption{Plot of the odd-frequency interaction amplitude $V_{\mathrm{odd}}(\epsilon_{1},\epsilon_{2})$ for $\epsilon_{1}\tau = 0.05$ (blue), $\epsilon_{1}\tau = 0.1$ (orange), and $\epsilon_{1}\tau = 0.3$ (green); $\omega_{\mathrm{c}2}\tau = 0.1$ and the magnetic field is: left $\omega_{\mathrm{c}}\tau = 0.15$, center $\omega_{\mathrm{c}}\tau=0.3$, and right $\omega_{\mathrm{c}}\tau=0.45$. Temperature in all cases is $T\tau = 0.01$. Parameter $4\rho\Gamma_{2}$ is set to be equal $1$.}

\label{fig:diagramCDfig2}
\end{figure}
\end{widetext}

\section{odd-frequency pairing instability}

One can easily see from Eq.~\eqref{equation9} that for a type of the odd-frequency interaction amplitude presented in
Fig.~\ref{fig:diagramCDfig1}, i.e., for an attractive $\Gamma_{\mathrm{s}}$,
the sign of the effective amplitude generated in
Fig.~\ref{fig:diagramCDtext} is in favor of the triplet instability (for more details see Appendix \ref{appendixC}).
Next, because of the extended frequency dependence of the amplitude $V_{\mathrm{odd}}(\epsilon_1,\epsilon_2)$ demonstrated in Fig.~\ref{fig:diagramCDfig1},
we arrive at the situation, which is very similar to the instability in the conventional singlet channel.
There, pairing at low enough temperatures is inevitable as long as the sign of the electron interaction is attractive.
Similarly, in the triplet channel the instability is also inevitable if $V_{\mathrm{s}}<0$.
We suggest that this instability may explain the observed pseudo-gap behavior in the disordered superconducting films.
Since the amplitude $V_{\mathrm{odd}}$ benefits from the divergence of $\Gamma_{\mathrm{s}}$, the pseudo-gap has to develop at temperatures larger than $T_{\mathrm{c}}$, and be larger than the gap in the singlet channel, $\Delta_{\mathrm{s}}$. However, the physical picture, is obscured by the effect of finite temperature and coexistence of two pairings.

In an attempt to elucidate the triplet instability in the triplet channel,
we will turn to the case when the singlet superconductivity is suppressed by the magnetic field, i.e. $H > H_{\mathrm{c}2}$.
Then the lowest temperatures are accessible for the analysis of the triplet odd-frequency instability.
As is well known, orbital motion introduces an energy gap $\omega_{\mathrm{c}}=D\ell_{\mathrm{H}}^{-2}$ into the Cooperon modes, where $\ell_{\mathrm{H}}$ is the cyclotron radius. The orbital effect does not depend on spins and it influences the Cooperons in Fig.~\ref{fig:EquationSingletTriplet} as well as those in Fig.~\ref{fig:diagramCDtext}. Besides, the magnetic field cuts off the effective interaction amplitude $\Gamma_{\mathrm{s}}$.

Now the actual form of the effective amplitude of the interaction in the Cooper channel
$\Gamma_{\mathrm{s}}(\omega, {\bf q}, H, T)$ has to be employed \cite{GalitskiLarkinPRB2001,MichaeliTikhonovFinkelsteinPRB2012} (see Appendix \ref{appendixD} for details).
The results for different magnetic fields are presented in Fig.~\ref{fig:diagramCDfig2}.
A comparison of Figs.~\ref{fig:diagramCDfig1} and \ref{fig:diagramCDfig2} reveals substantial changes in the behavior of the amplitude $V_{\mathrm{odd}}(\epsilon_1,\epsilon_2)$.
The jump disappears at $\epsilon_2=0$ and, instead, a finite slope at small energies evolves with the magnetic field which leads to a suppression of the low-frequency contribution.

\begin{figure}[h]
\centerline{
\includegraphics[width=0.45\textwidth,height=0.27\textheight]{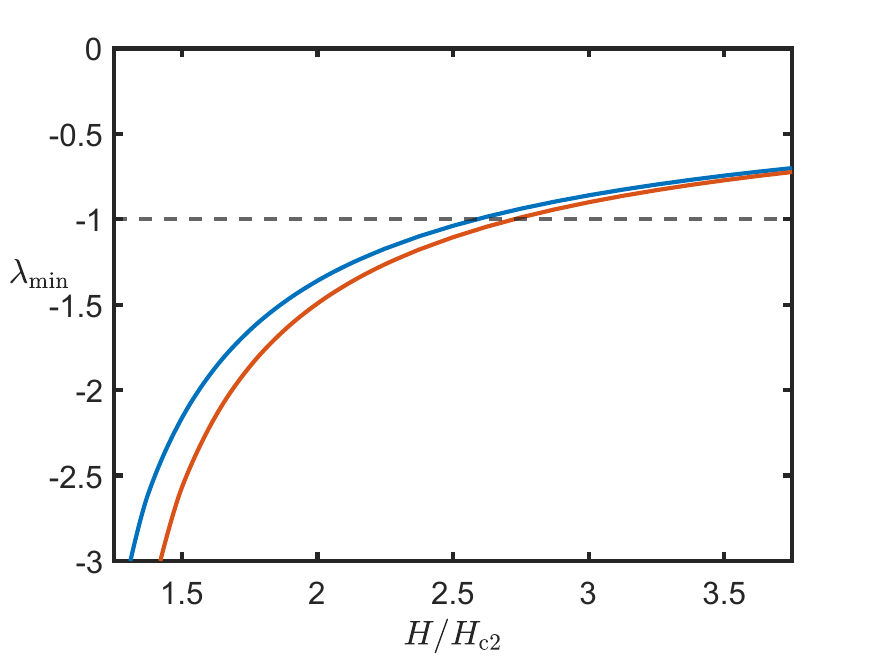}
}
\protect\caption{
Finding the odd-frequency pairing instability which corresponds to $\lambda_{\mathrm{min}}<-1$.
Critical magnetic fields $H_{\mathrm{c}2}$ was chosen to give ${\omega_{\mathrm{c}2}} \tau = 0.1$, and
parameter $\rho\Gamma_{2} = 0.05$. Blue $T\tau = 0.01$, red
$T\tau = 0.001$.
Matrix used for finding the instability is $200 \times 200$.
The instability extends up to the magnetic fields $H_{\mathrm{odd}}\approx 2.5 H_{\mathrm{c}2}$. }

\label{fig:MatLabCD4}

\end{figure}

Furthermore, the overall value of amplitude decreases with the magnetic field.
However, after the initial rapid drop, the further evolution of the amplitude $V_{\mathrm{odd}}(\epsilon_1,\epsilon_2)$ with the magnetic field develops relatively slowly.
This yields a chance to get a rather broad window of magnetic fields for which the odd-frequency pairing instability may occur. We have analyzed the instability by solving the equation of Fig.~\ref{fig:EquationSingletTriplet} in a matrix form.
The minimal eigenvalue $\lambda_{\mathrm{min}}$ of the kernel has to be less than $-1$ in order for the instability to develop.
For the parameters indicated in Fig.~\ref{fig:MatLabCD4}, the instability extends up to the magnetic field $H_{\mathrm{odd}}\approx 2.5 H_{\mathrm{c}2}$.
The plot also shows that for the chosen parameter values the temperature dependence of the instability
is almost saturated.

\section{Discussion}

No reliable mechanism for the spontaneous odd-frequency pairing instability in bulk materials has been demonstrated
so far (for a review see \cite{FominovPRB2015,LinderBalatsky2017}).
In this paper,
we proposed disordered electron liquid as the platform for the odd-frequency pairing.
Starting from the action of the NL$\sigma$M, we have generated a term in the \emph{e-e} interaction, which is relevant for the spin-triplet superconducting channel.
In a disordered system with strong electron momentum scattering by impurities, the $s$-wave odd-frequency pairing is left as the only pairing possibility in the spin triplet channel.

In order to elevate the effectiveness of the pairing, we considered the odd-frequency pairing in the superconducting films when the superconducting phase is destroyed by the magnetic field.
The obtained amplitude of the \emph{e-e} interaction is very peculiar.
The largest of the two external frequencies (i.e., incoming and outgoing) does not cut off automatically the amplitude .
In that sense the picture is very different from the logarithmic integrals encountered in the Renormalization
Group.
As a consequence of this peculiar feature, the development of the triplet pairing instability in the vicinity of the domain of the conventional superconductivity becomes inevitable.

We would like to comment that, paradoxically, while the odd-frequency pairing requires a pronounced time-dependence of the pairing potential, the main obstacle for such pairing is the time-dependent potential itself.
The point here is that the wave function renormalization can be rather strong for the retarded interactions, see e.g., \cite{Eliashberg1960,AbanovChubukovFinkelsteinEPL2001}.
This renormalization suppresses the pairing transition, and can even eliminate it completely \cite{AbrahamsBalatskySchriefferAllenPRB1993}.
This is, however, valid for an ordinary time-dependent potential $V(\epsilon)$, but not for the interaction amplitude $V_{\mathrm{odd}}(\epsilon_1,\epsilon_2)$ generated by the processes presented in Fig.~\ref{fig:diagramCDtext}.
This interaction does not contribute to the wave function renormalization.

There remains ordinary (i.e., not related to $V_{\mathrm{odd}}(\epsilon_1,\epsilon_2)$) wave-function renormalizations, which can be strong enough in a disordered system.
One can be concerned with the effect of renormalizations induced by the interaction $\Gamma_{\mathrm{s}}$ whose large value we exploited in this paper to overcome the smallness of the parameter $\rho$ induced by the disorder.
If this effect were so drastic, these corrections would have also eliminated superconducting transition in the singlet channel.
But by assumption we limited ourselves to the systems in which the $s$-wave superconductivity does develop.

Our calculations were performed in the framework of the Keldysh technique rather than using the Matsubara frequencies.
As a byproduct, this allowed us to avoid complications \cite{SolenovMartinMozirskyPRB2009,KusunoseFuseyaMiyakeJPSJ2011b} with the anomalous function $F$ written in the Matsubara frequencies.

We think that the two-gap structure observed in the disordered superconductors \cite{SacepeNatPhys2011,ChapelierNatPhys2019} can be a manifestation of the two distinct pairing mechanisms, i.e.
spin-singlet and odd-frequency spin-triplet, rather than two limits of the same $s$-wave spin-singlet pairing.
If the latter case was indeed true, one would expect a continuous transition of one gap into another as a function of temperature or of the rate of the disorder, but it does not happen, and the two gaps coexist.
In regards of any other non $s$-wave pairing, any $p$-wave or Fulde-Ferrell-Larkin-Ovchinnikov orders do not survive strong disorder.
Next, in clean systems it is common that if two gaps appear, they correspond to pairing on different parts of the Fermi surface.
Again, due to the impurity scattering all such peculiarities of the Fermi surface are expected to be smeared out in the disordered superconductors. 

Moreover, we believe that the mechanism of the odd-frequency pairing in the regime when the superconducting phase is destroyed by the magnetic field leads to opening of a gap in the spectrum of electrons but does not result in superconductivity.
Then, experimentally observed insulating behavior induced by the magnetic field in the vicinity of the superconducting domain \cite{GantmakherJETPletters2000, ShaharPRL2004, SteinerBoebingerKapitulnikPRL2005, BaturinaPRL2007} can naturally be explained.
At this point, this, of course, is only an educated guess.
The question of the phase coherence for the odd-frequency pairing, which is needed to confirm or disprove our proposition, has so far not received adequate attention,
especially in the presence of the magnetic field.
The phase fluctuations may depend crucially on the interaction amplitude in the spin-triplet pairing channel.
If so, the analysis of the phase fluctuations for the odd-frequency pairing can't be performed in the universal manner.
This question remains open for further studies.

\section{Conclusions}
In this paper we have proposed experimentally reliable and so far elusive mechanism of the \emph{spontaneous} odd-frequency spin-triplet pairing instability.
As is well known, fluctuations appearing at the phase transition lead to strong effects, and can even change the order of the phase transition \cite{LarkinPikin1969, LyuksyutovPokrovskii1975, BrazovskiiDzyaloshinskii1975}. Here, we mixed two fluctuations of different origin to generate an effective interaction amplitude suitable for the s-wave odd-frequency, spin-triplet pairing. Importantly, this analysis was performed for a conventional system without any modelling assumptions. Based on our theoretical calculations, we have made a claim that the odd-frequency paired state could have already been experimentally observed in the two-gap measurements of the disordered film superconductors \cite{SacepeNatPhys2011,ChapelierNatPhys2019}.
There, one of the gaps corresponds to the odd-frequency paired state.
Moreover, we stated that in the experiments where a disordered superconducting thin film is driven to an insulator with an external magnetic field \cite{GantmakherJETPletters2000, ShaharPRL2004, SteinerBoebingerKapitulnikPRL2005, BaturinaPRL2007}, the insulator corresponds to an odd-frequency paired state.

We would like to conclude with a remark that the
mechanism of the odd-frequency instability found in this paper, i.e. the one which develops on the "shoulders"
of a conventional phase transition due to mixing of the fluctuations which necessary accompany the transition, could also be searched for in other physical scenarios and systems, not only in a disordered electron liquid. 

\acknowledgements

We thank Ar. Abanov, Ya. Fominov, V. Mineev, G. Schwiete and K. Tikhonov for their interest to the work and valuable comments. We are especially grateful to M. Feigel'man and B. Sacepe for valuable discussions. VAZ thanks the Department of the Condensed Matter Physics at the Weizmann Institute of Science for hospitality during the summer months of 2017-2019, and especially for the support by a research grant from the Weston Nanophysics Challenge Fund during the Summer of 2019.
VAZ is also thankful to Pirinem School of Theoretical Physics for hospitality.
VAZ is supported by the Russian Foundation for Basic Research (grant No. 20-52-12
013), Deutsche Forschungsgemeinschaft (grant No. EV 30/14-1) cooperation, and by the Foundation for the Advancement of Theoretical Physics and Mathematics BASIS.
Most of this work was fulfilled under the support the U.S. Department
of Energy, Office of Basic Energy Sciences, Division
of Materials Sciences and Engineering under Award DESC0014154.

\appendix
\setcounter{equation}{0}
\setcounter{section}{0}

\begin{widetext}



\section{Technical details of the non-linear sigma model in Keldysh space}\label{appendixA}
\subsection{Keldysh contour and the action}

The Hamiltonian $\hat{H}$ density of non-interacting fermions consists of the kinetic part and the Gaussian distributed disorder $V_{\mathrm{d}}({\bf r})$, namely $\hat{H} = \frac{{\bf k}^2}{2m} + V_{\mathrm{d}}({\bf r})- \mu$, where $\mu$ is the chemical potential, $m$ is the fermion's mass, and ${\bf k}$ is the momentum. We assume the system to be three-dimensional for the sake of generality.
Since we will be looking at the time properties of the Cooper channel, we chose to work in the Keldysh formalism.
Construction of the Keldysh field theory below is along the lines of
\cite{Kamenev, FeigelmanLarkinSkvortsovPRB2000,SchwieteFinkel'steinPRB2014}.
In this formalism, the time dependence is defined on the Keldysh contour $\mathrm{C}$ which consists of a forward and backward parts.
The fermion action on the contour is formally written as
\begin{align}
iS_{0} = i\int_{\mathrm{C}} dt \int_{{\bf r}} {\bar \Psi}(t,{\bf r}) \left( i\tau_{3} \partial_{t}  - \tau_{0} H \right)\Psi(t,{\bf r}).
\end{align}
In its definition we also chose to write the fermion Grassmann fields in spin and time-reversal (Gor'kov-Nambu) spaces, namely
\begin{align}
\Psi = \frac{1}{2}
\left(\begin{array}{c}
\psi_{\uparrow}
\\
\psi_{\downarrow}
\\
\bar{\psi}_{\downarrow}^{\mathrm{T}_{\mathrm{F}}}
\\
-\bar{\psi}_{\uparrow}^{\mathrm{T}_{\mathrm{F}}}
\end{array} \right), ~~
{\bar \Psi} = \frac{1}{2}
\left( {\bar \psi}_{\uparrow},\bar{\psi}_{\downarrow}, -\psi_{\downarrow}^{\mathrm{T}_{\mathrm{F}}},\psi_{\uparrow}^{\mathrm{T}_{\mathrm{F}}} \right),
\end{align}
where the arrows denote real spin of fermions, and Pauli matrices $\tau_{3}$ and $\tau_{0}$ act in the Gor'kov-Nambu space.
Throughout the Supplemental Material matrices $\tau_{0},~\tau_{1},~\tau_{2},~\tau_{3}$, identity and three Pauli matrices correspondingly, will operate in the Gor'kov-Nambu space.
Here $\mathrm{T}_{\mathrm{F}}$ acts only on time, which is denoted by a $\mathrm{F}$ (frequency) abbreviation.
The spinors obey
\begin{align}
\bar{\Psi} = (-i\tau_{1}\sigma_{y} \Psi)^{\mathrm{T}_{\mathrm{FSN}}},
\end{align}
here $\mathrm{T}_{\mathrm{FSN}} $ acts on the spin (S), and Nambu (N) structure and time.
We transform the action in to
\begin{align}
iS_{0} = i\int_{t,{\bf r}} {\hat {\bar \Psi}}(t,{\bf r}) {\hat \sigma}_{3} \left( i\tau_{3}\partial_{t} - \tau_{0}  \hat{H} \right) {\hat \Psi}(t,{\bf r}),
\end{align}
where we expanded the space of the Grassmann fields to accomodate the two Keldysh contour parts (Keldysh space),
\begin{align}
&
{\hat \Psi} = \left( \begin{array}{c}\Psi_{+} \\ \Psi_{-} \end{array}\right), ~~
\hat{\bar{\Psi}} = \left( {\bar \Psi}_{+},~{\bar \Psi}_{-} \right),
\end{align}
 where $\pm$ correspond to the forward/backward contours.
To shorten the notations, we have introduced $\int_{{\bf r}}(..) \equiv \int d{\bf r}(..)$, $\int_{t}(..) \equiv \int_{-\infty}^{+\infty} dt(..)$, and later we will use $\int_{\bf q} (..) = \int \frac {d{\bf q}}{(2\pi)^3}(..)$ and $\int_{\epsilon}(..) \equiv \int \frac{d\epsilon}{2\pi}(..)$. 
These notations will be used throughout the Supplemental Material.
A Pauli matrix $\hat{\sigma}_{3}$, with a hat, acts in the Keldysh space.
Throughout the Supplemental Material the hat symbol will correspond to the Keldysh space only.
Overall, the $\hat{\Psi}$ is a spinor in time-reversal, spin, and Keldysh spaces.
Now, the spinors obey
\begin{align}
\hat{\bar{\Psi}} = (-i\tau_{1}\sigma_{y} \hat{\Psi})^{\mathrm{T}_{\mathrm{FKSN}}},
\end{align}
where now $\mathrm{T} \equiv \mathrm{T}_{\mathrm{FKSN}}$ acts on the whole space.
Here $\sigma_{y}$ is the Puali matrix in spin space.
Throughout the Supplemental Material matrices $\sigma_{0},~\sigma_{x},~\sigma_{y},~\sigma_{z}$, identity and three Pauli matrices, will operate in the spin space.
The Hamiltonian is $\hat{H} = \hat{\sigma}_{0}H$ in case it is the same on the both parts of the Keldysh contour, where $\hat{\sigma}_{0}$ is the identity matrix in Keldysh space. We note that another set of two Keldysh matrices will be introduced later when working with electron-electron interactions.
It is convenient to perform the Keldysh-Larkin-Ovchinnikov rotation with the help of a matrix
\begin{align}
{\hat L} = \frac{1}{\sqrt{2}} \left[ \begin{array} {cc}
1 & -1 \\
1 & 1
\end{array}\right]_{\mathrm{K}}, ~~
{\hat L}^{-1} = \frac{1}{\sqrt{2}} \left[ \begin{array} {cc}
1 & 1 \\
-1 & 1
\end{array}\right]_{\mathrm{K}}.
\end{align}
We then define
\begin{align}\label{rotationNambu}
\hat{\bar{\Phi}}
= \hat{\bar{\Psi}}
\left[ \begin{array} {cc}
\hat{L}^{-1} & 0 \\
0 & \hat{\sigma}_{3}\hat{L}^{-1}
\end{array}\right]_{\mathrm{N}},
~~
\hat{\Phi} =
\left[ \begin{array} {cc}
\hat{L}\hat{\sigma}_{3} & 0 \\
0 & \hat{L}
\end{array}\right]_{\mathrm{N}}
\hat{\Psi},
\end{align}
with a $\hat{\bar{\Psi}}\hat{\sigma_{3}}\hat{\Psi} =\hat{\bar{\Phi}}\hat{\Phi} $ property.
These new spinors also obey the same symmetry relation as the the $\hat{\bar{\Psi}}$ and $\hat{\Psi}$.
To show this, note that $\hat{L}^{-1} = \hat{L}^{\mathrm{T}_{\mathrm{K}}}$, and
\begin{align}
\left[ \begin{array} {cc}
\hat{L}^{-1} & 0 \\
0 & \hat{\sigma}_{3}\hat{L}^{-1}
\end{array}\right]_{\mathrm{N}}^{\mathrm{T}_{\mathrm{KN}}}
\tau_{1}
=
\left[ \begin{array} {cc}
\hat{L} & 0 \\
0 & \hat{L}\hat{\sigma}_{3}
\end{array}\right]_{\mathrm{N}}
\tau_{1}
=
\tau_{1}
\left[ \begin{array} {cc}
\hat{L}\hat{\sigma}_{3} & 0 \\
0 & \hat{L}
\end{array}\right]_{\mathrm{N}}
\end{align}
The Hamiltonian gets rotated as
\begin{align}
\hat{\cal{H}} = \hat{L} \hat{H}\hat{L}^{-1},
\end{align}
and the action becomes,
\begin{align}
iS_{0} = i \int_{t,{\bf r}} {\hat {\bar \Phi}}(t,{\bf r}) \left( i\tau_{3}\partial_{t} - \tau_{0}  \hat{\cal{H}} \right) {\hat \Phi}(t,{\bf r}).
\end{align}
Used above, and which will be used throughout the Supplemental Material, matrix notation
\begin{align}
\left[ \begin{array} {cc}
.. & .. \\
.. & ..
\end{array}\right]_{\mathrm{K}}
\end{align}
will be used to highlight that the matrix is in the Keldysh (K) space.
Similarly, a matrix in spin space will be denoted with $\mathrm{S}$ and in Gor'kov-Nambu space with $\mathrm{N}$ symbol.


\subsection{Disorder average and Hubbard-Stratonovich decoupling}
The disorder potential $V_{\mathrm{d}}({\bf r})$ is assumed a short-range delta-correlated, such that
\begin{align}
\langle V_{\mathrm{d}}({\bf r})V_{\mathrm{d}}({\bf r}^\prime) \rangle 
\equiv 
\int D[V_{\mathrm{d}}({\bf r})] V_{\mathrm{d}}({\bf r})V_{\mathrm{d}}({\bf r}^\prime) e^{-\pi \nu \tau \int_{\bf r} V_{\mathrm{d}}^{2}({\bf r})}  = \frac{1}{2 \pi \nu \tau}\delta({\bf r} - {\bf r}^\prime),
\end{align}
where $\nu$ is the three-dimensional density of fermionic states and $\tau$ is the impurty scattering time.
We then average the action over the disorder,
\begin{align}
\langle e^{iS_{0}}\rangle_{\mathrm{dis}} \equiv \int D[V] e^{-\pi \nu \tau \int_{\bf r} V^{2}({\bf r})} 
e^{-i\int_{t,{\bf r}} V({\bf r})  \hat{\bar{\Phi}}(t,{\bf r}) \hat{\Phi}(t,{\bf r}) }
= e^{-\frac{1}{4 \pi \nu \tau} \int_{\bf r}   \int_{t}\hat{\bar{\Phi}}(t,{\bf r})  \hat{\Phi}(t,{\bf r})   \int_{t^\prime} \hat{\bar{\Phi}}({\bf r},t^\prime)  \hat{\Phi}({\bf r},t^\prime) }.
\end{align}
We need to now decouple the resulting four-fermion term.
It can be shown that a Hubbard-Stratonovich field formed out of same time fermion field operators, will only redefine the chemical potential of fermions.
The Hubbard-Stratonovich field formed out of different time fermion fields is of particular importance. Introduce
\begin{align}
\int D[\hat{Q}] e^{- \frac{\pi \nu}{4\tau} \int_{{\bf r}} \mathrm{Tr}\left[ \hat{Q}({\bf r}) \hat{Q} ({\bf r})  \right] } = 1,
\end{align}
where $\hat{Q}$ is a matrix in spin (S), Gor'kov-Nambu (N), and Keldysh (K) spaces.
 The trace $\mathrm{Tr}$ here is over K, S, N spaces and also over the time, namely $\mathrm{Tr}[\hat{Q}\hat{Q}]=\int_{tt^\prime}\mathrm{Tr}_{\mathrm{KSN}}[\hat{Q}_{tt^\prime}\hat{Q}_{t^\prime t}]$. This notation will be used throughout the Supplemental Material, when working in the time domain. In frequency space, the trace over time turns in to a trace over the frequencies, whose notation will be introduced in the next subsection.
Under the exponent, we transform
\begin{align}
\mathrm{Tr}\left[ \hat{Q}_{tt^\prime}\hat{Q}_{t^\prime t}\right] 
+ \int_{tt^\prime}\frac{1}{\pi^2 \nu^2} \left[ \hat{\bar{\Phi}}(t)  \hat{\Phi}(t) \right] \left[ \hat{\bar{\Phi}}( t^\prime )  \hat{\Phi}(t^\prime) \right]
\rightarrow 
\mathrm{Tr}\left[ \hat{Q}_{tt^\prime}\hat{Q}_{t^\prime t}\right] 
- \int_{tt^\prime}\frac{2}{\pi \nu}  \hat{\bar{\Phi}}(t) \hat{Q}_{tt^\prime} \hat{\Phi}(t^\prime),
\end{align}
where in deriving it is useful to present $\hat{\bar{\Phi}}( t )  \hat{\Phi}(t)  = \sum_{n} \hat{\bar{\Phi}}^{(n)}(t)  \hat{\Phi}^{(n)}(t)$, where $n$ denotes spinor's element, and do the same for
$\mathrm{Tr}\hat{Q}_{tt^\prime}\hat{Q}_{t^\prime t} = \sum_{nm}\int_{tt^\prime}\hat{Q}^{nm}_{tt^\prime}\hat{Q}^{mn}_{t^\prime t}$ product.
Elements of the $\hat{Q}_{tt^\prime}$ matrix are in accord with the product of the spinors it decomposed, namely $\hat{Q}_{tt^\prime} \propto \Phi(t)\bar{\Phi}(t^\prime) $.


\subsection{Non-interacting action}\label{section3}
After we integrate the Grassmann fields out, we obtain an action for the ${\hat Q}$ matrix,
\begin{align}
iS_{0} = -\frac{\pi \nu}{8\tau}\int_{{\bf r}} \mathrm{Tr} [ \hat{Q}^2 ] + \frac{1}{2}\int_{{\bf r}} \mathrm{Tr} \ln [ \hat{G}^{-1} + \frac{i}{2\tau}\hat{Q}  ]
\end{align}
\begin{align}
\hat{Q}_{tt^\prime}({\bf r})  =  \frac{i}{\pi \nu}  [ \hat{G}^{-1} + \frac{i}{2\tau}\hat{Q}  ]^{-1}_{tt^{\prime},{\bf r}},
\end{align}
where $\hat{G}^{-1}$ is the inversed of the Green function of the system. Here, recall, $\mathrm{Tr}$ also stands for the trace over time variables as introduced in the previous subsection.
Because of the symmetry the $\hat{Q}$ matrix obeys, $\hat{Q} = \tau_{1} \sigma_{y} \hat{Q}^{\mathrm{T}} \sigma_{y}\tau_{1}$ (because of the chosen time-reversal space of the spinors, and due to the extra minus sign when interchanging the Grassmann fields when transposing the product of two Grassmann fields),
one gets the saddle-point solution
\begin{align}\label{noninteracting}
\hat{Q}_{tt^\prime} =
\left[\begin{array}{cc} \hat{\Lambda}(t-t^\prime) & 0  \\
0 & \hat{\Lambda}^{\mathrm{T}}(t-t^\prime)\end{array}\right]_{\mathrm{N}}
\equiv
\hat{\Lambda}^{[\mathrm{N}]}(t-t^\prime),
\end{align}
where in frequency domain,
\begin{align}
\hat{\Lambda}_{\epsilon}
= \left[ \begin{array}{cc}
\Lambda^{\mathrm{R}}_{\epsilon} & \Lambda^{\mathrm{K}}_{\epsilon} \\
0 & \Lambda^{\mathrm{A}}_{\epsilon}
\end{array}\right]_{\mathrm{K}}
=
\left[ \begin{array}{cc}
1 & 2\cal{F}_{\epsilon} \\
0 & -1
\end{array}\right]_{\mathrm{K}},
\end{align}
where ${\cal F}_{\epsilon} = \tanh\left( \frac{\epsilon}{2T} \right)$ is fermionic distribution function at equilibrium.
Gradient expansion around the saddle-point gives the following action describing fluctuations,
\begin{align}
iS_{0} = -\frac{\pi \nu }{8}
\int_{{\bf r}} 
\mathrm{Tr} \{ D[ \nabla \hat{Q}({\bf r}) ]^2
- 4\tau_{3}\partial_{t}\hat{Q}({\bf r}) \},
\end{align}
where $D = \frac{1}{3}v_{\mathrm{F}}^2\tau$ is the diffusion coefficient.
We note that the diffusion coefficient is that of a three-dimensional system.
In the Main Text a limit of thick film is taken.
Fourier convention is
\begin{align}
\hat{Q}_{\epsilon\epsilon^{\prime}}({\bf r}) = \int_{tt^\prime} \hat{Q}_{tt^\prime}({\bf r}) e^{i\epsilon t - i\epsilon^\prime t^\prime}.
\end{align}
The action is rewritten as
\begin{align}
iS_{0}
& 
= -\frac{\pi \nu }{8}
 \mathrm{Tr}_{\mathrm{KSN}}
\int_{{\bf r}}
 \int_{\epsilon\epsilon^\prime}
\{ D[ \nabla \hat{Q}_{\epsilon \epsilon^\prime}({\bf r}) ][ \nabla \hat{Q}_{\epsilon^\prime \epsilon}({\bf r}) ]
+ 4i\tau_{3}\hat{\epsilon}\hat{Q}_{\epsilon \epsilon^\prime}({\bf r})\delta_{\epsilon,\epsilon^{\prime}} \}
\\
&
\equiv
-\frac{\pi \nu }{8}
 \mathrm{Tr}
\int_{{\bf r}}
\{ D[ \nabla \hat{Q}({\bf r}) ]^2
+ 4i\tau_{3}\hat{\varepsilon}\hat{Q}({\bf r}) \}
\end{align}
where $\delta_{\epsilon,\epsilon^{\prime}} = 2\pi \delta(\epsilon - \epsilon^{\prime})$. Here we have introduced a trace over frequencies, $\mathrm{Tr}_{\mathrm{F}}$. Together with it, we have also introduced a notation $\mathrm{Tr}$ for the trace over the full space, namely $\mathrm{Tr} \equiv \mathrm{Tr}_{\mathrm{FKSN}}$ (note that this same notation was also used in the time domain). For example, for some operator $\hat{Y}_{\epsilon\epsilon^\prime}$ the trace over frequencies reads as $\mathrm{Tr}_{\mathrm{F}} \hat{Y} = \int_{\epsilon}\hat{Y}_{\epsilon\epsilon}$. For two operators, $\mathrm{Tr}_{\mathrm{F}}\left[ A_{\epsilon_{1}\epsilon_{2}}Y_{\epsilon_{2}\epsilon_{1}}\right] \equiv \int_{\epsilon_{1}\epsilon_{2}}\left[ A_{\epsilon_{1}\epsilon_{2}}Y_{\epsilon_{2}\epsilon_{1}}\right]$. Sometimes we will be switching from $\mathrm{Tr}_{\mathrm{F}}[..]$ to $\int_{\epsilon}[..]$ or vice versa.
We will be using this $\mathrm{Tr}_{\mathrm{F}}[..]$ notation throughout the rest of the Supplemental Material.
When introducing the trace over frequencies, we had to also introduce a $\hat{\varepsilon}\hat{Q}_{\epsilon \epsilon^\prime} = \epsilon\hat{Q}_{\epsilon \epsilon^\prime}$ operation.


\subsection{Interactions}
Now let us add electron-electron interactions.
Here we will follow lines of Refs. \cite{FinkelsteinReview1990,SchwieteFinkel'steinPRB2014} when including the interactions in to the theory.
\begin{align}
H_{\mathrm{int}}
&= \frac{1}{2} \int_{{\bf r}{\bf r}^{\prime}} \rho(t,{\bf r}) \tilde{V}_{\rho}({\bf r} - {\bf r}^{\prime}) \rho(t,{\bf r}^\prime)
- 2V_{\sigma}  \int_{{\bf r}} {\bf s}(t,{\bf r})  {\bf s}(t,{\bf r})
+ \frac{V_{\mathrm{s}}}{\nu} \sum_{\alpha \neq \beta} \int_{{\bf r}}
\bar{\Delta}_{\alpha\beta}(t,{\bf r})
\Delta_{\beta\alpha}(t,{\bf r}),
\end{align}
where
\begin{align}
&
\rho(t,{\bf r}) = \sum_{\alpha} \bar{\Psi}_{\alpha}(t,{\bf r})\Psi_{\alpha}(t,{\bf r}),
\\
&
{\bf s}(t,{\bf r}) = \frac{1}{2} \sum_{\alpha\beta} \bar{\Psi}_{\alpha}(t,{\bf r}){\bm \sigma}_{\alpha \beta}\Psi_{\beta}(t,{\bf r}),
\\
&
\Delta_{{\bf q},\beta\alpha}(t) = \sum_{\bf k} \Psi_{{\bf k}+{\bf q}\beta}(t)\Psi_{-{\bf k}\alpha}(t).
\end{align}
where $\rho(t,{\bf r})$ and ${\bf s}(t,{\bf r})$ are the charge and spin densities correspondingly, and $\Delta_{{\bf q},\beta\alpha}(t)$ is the spin-singlet Cooper pair density.
The part of the action corresponding to the interactions reads (we outline it for the sake of sign bookkeeping)
\begin{align}
e^{iS_{\mathrm{int}}}
= e^{-i\int_{\mathrm{C}}dt H_{\mathrm{int}}[\bar{\Psi},\Psi]}
= e^{-i\int_{t}dt \left( H_{\mathrm{int}}[\bar{\Psi}_{+},\Psi_{+}] - H_{\mathrm{int}}[\bar{\Psi}_{-},\Psi_{-}] \right) },
\end{align}
where in the second equality sign we have split the action in to forward and backward parts of the Keldysh contour.
The two contours do not get coupled because of the local in time interaction.
Hubbard-Stratonovich decoupling of the charge part of the interaction goes as,
\begin{align}
&
\frac{1}{2}\left[ \theta_{\pm}(t,{\bf r}^\prime) +   \tilde{V}_{\rho}({\bf r} - {\bf r}^{\prime})\rho_{\pm}(t,{\bf r}^\prime) \right]
\tilde{V}^{-1}_{\rho}({\bf r} - {\bf r}^{\prime})
\left[ \theta_{\pm}(t,{\bf r}) +  \tilde{V}_{\rho}({\bf r} - {\bf r}^{\prime}) \rho_{\pm}(t,{\bf r}) \right]
-\frac{1}{2}  \rho_{\pm}(t,{\bf r}) \tilde{V}_{\rho}({\bf r} - {\bf r}^{\prime}) \rho_{\pm}(t,{\bf r}^\prime)
\\
&
=
\frac{1}{2}\theta_{\pm}(t,{\bf r}^\prime) \tilde{V}^{-1}_{\rho}({\bf r} - {\bf r}^{\prime}) \theta_{\pm}(t,{\bf r})
 + \theta_{\pm}(t,{\bf r}) \rho_{\pm}(t,{\bf r}),
\end{align}
where $\rho_{\pm}(t,{\bf r}) =  \sum_{\alpha} \bar{\Psi}_{\pm;\alpha}(t,{\bf r})\Psi_{\pm;\alpha}(t,{\bf r})$ is the fermion's density on the $\pm$ contour.
We have introduced a short notation for summation over the contour parts, for example $\theta_{\pm}(t,{\bf r}) \rho_{\pm}(t,{\bf r}) \equiv \theta_{+}(t,{\bf r}) \rho_{+}(t,{\bf r}) + \theta_{-}(t,{\bf r}) \rho_{-}(t,{\bf r})$.
This notation will be used throughout the Supplemental Material.
Same Hubbard-Stratonovich transformation applies to the spin sector,
\begin{align}
&
-\frac{1}{2}\left[ \vec{\theta}_{\pm}(t,{\bf r}) +  2V_{\sigma} {\bf s}_{\pm}(t,{\bf r}^\prime) \right]
V^{-1}_{\sigma}
\left[ \vec{\theta}_{\pm}(t,{\bf r}) +  2V_{\sigma} {\bf s}_{\pm}(t,{\bf r}) \right]
+ 2 {\bf s}_{\pm}(t,{\bf r}) V_{\sigma} {\bf s}_{\pm}(t,{\bf r})
\\
&
=
-\frac{1}{2}\vec{\theta}_{\pm}(t,{\bf r}) V^{-1}_{\sigma} \vec{\theta}_{\pm}(t,{\bf r})
- 2\vec{\theta}_{\pm}(t,{\bf r}){\bf s}_{\pm}(t,{\bf r}).
\end{align}
In the Cooper channel we make a similar transformation,
\begin{align}
&
\left[ \bar{\theta}^{\mathrm{c}}_{\pm}(t,{\bf r}) +  \frac{V_{\mathrm{s}}}{\nu} \bar{\Delta}_{\pm,\uparrow\downarrow}(t,{\bf r}) \right]
\frac{\nu}{V_{\mathrm{s}}}
\left[ \theta^{\mathrm{c}}_{\pm}(t,{\bf r}) + \frac{V_{\mathrm{s}}}{\nu} \Delta_{\pm,\downarrow\uparrow}(t,{\bf r}) \right]
- 
\bar{\Delta}_{\pm,\uparrow\downarrow}(t,{\bf r})  \frac{V_{\mathrm{s}}}{\nu} \Delta_{\pm,\downarrow\uparrow}(t,{\bf r})
\\
&
=
 \bar{\theta}^{\mathrm{c}}_{\pm}(t,{\bf r}) \frac{\nu}{V_{\mathrm{s}}}\theta^{\mathrm{c}}_{\pm}(t,{\bf r})
+ \bar{\Delta}_{\pm,\uparrow\downarrow}(t,{\bf r})\theta^{\mathrm{c}}_{\pm}(t,{\bf r})
+ \bar{\theta}^{\mathrm{c}}_{\pm}(t,{\bf r})\Delta_{\pm,\downarrow\uparrow}(t,{\bf r}).
\end{align}

We split the electron-electron interaction action in to two parts, $iS_{\mathrm{int}} = iS_{\mathrm{int},1}+ iS_{\mathrm{int},2}$, where $iS_{\mathrm{int},1}$ describes coupling of the Grassmann operators with the Hubbard-Stratonovich fields, and $iS_{\mathrm{int},2}$ contains only the Hubbard-Stratonovich fields.
Overall, after splitting in to $\pm$ Keldysh contours and performing Keldysh-Larkin-Ovchinnikov rotation, the $ iS_{\mathrm{int},1}$ becomes
\begin{align}
iS_{\mathrm{int},1} = 
i\int_{t,{\bf r}}\hat{{\bar \Phi}}(t,{\bf r})
\left[ \begin{array} {cc}
  \hat{\theta} + \hat{\vec{\theta}}\vec{\sigma}  &
\hat{\theta}^{\mathrm{c}} \\
-\hat{\bar{\theta}}^{\mathrm{c}} &  \hat{\theta}^{\mathrm{T}} + \sigma_{y}(\hat{\vec{\theta}}\vec{\sigma})^{\mathrm{T}}\sigma_{y}
\end{array}\right]_{\mathrm{N}}
\hat{\Phi}(t,{\bf r})
\equiv
- i\int_{t,{\bf r}}\hat{{\bar \Phi}}(t,{\bf r})
\hat{H}_{\mathrm{int},\theta}
\hat{\Phi}(t,{\bf r}) ,
\end{align}
where, we repeat, we defined $ \hat{\bar{\Phi}} = \hat{\bar{\Psi}}\hat{L}^{-1}$ and $\hat{\Phi} = \hat{L}\hat{\sigma_{3}}\hat{\Psi}$, with a property $\hat{\bar{\Psi}}\hat{\sigma_{3}}\hat{\Psi} =\hat{\bar{\Phi}}\hat{\Phi}$.
Note, the way the Gor'kov-Nambu space is organized due to the time-reversal space properties.
Each Hubbard-Stratonovich field has a structure in Keldysh space. \begin{align}
&
\hat{\theta} = \hat{\gamma}^{1}\theta_{\mathrm{cl}} + \hat{\gamma}^{2}\theta_{\mathrm{q}}, \\ &
\hat{\theta}^{\mathrm{c}} = \hat{\gamma}^{1}\theta_{\mathrm{q}}^{\mathrm{c}} + \hat{\gamma}^{2}\theta_{\mathrm{cl}}^{\mathrm{c}},
\end{align} 
where $\theta_{\mathrm{cl}/\mathrm{q}} =\frac{1}{2}\left( \theta_{+} \pm \theta_{-} \right)$ are classical (cl) and quantum (q) components correspondingly. Same definitions apply to $\theta^{\mathrm{c}}$ fields.
The difference in the way diagonal and off-diagonal in Gor'kov-Nambu space fields rotate is due to the specifics of the rotation Eq. (\ref{rotationNambu}).
Here and throughout the Supplemental Material a new set of matrices acting in Keldysh space is introduced,
\begin{align}
\hat{\gamma}^{1} = \left[\begin{array}{cc} 1 & 0 \\ 0 & 1 \end{array}\right]_{\mathrm{K}}, ~~
\hat{\gamma}^{2} = \left[\begin{array}{cc} 0 & 1 \\ 1 & 0 \end{array}\right]_{\mathrm{K}}.
\end{align}
These matrices will only appear in the action describing the electron-electron interaction, thus distinct from existing $\hat{\sigma}_{0}$ and $\hat{\sigma}_{3}$ notations.
The spin structure of the Cooper channel interaction field is
\begin{align}
\hat{\theta}^{\mathrm{c}}_{\alpha\beta} =
\hat{\theta}^{\mathrm{c}} 
\left[ \begin{array}{cc}
 1& 0 \\
0 & 1
 \end{array} \right]_{\mathrm{S}} = \hat{\theta}^{\mathrm{c}} \sigma_{0},
\end{align}
and the same for the $\hat{\bar{\theta}}^{\mathrm{c}}$ field.

Action describing the Hubbard-Stratonovich, $ iS_{\mathrm{int},2}$, fields is
\begin{align}
iS_{\mathrm{int},2}
&=
\frac{i}{2} \int_{t,{\bf r}{\bf r}^\prime} \mathrm{Tr}_{\mathrm{KSN}}
\left[\hat{\theta}({\bf r}^\prime,t) \tilde{V}^{-1}_{\rho}({\bf r} - {\bf r}^{\prime})\hat{\gamma}^{2} \hat{\theta}(t,{\bf r}) \right]
\\
&
-\frac{i}{2} \int_{t,{\bf r}} \mathrm{Tr}_{\mathrm{KSN}}\left[
 \hat{\vec{\theta}}(t,{\bf r}) V^{-1}_{\sigma} \hat{\gamma}^{2} \hat{\vec{\theta}}(t,{\bf r}) \right]
\\
&
+ i  \int_{t,{\bf r}} \mathrm{Tr}_{\mathrm{KSN}}\left[ \hat{\bar{\theta}}^{\mathrm{c}}(t,{\bf r}) \frac{\nu}{V_{\mathrm{s}}} \hat{\gamma}^{2}
\hat{\theta}^{\mathrm{c}}(t,{\bf r}) \right].
\end{align}

After integrating fermions out, one gets an action
\begin{align}
 iS_{0}+iS_{\mathrm{int},1}
= -\frac{\pi \nu}{8\tau} \int_{{\bf r}} \mathrm{Tr}  [ \hat{Q}^2 ] + \frac{1}{2} \int_{{\bf r}} \mathrm{Tr} \ln [ \hat{G}^{-1} + \frac{i}{2\tau}\hat{Q} - \hat{H}_{\mathrm{int},\theta}  ] .
\end{align}
It is not clear now how to find the saddle-point of the action containing the interaction fields. We then take a perturbative route in which, first, the non-interacting saddle-point is derived and we know everything about the fluctuations around it (see Sec. \ref{section3}).
Next, the interactions are assumed as perturbations to the saddle-point, and one studies them by expanding the logarithm,
\begin{align}
 iS_{0}+iS_{\mathrm{int},1} =
&-\frac{\pi \nu }{8} \int_{{\bf r}}  \mathrm{Tr} \{ D[ \nabla \hat{Q}({\bf r}) ]^2 - 4\tau_{3}\partial_{t}\hat{Q}({\bf r}) \}
\\
&
- \frac{i\pi \nu}{2} \int_{{\bf r}}  \mathrm{Tr}  [ \underline{\hat{Q}({\bf r})}\hat{H}_{\mathrm{int},\theta}({\bf r}) ],
\end{align}
here by underscored $\underline{\hat{Q}}$ matrix we mean fluctuations around the saddle-point in Keldysh and Gor'kov-Nambu spaces.
It will not be important for the transformation which will be made in this section.
We will come back to this definition in the next section.

Now one has to integrate all interaction fields out and obtain an effective action for the $\hat{Q}$ matrix.
We will be using an identity based on Gaussian integration where for arbitrary vector $\hat{\chi}$, $\hat{\bar{\chi}}$, one writes a path integral over vectors $\hat{\bar{\phi}}$ and $\hat{\phi}$,
\begin{align}
\frac{1}{N_{\mathrm{A}}}\int D[\hat{\bar{\phi}},\hat{\phi}] e^{-\hat{\bar{\phi}}\hat{A}\hat{\phi}} e^{\hat{\bar{\phi}}\hat{\chi} + \hat{\bar{\chi}} \hat{\phi}}
= \frac{1}{N_{\mathrm{A}}} N_{\mathrm{A}} e^{ \hat{\bar{\chi}} \hat{A}^{-1} \hat{\chi} }
= e^{ \hat{\bar{\chi}} \hat{A}^{-1} \hat{\chi} },
\end{align}
where $N_{\mathrm{A}}$ is the normalization factor. Summation over all indexes was assumed in deriving the identity.

We outline the steps for the interaction in the charge sector.
Hamiltonian in the charge sector $\hat{H}_{\mathrm{int},\rho}$ (the one containing $\hat{\theta}$ fields) of the $\hat{H}_{\mathrm{int},\theta}$ Hamiltonian is
\begin{align}
-\frac{i\pi\nu}{2} \mathrm{Tr}_{\mathrm{KSN}} [ \underline{\hat{Q}}\hat{H}_{\mathrm{int},\rho} ]
= -i\pi\nu  \mathrm{Tr}_{\mathrm{KSN}} [ \underline{\hat{Q}}\tau_{+} \hat{\gamma}^{1} ] \theta^{\mathrm{cl}}
-i\pi\nu  \mathrm{Tr}_{\mathrm{KSN}} [ \underline{\hat{Q}}\tau_{+} \hat{\gamma}^{2} ] \theta^{\mathrm{q}}.
\end{align}
The action describing Hubbard-Stratonovich fields in charge sector is
\begin{align}
i\frac{1}{2} \tilde{V}_{\rho}^{-1} \mathrm{Tr}_{\mathrm{KSN}}[ \hat{\theta} \hat{\gamma}^{2} \hat{\theta}]
= 2i\tilde{V}_{\rho}^{-1}  \theta^{\mathrm{cl}}\theta^{\mathrm{q}}.
\end{align}
Summing the two and making a shift in the Hubbard-Stratonovich fields, we get
\begin{align}
&
-\frac{i\pi\nu}{2} \mathrm{Tr}_{\mathrm{KSN}} [ \underline{\hat{Q}}\hat{H}_{\mathrm{int},\rho} ]
+i\frac{1}{2} \tilde{V}_{\rho}^{-1} \mathrm{Tr}_{\mathrm{KSN}}[ \hat{\theta} \hat{\gamma}^{2} \hat{\theta}]
\\
&
=
2i \tilde{V}_{\rho}^{-1}  ( \theta^{\mathrm{cl}} -
\frac{1}{2} \tilde{V}_{\rho} \pi\nu
\mathrm{Tr}_{\mathrm{KSN}}[ \underline{\hat{Q}}\tau_{+} \hat{\gamma}^{2} ]  )
  (\theta^{\mathrm{q}}  - \frac{1}{2} \tilde{V}_{\rho} \pi\nu
\mathrm{Tr}_{\mathrm{KSN}}[ \underline{\hat{Q}}\tau_{+} \hat{\gamma}^{1} ])
-
i\frac{\pi^2\nu^2}{2} \tilde{V}_{\rho}
\mathrm{Tr}_{\mathrm{KSN}}[ \underline{\hat{Q}}\tau_{+} \hat{\gamma}^{2} ]
\mathrm{Tr}_{\mathrm{KSN}}[ \underline{\hat{Q}}\tau_{+} \hat{\gamma}^{1} ]
\\
&
=
2i \tilde{V}_{\rho}^{-1}
\tilde{\theta}^{\mathrm{cl}}
\tilde{\theta}^{\mathrm{q}}
-
i\frac{\pi^2\nu^2}{2} \tilde{V}_{\rho}
\mathrm{Tr}_{\mathrm{KSN}}[ \underline{\hat{Q}}\tau_{+} \hat{\gamma}^{2} ]
\mathrm{Tr}_{\mathrm{KSN}}[ \underline{\hat{Q}}\tau_{+} \hat{\gamma}^{1} ].
\end{align}
After integrating $\tilde{\theta}^{\mathrm{cl}} $ and $\tilde{\theta}^{\mathrm{q}} $ fields out, we will obtain action describing rescattering of charge parts of the $\hat{Q}$ matrices.

Generalizing to all combinations of matrices in Keldysh and Gor'kov-Nambu spaces, performing the same transformation as above for remaining spin density and Cooper parts of the interaction, we finally obtain the action $iS_{\mathrm{ee}}$ for the rescattering between different blocks of the $\hat{Q}$ matrix,
\begin{align}\label{newsumnotation}
iS_{\mathrm{ee}} =
&
-  i \frac{\pi^2 \nu^2}{8}
\int_{t,{\bf r} {\bf r}^\prime}
\tilde{V}_{\rho}({\bf r} - {\bf r}^\prime)
\mathrm{Tr}_{\mathrm{KSN}}[ \hat{\gamma}^{1/2} \tau_{\pm} \underline{\hat{Q}_{tt}({\bf r})} ]
\mathrm{Tr}_{\mathrm{KSN}}[ \hat{\gamma}^{2/1} \tau_{\pm} \underline{\hat{Q}_{tt}({\bf r}^\prime)} ]
\\
&
+ i\frac{\pi^2\nu^2}{8}
V_{\sigma}
\int_{t,{\bf r}}
\mathrm{Tr}_{\mathrm{KSN}}[ \hat{\gamma}^{1/2} \tau_{\pm} \bm{\sigma} \underline{\hat{Q}_{tt}({\bf r})} ]
\mathrm{Tr}_{\mathrm{KSN}}[ \hat{\gamma}^{2/1} \tau_{\pm} \bm{\sigma}  \underline{\hat{Q}_{tt}({\bf r})} ]
\label{spininteraction}
\\
&
+ i\frac{\pi^2\nu}{8}
V_{\mathrm{s}}
\int_{t,{\bf r}}
\mathrm{Tr}_{\mathrm{KSN}}[ \hat{\gamma}^{1/2} \tau^{\pm} \underline{\hat{Q}_{tt}({\bf r})} ]
\mathrm{Tr}_{\mathrm{KSN}}[ \hat{\gamma}^{2/1} \tau^{\mp}  \underline{\hat{Q}_{tt}({\bf r})} ],
\end{align}
where $\tau_{\pm} = \frac{1}{2}(\tau_{0} \pm \tau_{z})$ and $\tau^{\pm} = \frac{1}{2}(\tau_{x} \pm i\tau_{y})$.
Here we introduced summation notation $\hat{\gamma}^{1/2}\hat{A}\hat{\gamma}^{2/1}\hat{B} = \hat{\gamma}^{1}\hat{A}\hat{\gamma}^{2}\hat{B} + \hat{\gamma}^{2}\hat{A}\hat{\gamma}^{1}\hat{B}$, and the same for the Gor'kov-Nambu space, $\tau^{\pm}\hat{A} \tau^{\mp}\hat{B} = \tau^{+}\hat{A}\tau^{-}\hat{B} + \tau^{-}\hat{A}\tau^{+}\hat{B}$. Summations over Keldysh and Gor'kov-Nambu spaces, as in the expressions above, are independent of each other. Overall, there are four terms in the
$\mathrm{Tr}_{\mathrm{KN}}[ \hat{\gamma}^{1/2} \tau^{\pm} .. ]
\mathrm{Tr}_{\mathrm{KN}}[ \hat{\gamma}^{2/1} \tau^{\mp}  .. ]$ type sum.
In the line (\ref{spininteraction}), the product of the spin matrices reads as
\begin{align}
\mathrm{Tr}_{\mathrm{S}}[\bm{\sigma}..]\mathrm{Tr}_{\mathrm{S}}[\bm{\sigma}..]
=\mathrm{Tr}_{\mathrm{S}}[\sigma_{x}..]\mathrm{Tr}_{\mathrm{S}}[\sigma_{x}..]
+\mathrm{Tr}_{\mathrm{S}}[\sigma_{y}..]\mathrm{Tr}_{\mathrm{S}}[\sigma_{y}..]
+\mathrm{Tr}_{\mathrm{S}}[\sigma_{z}..]\mathrm{Tr}_{\mathrm{S}}[\sigma_{z}..].
\end{align}
Such notations will be used throughout the Supplemental Material to facilitate readability of the expressions.

Picking only the short-range parts of the interaction, $V_{\rho}({\bf r}) = V_{0,\rho}({\bf r}) + V_{1,\rho}\delta({\bf r})$, and recalling $\tilde{V}_{\rho}({\bf r}) = 2V_{\rho}({\bf r}) - V_{\sigma}$ we introduce
\begin{align}
&
Z = \nu(2V_{1,\rho} - V_{\sigma}) \equiv 2\Gamma_{1} - \Gamma_{2},
\\
&
\Gamma_{2} = \nu V_{\sigma},
\\
&
\Gamma_{0}({\bf r}) = 2\nu V_{0,\rho}({\bf r}).
\end{align}
We will omit the $\Gamma_{0}({\bf r})$ in the following.
We then rewrite the interaction
\begin{align}
iS_{\mathrm{ee}} \rightarrow iS_{\mathrm{ee}} &\equiv
-  i \frac{\pi^2 \nu}{8} Z
\int_{t,{\bf r}}
\mathrm{Tr}_{\mathrm{KSN}}[ \hat{\gamma}^{1/2} \tau_{\pm} \underline{\hat{Q}_{tt}({\bf r})} ]
\mathrm{Tr}_{\mathrm{KSN}}[ \hat{\gamma}^{2/1} \tau_{\pm} \underline{\hat{Q}_{tt}({\bf r})} ]
\\
&
+ i\frac{\pi^2\nu}{8} \Gamma_{2}
\int_{t,{\bf r}}
\mathrm{Tr}_{\mathrm{KSN}}[ \hat{\gamma}^{1/2} \tau_{\pm} \bm{\sigma} \underline{\hat{Q}_{tt}({\bf r})} ]
\mathrm{Tr}_{\mathrm{KSN}}[ \hat{\gamma}^{2/1} \tau_{\pm} \bm{\sigma} \underline{\hat{Q}_{tt}({\bf r})} ]
\\
&
+ i\frac{\pi^2\nu}{8} V_{\mathrm{s}}
\int_{t,{\bf r}}
\mathrm{Tr}_{\mathrm{KSN}}[ \hat{\gamma}^{1/2} \tau^{\pm} \underline{\hat{Q}_{tt}({\bf r})} ]
\mathrm{Tr}_{\mathrm{KSN}}[ \hat{\gamma}^{2/1} \tau^{\mp}  \underline{\hat{Q}_{tt}({\bf r})} ].
\end{align}
Fourier transform in time domain reads,
\begin{align}
iS_{\mathrm{ee}} =
&
-  i \frac{\pi^2 \nu^2}{8}
Z
\int_{\epsilon_{1},\epsilon_{1}^\prime,\epsilon_{2},\epsilon_{2}^\prime}
\int_{{\bf r}}
\mathrm{Tr}_{\mathrm{KSN}}[ \hat{\gamma}^{1/2} \tau_{\pm} \underline{\hat{Q}_{\epsilon_{1}\epsilon_{1}^\prime}({\bf r})} ]
\mathrm{Tr}_{\mathrm{KSN}}[ \hat{\gamma}^{2/1} \tau_{\pm} \underline{\hat{Q}_{\epsilon_{2}\epsilon_{2}^\prime}({\bf r})} ]
\delta_{\epsilon_{1} - \epsilon_{1}^\prime, \epsilon_{2}^\prime - \epsilon_{2}}
\\
&
+  i\frac{\pi^2\nu^2}{8}
\Gamma_{2}
\int_{\epsilon_{1},\epsilon_{1}^\prime,\epsilon_{2},\epsilon_{2}^\prime}
\int_{{\bf r}}
\mathrm{Tr}_{\mathrm{KSN}}[ \hat{\gamma}^{1/2} \tau_{\pm} \bm{\sigma} \underline{\hat{Q}_{\epsilon_{1}\epsilon_{1}^\prime}({\bf r})}) ]
\mathrm{Tr}_{\mathrm{KSN}}[ \hat{\gamma}^{2/1} \tau_{\pm} \bm{\sigma}  \underline{\hat{Q}_{\epsilon_{2}\epsilon_{2}^\prime}({\bf r})} ]
\delta_{\epsilon_{1} - \epsilon_{1}^\prime, \epsilon_{2}^\prime - \epsilon_{2}}
\\
&
+ i\frac{\pi^2\nu}{8}
V_{\mathrm{s}}
\int_{\epsilon_{1},\epsilon_{1}^\prime,\epsilon_{2},\epsilon_{2}^\prime}
\int_{{\bf r}}
\mathrm{Tr}_{\mathrm{KSN}}[ \hat{\gamma}^{1/2} \tau^{\pm} \underline{\hat{Q}_{\epsilon_{1}\epsilon_{1}^\prime}({\bf r})} ]
\mathrm{Tr}_{\mathrm{KSN}}[ \hat{\gamma}^{2/1} \tau^{\mp}  \underline{\hat{Q}_{\epsilon_{2}\epsilon_{2}^\prime}({\bf r})} ]
\delta_{\epsilon_{1} - \epsilon_{1}^\prime, \epsilon_{2}^\prime - \epsilon_{2}},
\end{align}
where $\delta_{\epsilon_{1} - \epsilon_{1}^\prime, \epsilon_{2}^\prime - \epsilon_{2}} \equiv 2\pi \delta(\epsilon_{1} - \epsilon_{1}^\prime - \epsilon_{2}^\prime + \epsilon_{2})$ is the energy conservation. Recall, $\int_{\epsilon}(..) \equiv \int_{-\infty}^{+\infty}\frac{d\epsilon}{2\pi}(..)$.

For the sake of convenience, to compactly describe action for the interaction, we will be using new auxiliary interaction fields.
We introduce
\begin{align}
&
\hat{\phi}_{\rho, \alpha\beta}(t,{\bf r}) =
\delta_{\alpha\beta} \tau_{+} \hat{\phi}_{ \rho,+}(t,{\bf r})+
\delta_{\alpha\beta} \tau_{-} \hat{\phi}_{ \rho,-}(t,{\bf r}),
\\
&
\hat{\bm{\phi}}_{\sigma, \alpha\beta}(t,{\bf r}) =
{\bm \sigma}_{\alpha\beta} \tau_{+} \hat{\phi}_{ \sigma,+}(t,{\bf r}) +
{\bm \sigma}_{\alpha\beta} \tau_{-} \hat{\phi}_{ \sigma,-}(t,{\bf r}),
\\
&
\hat{\phi}_{\mathrm{c}, \alpha\beta}(t,{\bf r})
=\delta_{\alpha\beta} \tau^{+}\hat{\phi}_{\mathrm{c}, +}(t,{\bf r})
+ \delta_{\alpha\beta}  \tau^{-}\hat{\phi}_{\mathrm{c}, -}(t,{\bf r}),
\end{align}
where $\pm$ index in $\tau^{\pm}$ in $\hat{\phi}_{\rho,\sigma,\mathrm{c}, \pm}(t,{\bf r})$ refers to the component of the Gor'kov-Nambu space.
Because of the $\mathrm{SU}(2)$ invariance of the system, we did not specify spin components of the $\hat{\bm{\phi}}_{\sigma, \alpha\beta}(t,{\bf r}) $ with their own index.
Each field is also decomposed in Keldysh space as
\begin{align}
&
\hat{\phi}_{\rho,\pm}(t,{\bf r})
=
\hat{\gamma}^{1} \phi^{(1)}_{\rho,\pm}(t,{\bf r}) +
\hat{\gamma}^{2} \phi^{(2)}_{\rho,\pm}(t,{\bf r}),
\\
&
\hat{\phi}_{\sigma,\pm}(t,{\bf r})
=
\hat{\gamma}^{1}\phi^{(1)}_{\sigma,\pm}(t,{\bf r}) +
\hat{\gamma}^{2}\phi^{(2)}_{\sigma,\pm}(t,{\bf r}),
\\
&
\hat{\phi}_{\mathrm{c},\pm}(t,{\bf r})
=
\hat{\gamma}^{1} \phi^{(1)}_{\mathrm{c},\pm}(t,{\bf r}) +
\hat{\gamma}^{2} \phi^{(2)}_{\mathrm{c},\pm}(t,{\bf r}).
\end{align}
We rewrite the three electron-electron interaction terms as (we omit spin indexes)
\begin{align}
iS_{\mathrm{ee}}
&
= -\frac{\pi^2 \nu^2 }{4} \sum_{m=\rho,\sigma,\mathrm{c}}
\int_{tt^\prime, {\bf r}{\bf r}^\prime}
\mathrm{Tr}_{\mathrm{KSN}}\left[ \hat{\phi}_{m}(t,{\bf r})\underline{\hat{Q}_{tt}({\bf r})}\right]
\mathrm{Tr}_{\mathrm{KSN}}\left[ \hat{\phi}_{m}({\bf r}^\prime,t^\prime)\underline{\hat{Q}_{t^\prime t^\prime}({\bf r}^\prime)}\right]
\\
&
= - \frac{\pi^2 \nu^2 }{4} \sum_{m=\rho,\sigma,\mathrm{c}}
\int_{\epsilon_{1},\epsilon_{1}^\prime,\epsilon_{2},\epsilon_{2}^\prime}
\int_{{\bf r}{\bf r}^\prime}
\mathrm{Tr}_{\mathrm{KSN}}\left[ \hat{\phi}_{m;\epsilon_{1}\epsilon_{1}^\prime}({\bf r})\underline{\hat{Q}_{\epsilon_{1}^\prime\epsilon_{1}}({\bf r})}\right]
\mathrm{Tr}_{\mathrm{KSN}}\left[ \hat{\phi}_{m;\epsilon_{2}\epsilon_{2}^\prime}({\bf r}^\prime,t^\prime)\underline{\hat{Q}_{\epsilon_{2}^\prime\epsilon_{2}}({\bf r}^\prime)}\right]
\\
&
\equiv
- \frac{\pi^2 \nu^2 }{4} \sum_{m=\rho,\sigma,\mathrm{c}}
\int_{{\bf r}{\bf r}^\prime}
\mathrm{Tr}\left[ \hat{\phi}_{m;\epsilon_{1}\epsilon_{1}^\prime}({\bf r})\underline{\hat{Q}_{\epsilon_{1}^\prime\epsilon_{1}}({\bf r})}\right]
\mathrm{Tr}\left[ \hat{\phi}_{m;\epsilon_{2}\epsilon_{2}^\prime}({\bf r}^\prime,t^\prime)\underline{\hat{Q}_{\epsilon_{2}^\prime\epsilon_{2}}({\bf r}^\prime)}\right].
\end{align}
In the last line, to remind, $\mathrm{Tr} \equiv \mathrm{Tr}_{\mathrm{FKSN}}$ was used, where $\mathrm{Tr}_{\mathrm{F}}$ is a trace over the frequency variable. 
We note that for the spin sector, $m=\sigma$, the fields turn in to scalar product of $\phi$ vectors, namely
\begin{align}
iS_{\mathrm{ee},\sigma} =
-\frac{\pi^2 \nu^2 }{4}
\int_{tt^\prime, {\bf r}{\bf r}^\prime}
\mathrm{Tr}_{\mathrm{KSN}}\left[ \hat{\bm{\phi}}_{\sigma}(t,{\bf r})\underline{\hat{Q}_{tt}({\bf r})}\right]
\mathrm{Tr}_{\mathrm{KSN}}\left[ \hat{\bm{\phi}}_{\sigma}({\bf r}^\prime,t^\prime)\underline{\hat{Q}_{t^\prime t^\prime}({\bf r}^\prime)}\right].
\end{align}

In frequency space, which we will be working below, the fields are correlated as
\begin{align}
\langle
\phi^{(i)}_{\rho; n, \epsilon_{1}\epsilon_{1}^\prime}({\bf r}_{1})
\phi^{(j)}_{\rho; m, \epsilon_{2}\epsilon_{2}^\prime}({\bf r}_{2})
\rangle_{\phi}
= \frac{i}{2\nu} Z
\hat{\gamma}^{2}_{ij}
\delta_{n,m}
\delta({\bf r}_{1} - {\bf r}_{2})
\delta_{\epsilon_{1} - \epsilon_{1}^\prime, \epsilon_{2}^\prime - \epsilon_{2}} ,
\end{align}
here $\gamma^{2}$ is the second Pauli matrix referring to the Keldysh space.
\begin{align}
\langle
\phi^{(i)}_{\sigma; n, \epsilon_{1}\epsilon_{1}^\prime}({\bf r}_{1})
\phi^{(j)}_{\sigma; m, \epsilon_{2}\epsilon_{2}^\prime}({\bf r}_{2})
\rangle_{\phi}
= - \frac{i}{2\nu} \Gamma_{2}
\hat{\gamma}^{2}_{ij}
\delta_{n,m}
\delta({\bf r}_{1} - {\bf r}_{2})
\delta_{\epsilon_{1} - \epsilon_{1}^\prime, \epsilon_{2}^\prime - \epsilon_{2}}.
\end{align}
Interaction in the Cooper channel
\begin{align}
\langle
\phi^{(i)}_{\mathrm{c}; n, \epsilon_{1}\epsilon_{1}^\prime}({\bf r}_{1})
\phi^{(j)}_{\mathrm{c}; m, \epsilon_{2}\epsilon_{2}^\prime}({\bf r}_{2})
\rangle_{\phi}
=
-\frac{i}{2\nu}V_{\mathrm{s}}
\hat{\gamma}^{2}_{ij}
\delta_{n,- m}
\delta({\bf r}_{1}- {\bf r}_{2})   
\delta_{\epsilon_{1} - \epsilon_{1}^\prime, \epsilon_{2}^\prime - \epsilon_{2}} .
\end{align}
Here everywhere $\delta_{n,m}$ is the Kronecker symbol and not a delta-function like in the frequency space.
It can also be shown that
\begin{align}
\langle
\phi^{(i)}_{\rho; n, \epsilon_{1}\epsilon_{1}^\prime}({\bf r}_{1})
\phi^{(j)}_{\rho; m, \epsilon_{2}\epsilon_{2}^\prime}({\bf r}_{2})
\rangle_{\phi}
= \frac{i}{2\nu} Z
\hat{\gamma}^{2}_{ij}
\delta_{n,-m}
\delta({\bf r}_{1} - {\bf r}_{2})
\delta_{\epsilon_{1} - \epsilon_{1}^\prime, \epsilon_{2}^\prime - \epsilon_{2}} ,
\end{align}
and the same for $\sigma$ part. Namely, the two charge and spin parts of different Gor'kov-Nambu sectors are connected with the same interaction as within the same sector.

\subsection{Matrix ${\hat Q}$ parametrization}
We parametrize the saddle-point of the matrix $\hat{\underline{Q}}$ and fluctuations around the saddle-point as
\begin{align}
\hat{\underline{Q}} = \hat{{\cal U}} \circ \hat{U} \circ \hat{\sigma}_{z} \circ \hat{\bar{U}} \circ \hat{{\cal U}}^{-1}
\equiv \hat{{\cal U}} \circ \hat{Q} \circ \hat{{\cal U}}^{-1}
\end{align}
here $\circ$ is time convolution, and
\begin{align}
\hat{{\cal U}}_{tt^\prime}
= \left[
\begin{array} {cc}
\hat{u}_{tt^\prime} & 0 \\
0 & \hat{u}_{tt^\prime}^{\mathrm{T}}
 \end{array}
\right]_{\mathrm{N}},
\end{align}
where
\begin{align}
\hat{u}_{tt^\prime} = \left[
\begin{array}{cc}
1 & {\cal F}_{tt^\prime} \\
0 & -1
\end{array}
\right]_{\mathrm{K}},
\end{align}
where ${\cal F}_{tt^\prime}(\bf r)$ is the distribution function at, in general, non-equilibrium.
Rotation matrix is parametrized as $\hat{\bar{U}} = \exp ( \hat{W}/2 )$.
It must have a property $\hat{W}\hat{\sigma}_{3}+\hat{\sigma}_{3}\hat{W} = 0$.
From the $\hat{Q}_{\epsilon \epsilon^\prime}^{\mathrm{T}} = \sigma_{y}\tau_{1}\hat{Q}_{-\epsilon^\prime, -\epsilon}\tau_{1}\sigma_{y}$ identity, one can deduce another property, $\hat{W}_{\epsilon \epsilon^\prime}^{\mathrm{T}} = - \sigma_{y}\tau_{1}\hat{W}_{-\epsilon^\prime, -\epsilon}\tau_{1}\sigma_{y}$.
In TR basis, we write the matrix as
\begin{align}
\hat{W} =
\left[\begin{array}{cc} \hat{P} & \hat{\tilde{B}}_{1} \\
\hat{\tilde{B}}_{2} & -\sigma_{y}\hat{P}^{\mathrm{T}}\sigma_{y} \end{array}\right]_{\mathrm{N}}.
\end{align}
It is very important to rewrite off-diagonal elements (particle-particle part) of the Gor'kov-Nambu space as
\begin{align}
\hat{W} =
\left[\begin{array}{cc} \hat{P} & \hat{B}_{1} (i\sigma_{y}) \\
(-i\sigma_{y}) \hat{B}_{2} & -\sigma_{y}\hat{P}^{\mathrm{T}}\sigma_{y} \end{array}\right]_{\mathrm{N}},
\end{align}
in this way, one will connect rescattering in the particle-hole channel (denoted by $\hat{P}$ matrix) with the particle-particle channel (denoted by $\hat{B}_{1}$ and $\hat{B}_{2}$) in a straightforward manner.
In particle-hole channel one has
\begin{align}
\hat{P} = \left[\begin{array}{cc} 0 & d^{\mathrm{cl}} \\
                                                      d^{\mathrm{q}} & 0 \end{array} \right]_{\mathrm{K}} ,
\end{align}
where also $d^{\mathrm{cl}}$ and $d^{\mathrm{q}}$ have their own spin structure.
We will go over the spin structure in the next section.
In particle-particle channel the matrices are
\begin{align}
&
\hat{\tilde{B}}_{1/2}^{\mathrm{T}} = - \sigma_{y}  \hat{\tilde{B}}_{1/2} \sigma_{y} ,
\\
&
\hat{B}_{1/2}^{\mathrm{T}} = \hat{B}_{1/2},
\end{align}
this results in the following parametrization of the $\hat{B}_{1/2}$ matrices in spin basis,
\begin{align}
\hat{B}_{1/2} = \left[ \begin{array}{cc} \hat{A}_{1/2} & \hat{R}_{1/2} \\
 \hat{S}_{1/2} & \hat{D}_{1/2}\end{array}\right]_{\mathrm{S}}
\end{align}
with relations $\hat{R}^{\mathrm{T}_{\mathrm{K}}}_{1/2} = \hat{S}_{1/2}$, $\hat{A}_{1/2}^{\mathrm{T}_{\mathrm{K}}} = \hat{A}_{1/2}$, and $ \hat{D}_{1/2}^{\mathrm{T}_{\mathrm{K}}} =  \hat{D}_{1/2}$.
We will go over the parametrization in the following two subsections.

\subsection{Example $\#$1 of working in frequency domain. Absence of Cooper channel.}\label{example1}
Consider only the particle-hole sector of the $\underline{\hat{Q}}$ matrix by completely disregarding the particle-particle one.
In this case the $\hat{\underline{Q}}$ matrix is parametrized as
\begin{align}
\hat{\underline{Q}} = \hat{u} \circ \hat{U} \circ \hat{\sigma}_{3} \circ \hat{\bar{U}} \circ \hat{u}^{-1} \equiv  \hat{u} \circ \hat{Q} \circ \hat{u}^{-1},
\end{align}
with
\begin{align}
\hat{u}_{tt^\prime} = \left[
\begin{array}{cc}
1 & {\cal F}_{tt^\prime}(\bf r) \\
0 & -1
\end{array}
\right]_{\mathrm{K}},
~~
\hat{u}_{\epsilon} = \left[
\begin{array}{cc}
1 & {\cal F}_{\epsilon}(\bf r) \\
0 & -1
\end{array}
\right]_{\mathrm{K}}.
\end{align}
Fluctuations close to the saddle-point are described by a rotation matrix $\hat{U}=\exp(-\hat{P}/2) $, where $\hat{P}\hat{\sigma}_{3} = -\hat{\sigma}_{3}\hat{P}$, and is
\begin{align}
\hat{P}_{\alpha\beta} = \left[
\begin{array} {cc}
0 & d_{\alpha\beta}^{\mathrm{cl}} \\
d_{\alpha\beta}^{\mathrm{q}} & 0
\end{array}
\right]_{\mathrm{K}}.
\end{align}
With that, we get $\hat{Q} = \hat{\sigma}_{3}\exp(\hat{P})$.

The non-interacting part of the action, $iS_{0} \rightarrow iS_{0}^{\mathrm{D}}$ where index $\mathrm{D}$ stands for diffusion, is derived up to second power of $\hat{P}$,
\begin{align}
iS_{0}^{\mathrm{D}}
&
= - \frac{\pi \nu }{4} \int_{{\bf r}} \mathrm{Tr}[ D(\nabla \hat{Q})^2 + 4i\hat{\varepsilon}\hat{Q} ]
\\
&
\approx  - \frac{\pi \nu }{4} \int_{{\bf r}} \mathrm{Tr}[ - D(\nabla \hat{P})^2 + 2 i\hat{\varepsilon}\hat{\sigma}_{3}\hat{P}^2 ]
\\
&
=  - \frac{\pi \nu }{4} \int_{{\bf q}}\int_{\epsilon \epsilon^\prime}
d^{\mathrm{q}}_{\alpha\beta; \epsilon\epsilon^\prime}({\bf q})[Dq^2 -  i(\epsilon - \epsilon^\prime )]
d^{\mathrm{cl}}_{\beta\alpha; \epsilon^\prime \epsilon}(-{\bf q})
\\
&
-
\frac{\pi \nu }{4} \int_{{\bf q}}\int_{\epsilon \epsilon^\prime}
d^{\mathrm{cl}}_{\alpha\beta; \epsilon\epsilon^\prime}({\bf q})[Dq^2 + i(\epsilon - \epsilon^\prime)]
d^{\mathrm{q}}_{\beta\alpha; \epsilon^\prime \epsilon}(-{\bf q}).
\end{align}
This part describes the diffusion of the density modes. Here $\int_{{\bf q}} (..) \equiv \int \frac{d{\bf q}}{(2\pi)^3}(..)$.
Correlators in diffuson sector are
\begin{align}
&
\langle d^{\mathrm{cl/q}}_{\alpha\beta; \epsilon_{1}\epsilon_{1}^{\prime}}
({\bf q}) d^{\mathrm{q/cl}}_{\mu\eta; \epsilon_{2}\epsilon_{2}^{\prime}}(-{\bf q})  \rangle
=
 - \frac{2}{\pi \nu} \delta_{\epsilon_{1}\epsilon_{2}^\prime}\delta_{\epsilon_{1}^\prime \epsilon_{2}}
\delta_{\alpha\eta}\delta_{\beta\mu}
{\cal D}^{\mathrm{R/A}}(\epsilon_{1} - \epsilon_{1}^{\prime}, {\bf q}),
\\
&
\langle  d^{\mathrm{cl/q}}_{\alpha\beta;  \epsilon_{1}\epsilon_{1}^{\prime}}({\bf q})
[ d^{\mathrm{q/cl}}_{\mu\eta; \epsilon_{2}\epsilon_{2}^{\prime}}(-{\bf q}) ]^{\mathrm{T}} \rangle
=
 - \frac{2}{\pi \nu} \delta_{\epsilon_{1},-\epsilon_{2}}\delta_{\epsilon_{1}^\prime, -\epsilon_{2}^\prime}
\delta_{\alpha\mu}\delta_{\beta\eta}
{\cal D}^{\mathrm{R/A}}(\epsilon_{1} - \epsilon_{1}^{\prime}, {\bf q} ),
\\
&
\langle d^{\mathrm{q}}_{\alpha\beta;  \epsilon_{1}\epsilon_{1}^{\prime}}({\bf q})
d^{\mathrm{q}}_{\mu\eta; \epsilon_{2}\epsilon_{2}^{\prime}}(-{\bf q})  \rangle =
\langle d^{\mathrm{cl}}_{\alpha\beta; \epsilon_{1}\epsilon_{1}^{\prime}}({\bf q})
d^{\mathrm{cl}}_{\mu\eta; \epsilon_{2}\epsilon_{2}^{\prime}}(-{\bf q})  \rangle =0,
\end{align}
where
\begin{align}
{\cal D}^{\mathrm{R/A}}(\omega, {\bf q} ) = \frac{1}{Dq^2 \mp i\omega}
\end{align}
is called the diffuson mode, a mode which describes a diffusion of fermionic charge and spin densities.

\subsection{Example $\#$2 of working in frequency domain. Presence of Cooper channel.}
Let us now study the Cooper channel. Again, we write
\begin{align}
\hat{\underline{Q}} = \hat{\cal{U}} \circ \hat{U} \circ \hat{\sigma}_{3} \circ \hat{\bar{U}} \circ \hat{\cal{U}}^{-1}
\equiv  \hat{\cal{U}} \circ \hat{Q} \circ \hat{\cal{U}}^{-1},
\end{align}
here $\hat{U} = \exp(-\hat{W}/2)$ and $\hat{\bar{U}} = \exp(\hat{W}/2)$.
We recall that the matrix is parametrized as
\begin{align}
\hat{W}_{\epsilon\epsilon^\prime} =
 \left[ \begin{array}{cc} \hat{P} & \hat{B}_{1}(i\sigma_{y}) \\ (-i\sigma_{y})\hat{B}_{2} & - \sigma_{y}\hat{P}^{\mathrm{T}}\sigma_{y} \end{array}  \right]_{\epsilon\epsilon^\prime}.
\end{align}
In spin space it is convenient to present the matrices in singlet and triplet basis, namely
\begin{align}
\hat{B}_{1/2;\alpha\beta} = \frac{1}{\sqrt{2}} {\bm \sigma}_{\alpha\beta}\hat{{\bf b}}_{1/2},
\end{align}
where components in Keldysh space are
\begin{align}
&
\hat{b}_{1;0} =
\left[ \begin{array}{cc} 0 & c_{1;0} \\
 c^{\mathrm{T}_{\mathrm{F}}}_{1;0} & 0\end{array}\right]_{\mathrm{K}}, ~~
\hat{b}_{1;x} =
\left[ \begin{array}{cc} 0 & c_{1;x} \\
 c^{\mathrm{T}_{\mathrm{F}}}_{1;x} & 0\end{array}\right]_{\mathrm{K}}, ~~
\hat{b}_{1;y} =
\left[ \begin{array}{cc} 0 & - c_{1;y} \\
 c_{1;y}^{\mathrm{T}_{\mathrm{F}}} & 0\end{array}\right]_{\mathrm{K}}, ~~
\hat{b}_{1;z} =
\left[ \begin{array}{cc} 0 & c_{1;z} \\
 c_{1;z}^{\mathrm{T}_{\mathrm{F}}} & 0\end{array}\right]_{\mathrm{K}},
\\
&
\hat{b}_{2;0} =
\left[ \begin{array}{cc} 0 & c_{2;0}^{\mathrm{T}_{\mathrm{F}}} \\
 c_{2;0} & 0\end{array}\right]_{\mathrm{K}}, ~~
\hat{b}_{2;x} =
\left[ \begin{array}{cc} 0 & c_{2;x}^{\mathrm{T}_{\mathrm{F}}} \\
 c_{2;x} & 0\end{array}\right]_{\mathrm{K}}, ~~
\hat{b}_{2;y} =
\left[ \begin{array}{cc} 0 &  c_{2;y}^{\mathrm{T}_{\mathrm{F}}} \\
 -c_{2;y} & 0\end{array}\right]_{\mathrm{K}}, ~~
\hat{b}_{2;z} =
\left[ \begin{array}{cc} 0 & c_{2;z}^{\mathrm{T}_{\mathrm{F}}} \\
 c_{2;z} & 0\end{array}\right]_{\mathrm{K}}.
\end{align}
With the transposition rule
\begin{align}
\hat{b}^{\mathrm{T}_{\mathrm{K}}}_{1;0;-m,-n} =
\left[ \begin{array}{cc} 0 & c_{1;0;-m,-n} \\
 c^{\mathrm{T}_{\mathrm{F}}}_{1;0;-m,-n} & 0\end{array}\right]^{\mathrm{T}_{\mathrm{K}}}_{\mathrm{K}}
=
\left[ \begin{array}{cc} 0 &  c^{\mathrm{T}_{\mathrm{F}}}_{1;0;-m,-n} \\
 c_{1;0;-m,-n} & 0\end{array}\right]_{\mathrm{K}}
=
\left[ \begin{array}{cc} 0 &  c_{1;0;n,m} \\
 c_{1;0;n,m}^{\mathrm{T}_{\mathrm{F}}} & 0\end{array}\right]_{\mathrm{K}}
=
\hat{b}_{1;0;n,m}
\end{align}
it is straightforward to show $\hat{B}_{1/2}^{\mathrm{T}} = \hat{B}_{1/2}$.

Another matrix is
\begin{align}
\hat{{\cal U}}_{tt^\prime}
= \left[
\begin{array} {cc}
\hat{u}_{tt^\prime} & 0 \\
0 & \hat{u}_{tt^\prime}^{\mathrm{T}}
 \end{array}
\right]_{\mathrm{N}},
\end{align}
where
\begin{align}
\hat{u}_{tt^\prime} = \left[
\begin{array}{cc}
1 & {\cal F}_{tt^\prime}(\bf r) \\
0 & -1
\end{array}
\right]_{\mathrm{K}},
~~ \hat{u}_{\epsilon } = \left[
\begin{array}{cc}
1 & {\cal F}_{\epsilon}(\bf r) \\
0 & -1
\end{array}
\right]_{\mathrm{K}},
~~
\hat{u}_{\epsilon}^{\mathrm{T}} = \left[ \begin{array}{cc} 1 & 0 \\ -{\cal F}_{\epsilon}({\bf r}) & -1 \end{array}  \right]_{\mathrm{K}}.
\end{align}
Non-interacting action reads
\begin{align}
iS_{0} = -\frac{\pi \nu }{8}\int_{{\bf r}} \mathrm{Tr} \{ D[ \nabla \hat{Q}({\bf r}) ]^2
- 4\tau_{3}\partial_{t}\hat{Q}({\bf r}) \}.
\end{align}
Fourier convention is
\begin{align}
\hat{Q}_{\epsilon\epsilon^{\prime}}({\bf r}) = \int_{tt^\prime} \hat{Q}_{tt^\prime}({\bf r}) e^{i\epsilon t - t\epsilon^\prime t^\prime}.
\end{align}
The action is rewritten as
\begin{align}
iS_{0} = -\frac{\pi \nu }{8}
 \mathrm{Tr}_{\mathrm{KSN}}
\int_{{\bf r}}
 \int_{\epsilon\epsilon^\prime}
\{ D[ \nabla \hat{Q}_{\epsilon \epsilon^\prime}({\bf r}) ][ \nabla \hat{Q}_{\epsilon^\prime \epsilon}({\bf r}) ]
+ 4i\tau_{3}\hat{\varepsilon}\hat{Q}_{\epsilon \epsilon^\prime}({\bf r})\delta_{\epsilon,\epsilon^{\prime}} \},
\end{align}
where $\delta_{\epsilon,\epsilon^{\prime}} = 2\pi \delta(\epsilon - \epsilon^{\prime})$.
This action contains studied charge and spin density sectors - diffusons, also it now contains a Cooper channel.
We have already studied the diffuson modes, now let us focus on the Cooper channel.
Rule for the frequency matrix is
\begin{align}
&
\hat{\varepsilon} c_{1/2;j;\epsilon\epsilon^\prime} = \epsilon c_{1/2;j;\epsilon\epsilon^\prime},
\\
&
\hat{\varepsilon} c^{\mathrm{T}}_{1/2;j;\epsilon\epsilon^\prime} = \epsilon c^{\mathrm{T}}_{1/2;j;\epsilon\epsilon^\prime}.
\end{align}
We expand around the saddle point,
\begin{align}
\frac{1}{2}
\mathrm{Tr}_{\mathrm{FKSN}}\left[ \hat{\sigma}_{3}\tau_{3} \epsilon \hat{W}^2_{\epsilon\epsilon} \right]
&
=
\frac{1}{2}
\mathrm{Tr}_{\mathrm{FK}}
\left[ \epsilon \hat{\sigma}_{3} \left( \hat{B}_{1,\epsilon\epsilon^\prime} \hat{B}_{2,\epsilon^\prime\epsilon}
- \hat{B}_{2,\epsilon\epsilon^\prime}\hat{B}_{1,\epsilon^\prime\epsilon} \right)\right]
\\
&
=
\frac{1}{2}
\mathrm{Tr}_{\mathrm{FK}}
\left[ \epsilon
\hat{\sigma}_{3}
 \left( \hat{{\bf b}}_{1,\epsilon\epsilon^\prime} \hat{{\bf b}}_{2,\epsilon^\prime\epsilon}
- \hat{{\bf b}}_{2,\epsilon\epsilon^\prime}\hat{{\bf b}}_{1,\epsilon^\prime\epsilon} \right)\right]
\\
&
=  \sum_{j=0}^{3} \mathrm{Tr}_{\mathrm{F}}
\left[
(\epsilon + \epsilon^\prime) c_{1; j;\epsilon\epsilon^\prime}c_{2; j;\epsilon^\prime\epsilon}
\right],
\end{align}
The Gor'kov-Nambu part $iS_{0}^{\mathrm{GN}}$ of the non-interacting action, $iS_{0} = iS_{0}^{\mathrm{D}}+ iS_{0}^{\mathrm{GN}}$ where $iS_{0}^{\mathrm{D}}$ part has already been studied in Sec. (\ref{example1}), is
\begin{align}
iS_{0}^{\mathrm{GN}} = - \frac{ \pi \nu }{2}  \sum_{{\bf q};j} \mathrm{Tr}\left[  Dq^2  +  i(\epsilon + \epsilon^\prime)  \right]
c_{1;j;\epsilon\epsilon^\prime}(-{\bf q})c_{2;j;\epsilon^\prime\epsilon}({\bf q}).
\end{align}
We use this action to calculate correlation function.
Correlators in the Cooper channel written in frequency domain and momentum are
\begin{align}
&
\langle c_{2;i;\epsilon_{1}^{\prime}\epsilon_{1}}({\bf q}) c_{1;j;\epsilon_{2}^{\prime}\epsilon_{2}}(-{\bf q})  \rangle
=  - \frac{2}{\pi \nu}\delta_{ij} \delta_{\epsilon_{1}\epsilon_{2}^\prime}\delta_{\epsilon_{1}^\prime \epsilon_{2}}
C^{\mathrm{A}}(\epsilon_{1}^\prime + \epsilon_{1}, {\bf q}),
\\
&
\langle c_{1;i;\epsilon_{1}^{\prime}\epsilon_{1}}({\bf q}) c_{2;j;\epsilon_{2}^{\prime}\epsilon_{2}}(-{\bf q})  \rangle
= - \frac{2}{\pi \nu} \delta_{ij}\delta_{\epsilon_{1}\epsilon_{2}^\prime}\delta_{\epsilon_{1}^\prime \epsilon_{2}}
C^{\mathrm{A}}(\epsilon_{1} + \epsilon_{1}^\prime, {\bf q}),
\\
&
\langle c^{\mathrm{T}}_{2;i;\epsilon_{1}^{\prime}\epsilon_{1}}({\bf q}) c^{\mathrm{T}}_{1;j;\epsilon_{2}^{\prime}\epsilon_{2}}(-{\bf q})  \rangle
=  - \frac{2}{\pi \nu} \delta_{ij}
\delta_{\epsilon_{1}\epsilon_{2}^\prime}
\delta_{\epsilon_{1}^\prime \epsilon_{2}}
C^{\mathrm{A}}(-\epsilon_{1}^\prime - \epsilon_{1}, {\bf q})
=  - \frac{2}{\pi \nu} \delta_{ij}
\delta_{\epsilon_{1}\epsilon_{2}^\prime}
\delta_{\epsilon_{1}^\prime \epsilon_{2}}
 C^{\mathrm{R}}(\epsilon_{1}^\prime + \epsilon_{1}, {\bf q}),
\\
&
\langle c_{2;i;\epsilon_{1}^{\prime}\epsilon_{1}}({\bf q}) c^{\mathrm{T}}_{1;j;\epsilon_{2}^{\prime}\epsilon_{2}}(-{\bf q})  \rangle
= \langle c_{2;i;\epsilon_{1}^{\prime}\epsilon_{1}}({\bf q}) c_{1;j;-\epsilon_{2}-\epsilon_{2}^{\prime}}(-{\bf q})  \rangle
=  - \frac{2}{\pi \nu}\delta_{ij} \delta_{\epsilon_{1}, -\epsilon_{2}}\delta_{\epsilon_{1}^\prime , -\epsilon_{2}^\prime}
C^{\mathrm{A}}(\epsilon_{1}^\prime + \epsilon_{1}, {\bf q}),
\\
&
\langle c_{1;i;\epsilon_{1}^{\prime}\epsilon_{1}}({\bf q}) c^{\mathrm{T}}_{2;j;\epsilon_{2}^{\prime}\epsilon_{2}}(-{\bf q})  \rangle
= - \frac{2}{\pi \nu}\delta_{ij} \delta_{\epsilon_{1}, -\epsilon_{2}}\delta_{\epsilon_{1}^\prime , -\epsilon_{2}^\prime}
C^{\mathrm{A}}(\epsilon_{1}^\prime + \epsilon_{1}, {\bf q}),
\end{align}
where
\begin{align}
C^{\mathrm{R}/\mathrm{A}}(\omega , {\bf q}) = \frac{1}{Dq^2 \mp i\omega}
\end{align}
is called the Cooperon, which described weak localization (quantum interfernce) of fermions.

\subsection{Contraction}
We consider a correlator which will be heavily used in the perturbation theory,
\begin{align}
\langle \mathrm{Tr}[ \hat{A}({\bf r})\hat{W}({\bf r}) ] \mathrm{Tr}[ \hat{Y}({\bf r}^\prime)\hat{W}({\bf r}^\prime) ] \rangle_{W},
\end{align}
where $\hat{A}$ and $\hat{Y}$ are some fields in spin, Gor'kov-Nambu, and Keldysh spaces.
The correlator is calculated to be
\begin{align}
&
\langle \mathrm{Tr}[ \hat{A}_{\epsilon_{1}\epsilon_{1}^\prime}({\bf r})\hat{W}_{\epsilon_{1}^\prime \epsilon_{1}}({\bf r}) ]
 \mathrm{Tr}[ \hat{Y}_{\epsilon_{2}\epsilon_{2}^\prime}({\bf r}^\prime)\hat{W}_{\epsilon_{2}^\prime \epsilon_{2}}({\bf r}^\prime) ] \rangle_{W}
\\
=
&\label{contractionA_line1}
-\frac{2}{\pi \nu} \mathrm{Tr}
[\hat{A}({\bf r})]_{\alpha\beta; \perp \mathrm{K}; \parallel \mathrm{N};\epsilon_{1}\epsilon_{1}^\prime}
\hat{\Pi}_{\mathrm{N}}(\epsilon_{1} - \epsilon_{1}^\prime ,  {\bf r} - {\bf r}^\prime)
[\hat{Y}({\bf r}^\prime)]_{\beta\alpha;\perp \mathrm{K}; \parallel \mathrm{N};\epsilon_{2}\epsilon_{2}^\prime}
\delta_{\epsilon_{1},\epsilon_{2}^{\prime}} \delta_{\epsilon_{1}^{\prime},\epsilon_{2}}
\\
&\label{contractionA_line2}
+\frac{2}{\pi \nu} \mathrm{Tr}
[\tau_{1}\sigma_{y}\hat{A}({\bf r})\sigma_{y}\tau_{1}]^{\mathrm{T}}_{\alpha\beta; \perp \mathrm{K}; \parallel \mathrm{N};\epsilon_{1}\epsilon_{1}^\prime}
\hat{\Pi}_{\mathrm{N}}(\epsilon_{1} - \epsilon_{1}^\prime , {\bf r} - {\bf r}^\prime)
[\hat{Y}({\bf r}^\prime) ]_{\beta\alpha;\perp \mathrm{K}; \parallel \mathrm{N};\epsilon_{2}\epsilon_{2}^\prime}
\delta_{\epsilon_{1},\epsilon_{2}^{\prime}} \delta_{\epsilon_{1}^{\prime},\epsilon_{2}}
\\
&\label{contractionA_line3}
-\frac{2}{\pi \nu} \mathrm{Tr}
[\hat{A}({\bf r})]_{\alpha\beta;\perp \mathrm{K}; \perp \mathrm{N};\epsilon_{1}\epsilon_{1}^\prime}
\hat{C}_{\mathrm{N}}(\epsilon_{1} + \epsilon_{1}^\prime , {\bf r} - {\bf r}^\prime)
[\hat{Y}({\bf r}^\prime)]_{\beta\alpha;\perp \mathrm{K}; \perp \mathrm{N};\epsilon_{2}\epsilon_{2}^\prime}
\delta_{\epsilon_{1},\epsilon_{2}^{\prime}} \delta_{\epsilon_{1}^{\prime},\epsilon_{2}}
\\
&\label{contractionA_line4}
+\frac{2}{\pi \nu} \mathrm{Tr}
[\tau_{1}\sigma_{y}\hat{A}({\bf r})\sigma_{y}\tau_{1}]^{\mathrm{T}}_{\alpha\beta;\perp \mathrm{K}; \perp \mathrm{N};\epsilon_{1}\epsilon_{1}^\prime}
\hat{C}_{\mathrm{N}}(\epsilon_{1} + \epsilon_{1}^\prime , {\bf r} - {\bf r}^\prime)
[\hat{Y}({\bf r}^\prime)]_{\beta\alpha;\perp \mathrm{K}; \perp \mathrm{N};\epsilon_{2}\epsilon_{2}^\prime}
\delta_{\epsilon_{1},\epsilon_{2}^{\prime}} \delta_{\epsilon_{1}^{\prime},\epsilon_{2}},
\end{align}
where the diffusion part is
\begin{align}
&
\hat{\Pi}_{\mathrm{N}}(\omega, {\bf q}) =
\left[ \begin{array}{cc} \hat{\Pi}(\omega, {\bf q}) & 0 \\
0 & \hat{\Pi}^{\dag}(\omega, {\bf q}) \end{array}
\right]_{\mathrm{N}}, ~~
\hat{\Pi}(\omega, {\bf q}) =
\left[ \begin{array}{cc} {\cal D}^{\mathrm{A}}(\omega, {\bf q}) & 0 \\
0 & {\cal D}^{\mathrm{R}}(\omega, {\bf q}) \end{array}
\right]_{\mathrm{K}},
\\
&
\hat{\Pi}^{\dag}(\omega, {\bf q}) = \hat{\Pi}(-\omega, {\bf q}),
~~
\hat{\Pi}(\omega, {\bf q}) = \hat{\Pi}^{\mathrm{T}}(\omega, {\bf q}),
\end{align}
(note that $\hat{\Pi}_{\epsilon+\frac{\omega}{2},\epsilon-\frac{\omega}{2}}({\bf q})$ defined in Ref.\cite{SchwieteFinkel'steinPRB2014} is identical to $\hat{\Pi}(-\omega, {\bf q})$ defined above such that the results are in agreement with each other)
and the Cooperon part is
\begin{align}
&
\hat{ C}_{\mathrm{N}}(\omega, {\bf q})
= \left[ \begin{array}{cc} \hat{C}(\omega, {\bf q}) & 0 \\
0 & \hat{C}^{\dag}(\omega, {\bf q})
\end{array}\right]_{\mathrm{N}},
~~
\hat{C}(\omega, {\bf q}) = \left[ \begin{array}{cc} C^{\mathrm{A}}(\omega, {\bf q}) & 0 \\
0 & C^{\mathrm{R}}(\omega, {\bf q})
\end{array}\right]_{\mathrm{K}},
\\
&
\hat{C}^{\dag}(\omega, {\bf q}) = \hat{C}^{\mathrm{T}}(\omega, {\bf q}) = \hat{C}(-\omega, {\bf q}),
\end{align}
where
\begin{align}
{\cal D}^{\mathrm{R}/\mathrm{A}}(\omega,{\bf q}) = \frac{1}{Dq^2 \mp i\omega}, ~~ C^{\mathrm{R}/\mathrm{A}}(\omega, {\bf q}) = \frac{1}{Dq^2 \mp i\omega}.
\end{align}
We remind that $\mathrm{T}_{\mathrm{F}\mathrm{K}\mathrm{S}\mathrm{N}} \equiv \mathrm{T}$ is the transposition in frequency, Keldysh, spin, and Gor'kov-Nambu spaces correspondingly.
One can rewrite the diffusion propagator,
\begin{align}
\hat{\Pi}_{\mathrm{N}}(\omega, {\bf q})  = \frac{Dq^2}{(Dq^2)^2 + \omega^2} - \frac{i\omega}{(Dq^2)^2 + \omega^2} \hat{\sigma}_{3}\tau_{3}
\equiv
\underline{\Pi}_{\mathrm{N}}(\omega, {\bf q})  + \bar{\Pi}_{\mathrm{N}}(\omega, {\bf q})   \hat{\sigma}_{3}\tau_{3},
\end{align}
and the Cooperon matrix,
\begin{align}
\hat{C}_{\mathrm{N}}(\omega, {\bf q})  =
\frac{Dq^2}{(Dq^2)^2 + \omega^2} - \frac{i\omega}{(Dq^2)^2 + \omega^2} \hat{\sigma}_{3}\tau_{3}
\equiv
\underline{C}_{\mathrm{N}}(\omega, {\bf q})  + \bar{C}_{\mathrm{N}}(\omega, {\bf q})   \hat{\sigma}_{3}\tau_{3}.
\end{align}
We have introduced the following symbols
\begin{align}
&
\hat{A}_{\parallel\mathrm{N}} = \frac{1}{2} ( \hat{A} + \tau_{3} \hat{A} \tau_{3} ), ~~
\hat{A}_{\perp\mathrm{N}} = \frac{1}{2} ( \hat{A} - \tau_{3} \hat{A} \tau_{3} ),
\\
&
\hat{A}_{\perp\mathrm{K}} = \frac{1}{2} ( \hat{A} - \hat{\sigma}_{3} \hat{A} \hat{\sigma}_{3} ),
\end{align}
which will be used throughout the notes.
In deriving the contraction we have used following identities,
\begin{align}
&
(\hat{Y})^{\mathrm{T}_{\mathrm{F}\mathrm{K}\mathrm{S}}}_{\perp \mathrm{N}} = (\tau_{1}\hat{Y}\tau_{1})^{\mathrm{T}_{\mathrm{F}\mathrm{K}\mathrm{S}\mathrm{N}}}_{\perp \mathrm{N}} \equiv (\tau_{1}\hat{Y}\tau_{1})^{\mathrm{T}}_{\perp \mathrm{N}},
\\
&
\sum_{n=0,x,y,z}\mathrm{Tr}_{\mathrm{S}}(\hat{A}\sigma_{y}\sigma_{n})\mathrm{Tr}_{\mathrm{S}}(\sigma_{y}\hat{Y}\sigma_{n})
= 2 \mathrm{Tr}_{\mathrm{S}}(\hat{A}\sigma_{y}\sigma_{y} \hat{Y})
= 2 \mathrm{Tr}_{\mathrm{S}}(\hat{A} \hat{Y}),
\\
&
\label{identity2}
\sum_{n=0,x,z}\mathrm{Tr}_{\mathrm{S}}(\hat{A}\sigma_{y}\sigma_{n})\mathrm{Tr}_{\mathrm{S}}(\sigma_{y}\hat{Y}\sigma_{n})
-\mathrm{Tr}_{\mathrm{S}}(\hat{A}\sigma_{y}\sigma_{y})\mathrm{Tr}_{\mathrm{S}}(\sigma_{y}\hat{Y}\sigma_{y})
= -2 \mathrm{Tr}_{\mathrm{S}}[ (\sigma_{y}\hat{A}\sigma_{y})^{\mathrm{T}_{\mathrm{S}}} \hat{Y} ].
\end{align}
An extra minus sign in Eqs. (\ref{contractionA_line2}) and (\ref{contractionA_line4}) as compared with Eqs. (\ref{contractionA_line1}) and (\ref{contractionA_line3}) correspondingly is exactly due to the identity Eq. (\ref{identity2})

An equivalent, and useful, form of the correlator can be obtained utilizing the following identity,
\begin{align}
\mathrm{Tr} \hat{A}_{\perp\mathrm{K}\perp\mathrm{N}} \hat{C}_{\mathrm{N}} \hat{Y}_{\perp\mathrm{K}\perp\mathrm{N}}
=
\mathrm{Tr}\hat{C}_{\mathrm{N}}^{\dag} \hat{A}_{\perp\mathrm{K}\perp\mathrm{N}}  \hat{Y}_{\perp\mathrm{K}\perp\mathrm{N}}.
\end{align}


\section{s-wave Cooper channel}\label{appendixB}
In this Appendix we first derive propagator in the spin-singlet s-wave part of the Cooper channel when the interaction $V_{\mathrm{s}}$ is constant as a function of frequencies. This corresponds to the BCS model, and in most literature $V_{\mathrm{s}} = \Gamma_{\mathrm{c}}$ notation is used in this case.
Then, for a general form of the frequency-dependent interaction, we derive equation on the interaction amplitude in both spin-singlet and spin-triplet s-wave Cooper channels.

\subsection{Propagator in the spin-singlet part of the Cooper channel}
Here we derive dynamically dressed interaction in the Cooper channel - propagator in the Cooper channel \cite{FeigelmanLarkinSkvortsovPRB2000} for the case when $V_{\mathrm{s}} = \mathrm{const}$ as a function of frequencies.
It will appear in the following considerations and will simply substitute bare $V_{\mathrm{s}}$ interaction in the spin-singlet Cooper channel with the renormalized, $\Gamma_{\mathrm{s}}(\epsilon_{1},\epsilon_{1}^\prime , {\bf q})$ interaction amplitude.
We expand the interaction to second order in $\hat{W}$ matrices,
\begin{align}
iS_{\mathrm{ee},\mathrm{c}}^{[2]}
= -\frac{\pi^2 \nu^2 }{4}
\int_{{\bf r},{\bf r}^\prime}
\mathrm{Tr}\left[ \underline{ \hat{\phi}_{\mathrm{c},\epsilon_{1}\epsilon_{1}^\prime} ({\bf r})}  \hat{\sigma}_{3}\tau_{0}
\hat{W}_{\epsilon_{1}^\prime\epsilon_{1}}({\bf r}) \right]
\mathrm{Tr}\left[ \underline{ \hat{\phi}_{\mathrm{c},\epsilon_{2}\epsilon_{2}^\prime} ({\bf r}^\prime)} \hat{\sigma}_{3}\tau_{0}
\hat{W}_{\epsilon_{2}^\prime\epsilon_{2}}({\bf r}^\prime)  \right].
\end{align}
Here the interaction fields $\hat{\phi}_{\mathrm{c}}$ are correlated as (see SM for more details)
\begin{align}
\langle
\phi^{(i)}_{\mathrm{c}; n, \epsilon_{1}\epsilon_{1}^\prime}({\bf r}_{1})
\phi^{(j)}_{\mathrm{c}; m, \epsilon_{2}\epsilon_{2}^\prime}({\bf r}_{2})
\rangle_{\phi}
=
-\frac{i}{2\nu}V_{\mathrm{s}}
\hat{\gamma}^{2}_{ij}
\delta({\bf r}_{1}- {\bf r}_{2})   \delta_{n,- m}
\delta_{\epsilon_{1} - \epsilon_{1}^\prime, \epsilon_{2}^\prime - \epsilon_{2}} ,
\end{align}
where $V_{\mathrm{s}}$ is the bare interaction corresponding to attraction, that is $V_{\mathrm{s}}<0$.
 Up to second order in interaction action $iS_{\mathrm{ee}}^{[2]}$,
\begin{align}
&
\frac{1}{2}
\langle (iS_{\mathrm{ee},\mathrm{c}}^{[2]})(  iS_{\mathrm{ee},\mathrm{c}}^{[2]})\rangle_{W}
=
\frac{1}{2}
\left( -i\frac{\pi^2 \nu}{8} \right)^2
V_{\mathrm{s}}^2
\int_{\epsilon_{1},\epsilon_{1}^\prime,\epsilon_{2},\epsilon_{2}^\prime}
\int_{\epsilon_{3},\epsilon_{3}^\prime,\epsilon_{4},\epsilon_{4}^\prime}
\delta_{\epsilon_{1}-\epsilon_{1}^\prime, \epsilon_{2}^\prime-\epsilon_{2}}
\delta_{\epsilon_{3}-\epsilon_{3}^\prime, \epsilon_{4}^\prime-\epsilon_{4}}
\\
&
\langle
\mathrm{Tr}_{\mathrm{KSN}}[\hat{\gamma}^{[1/2]_{\mathrm{A}1}} \tau^{[\pm]_{\mathrm{A}2}}\underline{\hat{\sigma}_{3}\hat{W}_{\epsilon_{1}\epsilon_{1}^\prime}}]
\mathrm{Tr}_{\mathrm{KSN}}[\hat{\gamma}^{[2/1]_{\mathrm{A}1}}\tau^{[\mp]_{\mathrm{A}2}}\underline{\hat{\sigma}_{3}\hat{W}_{\epsilon_{2}\epsilon_{2}^\prime}}]
\mathrm{Tr}_{\mathrm{KSN}}[\hat{\gamma}^{[2/1]_{\mathrm{B}1}}\tau^{[\mp]_{\mathrm{B}2}}\underline{\hat{\sigma}_{3}\hat{W}_{\epsilon_{3}\epsilon_{3}^\prime}}]
\mathrm{Tr}_{\mathrm{KSN}}[\hat{\gamma}^{[1/2]_{\mathrm{B}1}} \tau^{[\pm]_{\mathrm{B}2}}\underline{\hat{\sigma}_{3}\hat{W}_{\epsilon_{4}\epsilon_{4}^\prime}}]
\rangle_{W}
\\
&
=
\left( -i\frac{\pi^2 \nu}{8} \right)^2
V_{\mathrm{s}}^2
\int_{\epsilon_{1},\epsilon_{1}^\prime,\epsilon_{4},\epsilon_{4}^\prime}
\mathrm{Tr}_{\mathrm{KSN}}[\hat{\gamma}^{[1/2]_{\mathrm{C}1}} \tau^{[\pm]_{\mathrm{C}2}}\underline{\hat{\sigma}_{3}\hat{W}_{\epsilon_{1}\epsilon_{1}^\prime}}]
\mathrm{Tr}_{\mathrm{KSN}}[\hat{\gamma}^{[2/1]_{\mathrm{C}1}} \tau^{[\mp]_{\mathrm{C}2}}\underline{\hat{\sigma}_{3}\hat{W}_{\epsilon_{4}\epsilon_{4}^\prime}}]
\left[
\int_{\epsilon_{2}}
I_{\epsilon_{2},\epsilon_{2}+\epsilon_{1}-\epsilon_{1}^\prime}
\right]
\delta_{\epsilon_{1}-\epsilon_{1}^\prime, \epsilon_{4}^\prime-\epsilon_{4}},
\end{align}
where $\langle ... \rangle_{W}$ must be understood as a contraction appearing in the correlator defining the whatever diagramm under study.
A factor of $2$ is due to two ways one could contract the traces. This factor is in agreement with the slow and fast fields decomposition used in the Renormalization Group procedure, see \cite{SchwieteFinkel'steinPRB2014}.
Summation notation over the $\mathrm{A}1(2)$, $\mathrm{B}1(2)$, and $\mathrm{C}1(2)$ indexes was introduced after Eq. (\ref{eesinglet}) in the MT (main text).
We have defined expression $I_{\epsilon_{2},\epsilon_{2}+\epsilon_{1}-\epsilon_{1}^\prime}$ for compactness of the already loaded expressions,
\begin{align}
\int_{\epsilon_{2}}
I_{\epsilon_{2},\epsilon_{2}+\epsilon_{1}-\epsilon_{1}^\prime}
&
=
\frac{8}{\pi\nu}\int_{\epsilon_{2}} \bar{C}_{\mathrm{N}}^{\dag}(\epsilon_{2}+\epsilon_{2} + \omega, {\bf q})
({\cal F}_{\epsilon_{2}} + {\cal F}_{\epsilon_{2} + \omega})
\\
&
\equiv
{\cal C}(\omega,{\bf q})
\end{align}
where $\omega = \epsilon_{1}-\epsilon_{1}^\prime$ and
\begin{align}
\bar{C}_{\mathrm{N}}(\omega, {\bf q}) = - \frac{i\omega}{(Dq^2)^2+\omega^2}.
\end{align}
It is instructive to note that if the sign of $\bar{C}_{\mathrm{N}}$ was different, we would have had a case when repulsion resulted in attraction in the s-wave spin-singlet Cooper ladder.
Integration over the frequency gives
\begin{align}
i\frac{\pi^2 \nu}{8}
{\cal C}(\omega,{\bf q})
&
\approx
-
\frac{1}{4}
\left\{
\ln\left[ \frac{\Lambda^2}{T^2 +  \left( \frac{Dq^2 - i\omega}{2}\right)^2 } \right]
+
2\ln\left[ \frac{2\gamma}{\pi}\right]
\right\}
-
\frac{1}{4}
\left\{
\ln\left[ \frac{\Lambda^2}{T^2 +  \left( \frac{Dq^2 + i\omega}{2}\right)^2 } \right]
+
2\ln\left[ \frac{2\gamma}{\pi}\right]
\right\}
\\
&
\equiv
-M(\omega,{\bf q}),
\end{align}
where $\ln \gamma = C \approx 0.577$ is the Euler's constant.
Again, schematically
\begin{align}
\frac{1}{2}\langle (iS_{\mathrm{ee}}^{[2]} )( iS_{\mathrm{ee}}^{[2]})\rangle_{W}
=
\left( i\frac{\pi^2 \nu}{8} \right)
V_{\mathrm{s}}
\int_{\epsilon_{1},\epsilon_{1}^\prime,\epsilon_{4},\epsilon_{4}^\prime}
&
\mathrm{Tr}_{\mathrm{KSN}}[\hat{\gamma}^{[1/2]_{\mathrm{C}1}} \tau^{[\pm]_{\mathrm{C}2}}\underline{\hat{\sigma}_{3}\hat{W}_{\epsilon_{1}\epsilon_{1}^\prime}}]
\mathrm{Tr}_{\mathrm{KSN}}[\hat{\gamma}^{[2/1]_{\mathrm{C}1}} \tau^{[\mp]_{\mathrm{C}2}}\underline{\hat{\sigma}_{3}\hat{W}_{\epsilon_{4}\epsilon_{4}^\prime}}]
\\
&
V_{\mathrm{s}}
[-M(\omega, {\bf q}) ]
\delta_{\epsilon_{1}-\epsilon_{1}^\prime, \epsilon_{4}^\prime-\epsilon_{4}},
\end{align}
therefore, effective interaction considered to all orders reads
\begin{align}
&
V_{\mathrm{s}}
+
V_{\mathrm{s}}V_{\mathrm{s}}
[-M(\omega, {\bf q}) ]
+ ...
=
\frac{V_{\mathrm{s}}}{1+ V_{\mathrm{s}}
M(\omega, {\bf q})}.
\end{align}
As a check, for $\omega = 0$ and $q=0$ case, setting $\Lambda = \frac{1}{\tau}$,
\begin{align}
M(0, 0) = \ln\left[ \frac{\gamma}{\pi T\tau} \right],
\end{align}
which is consistent with the renormalization group results.
Finally in the action for the interaction in the Cooper channel
\begin{align}
iS_{\mathrm{ee},\mathrm{c}}
=
 i\frac{\pi^2\nu}{8}
\int_{\epsilon_{1},\epsilon_{1}^\prime,\epsilon_{2},\epsilon_{2}^\prime}
\int_{\bf q}
\Gamma_{\mathrm{s}}(\epsilon_{1}-\epsilon_{1}^\prime, {\bf q})
\mathrm{Tr}[\hat{\gamma}^{1/2}\tau^{\pm}\underline{\hat{Q}_{\epsilon_{1}\epsilon_{1}^\prime}({\bf q})} ]
\mathrm{Tr}[\hat{\gamma}^{2/1}\tau^{\mp}\underline{\hat{Q}_{\epsilon_{2}\epsilon_{2}^\prime}(-{\bf q})} ]
\delta_{ \epsilon_{1}-\epsilon_{1}^\prime, \epsilon_{2}^\prime -\epsilon_{2} }
\end{align}
we have made the following substitution
\begin{align}
V_{\mathrm{s}} \rightarrow
\Gamma_{\mathrm{s}}(\epsilon_{1}-\epsilon_{1}^\prime, {\bf q})
=
 \frac{1}{ V_{\mathrm{s}}^{-1} +
M(\epsilon_{1} - \epsilon_{1}^\prime, {\bf q})}.
\end{align}
This interaction will be of use when we will be considering amplitudes in the triplet part of the Cooper channel.
Note that, as derived, $\Gamma_{\mathrm{s}}$ does not depend on the $\epsilon_{2}$ and $\epsilon_{2}^\prime$ frequencies, but only on the $\epsilon_{1}-\epsilon_{1}^\prime = \epsilon_{2}^\prime-\epsilon_{2}$ difference.
Now, the interaction fields must be updated,
\begin{align}
\langle
\phi^{(i)}_{\mathrm{c}; n, \epsilon_{1}\epsilon_{1}^\prime}({\bf r}_{1})
\phi^{(j)}_{\mathrm{c}; m, \epsilon_{2}\epsilon_{2}^\prime}({\bf r}_{2})
\rangle_{\phi}
=
-\frac{i}{2\nu}\Gamma_{\mathrm{s}}(\epsilon_{1} - \epsilon_{1}^\prime,{\bf r}_{1}-{\bf r}_{2})
\hat{\gamma}^{2}_{ij}
\delta({\bf r}_{1}- {\bf r}_{2})   \delta_{n,- m}
\delta_{\epsilon_{1} - \epsilon_{1}^\prime, \epsilon_{2}^\prime - \epsilon_{2}},
\end{align}
which will be used in the following.
In Appendix \ref{appendixD} we will specify structure of $\Gamma_{\mathrm{s}}$ in the presence of the external magnetic field.

\subsection{Equation for the interaction amplitude}
Here we generalize the calculations in the preceding subsection to the frequency dependent electron interaction in the Cooper channel. In particular, we consider the case when $V_{\mathrm{s}}$ is a function of frequencies, $V_{\mathrm{s}}(\epsilon_{1},\epsilon_{2})$. Furthermore, we outline the equation on the interaction amplitude in spin-triplet part of the Cooper channel, assuming some $V_{\mathrm{t}}(\epsilon_{1},\epsilon_{2})$ interaction in the spin-triplet part of the Cooper channel.
The action in the singlet part of the Cooper channel is
\begin{align}
iS_{\mathrm{singlet}} =
 i\frac{\pi^2\nu}{8}
\int_{\epsilon_{1},\epsilon_{1}^\prime,\epsilon_{2},\epsilon_{2}^\prime}
V_{\mathrm{s}}(\epsilon_{1},\epsilon_{2})
\int_{{\bf r}}
\mathrm{Tr}_{\mathrm{KSN}}[ \hat{\gamma}^{1/2} \tau^{\pm} \underline{\hat{Q}_{\epsilon_{1}\epsilon_{1}^\prime}({\bf r})} ]
\mathrm{Tr}_{\mathrm{KSN}} [ \hat{\gamma}^{2/1} \tau^{\mp}  \underline{\hat{Q}_{\epsilon_{2}\epsilon_{2}^\prime}({\bf r})} ]
\delta_{\epsilon_{1} - \epsilon_{1}^\prime, \epsilon_{2}^\prime - \epsilon_{2}},
\end{align}
where $V_{\mathrm{s}}(\epsilon_{1},\epsilon_{2})$ is some interaction, a function of the frequencies from different trace blocks.
Also the interaction in the triplet part of the Cooper channel is
\begin{align}
iS_{\mathrm{triplet}} =
 i\frac{\pi^2\nu}{8}
\int_{\epsilon_{1},\epsilon_{1}^\prime,\epsilon_{2},\epsilon_{2}^\prime}
V_{\mathrm{t}}(\epsilon_{1},\epsilon_{2})
\int_{{\bf r}}
\mathrm{Tr}_{\mathrm{KSN}}[ \hat{\gamma}^{1/2} \tau^{\pm} {\bm \sigma} \underline{\hat{Q}_{\epsilon_{1}\epsilon_{1}^\prime}({\bf r})} ]
\mathrm{Tr}_{\mathrm{KSN}} [ \hat{\gamma}^{2/1} \tau^{\mp} {\bm \sigma} \underline{\hat{Q}_{\epsilon_{2}\epsilon_{2}^\prime}({\bf r})} ]
\delta_{\epsilon_{1} - \epsilon_{1}^\prime, \epsilon_{2}^\prime - \epsilon_{2}},
\end{align}
where again $V_{\mathrm{t}}(\epsilon_{1},\epsilon_{2})$ is some interaction dependent on frequencies from different trace blocks.
In both cases we do not know explicit expression for $V_{\mathrm{s}}(\epsilon_{1},\epsilon_{2})$ and $V_{\mathrm{t}}(\epsilon_{1},\epsilon_{2})$.
For example, in the cosidered above example of the BCS model, in singlet part bare the interaction is $V_{\mathrm{s}}(\epsilon_{1},\epsilon_{2}) = \mathrm{const}$.
In the triplet part bare interaction is absent, but some effective interaction can be dynamically generated in rescattering processes.

In order to construct a ladder in the Cooper channel, we expand the $\hat{Q}$ matrices to the first order in $\hat{W}$, and get for the action
\begin{align}
iS_{\mathrm{singlet}} + iS_{\mathrm{triplet}} \approx
&
 i\frac{\pi^2\nu}{8}
\int_{\epsilon_{1},\epsilon_{1}^\prime,\epsilon_{2},\epsilon_{2}^\prime}
V_{\mathrm{s}}(\epsilon_{1},\epsilon_{2})
\int_{{\bf r}}
\mathrm{Tr}_{\mathrm{KSN}}[ \hat{\gamma}^{1/2} \tau^{\pm} \underline{\hat{\sigma}_{3}\hat{W}_{\epsilon_{1}\epsilon_{1}^\prime}({\bf r})} ]
\mathrm{Tr}_{\mathrm{KSN}} [ \hat{\gamma}^{2/1} \tau^{\mp}  \underline{\hat{\sigma}_{3}\hat{W}_{\epsilon_{2}\epsilon_{2}^\prime}({\bf r})} ]
\delta_{\epsilon_{1} - \epsilon_{1}^\prime, \epsilon_{2}^\prime - \epsilon_{2}}
\\
&
+
 i\frac{\pi^2\nu}{8}
\int_{\epsilon_{1},\epsilon_{1}^\prime,\epsilon_{2},\epsilon_{2}^\prime}
V_{\mathrm{t}}(\epsilon_{1},\epsilon_{2})
\int_{{\bf r}}
\mathrm{Tr}_{\mathrm{KSN}}[ \hat{\gamma}^{1/2} \tau^{\pm} {\bm \sigma} \underline{\hat{\sigma}_{3}\hat{W}_{\epsilon_{1}\epsilon_{1}^\prime}({\bf r})} ]
\mathrm{Tr}_{\mathrm{KSN}} [ \hat{\gamma}^{2/1} \tau^{\mp} {\bm \sigma} \underline{\hat{\sigma}_{3}\hat{W}_{\epsilon_{2}\epsilon_{2}^\prime}({\bf r})} ]
\delta_{\epsilon_{1} - \epsilon_{1}^\prime, \epsilon_{2}^\prime - \epsilon_{2}}.
\end{align}

\begin{figure}[h] \centerline{\includegraphics[clip, width=0.7  \columnwidth]{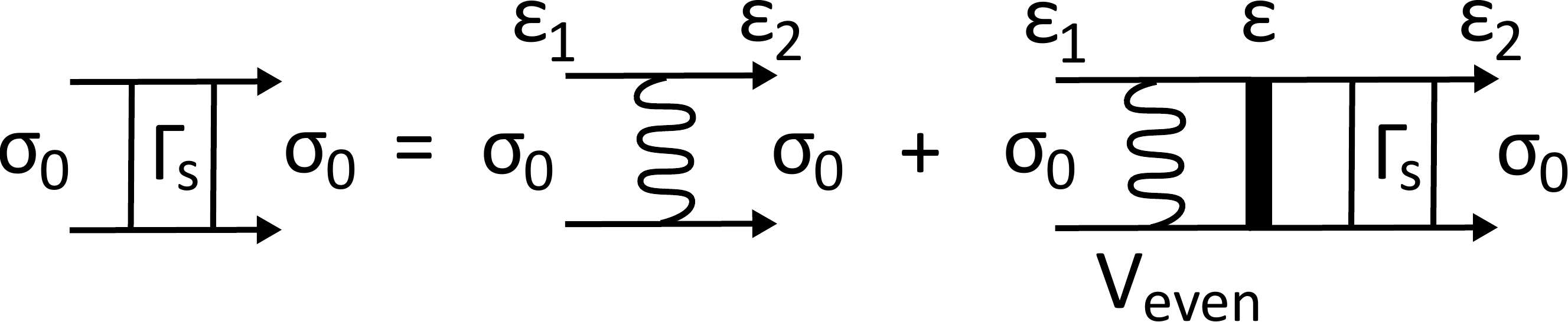}}

\protect\caption{Equation for the interaction in the spin-singlet part of the Cooper channel.
Here $\sigma_{0}$ corresponds to the spin identity matrix selecting the spin-singlet part of the Cooper channel.}

\label{fig:EquationSingletSM}

\end{figure}

The equation for the interaction amplitude in singlet part of the Cooper channel, shown in Fig. \ref{fig:EquationSingletSM}, is derived to be
\begin{align}
\Gamma_{\mathrm{s}}(\epsilon_{1},\epsilon_{2})
=
\frac{1}{2}\left[ V_{\mathrm{s}}(\epsilon_{1},\epsilon_{2})  +   V_{\mathrm{s}}(\epsilon_{1}, -\epsilon_{2}) \right]
- \pi \int_{\epsilon} \frac{1}{2}\left[ V_{\mathrm{s}}(\epsilon_{1},\epsilon)  +   V_{\mathrm{s}}(\epsilon_{1}, -\epsilon) \right]\frac{{\cal F}_{\epsilon}}{ \epsilon}
\Gamma_{\mathrm{s}}(\epsilon,\epsilon_{2}),
\end{align}
and in the triplet channel, shown in Fig.~\ref{fig:EquationTripletSM}, is
\begin{align}
\Gamma_{\mathrm{t}}(\epsilon_{1},\epsilon_{2})
=
\frac{1}{2}\left[ V_{\mathrm{t}}(\epsilon_{1},\epsilon_{2})  -   V_{\mathrm{t}}(\epsilon_{1}, -\epsilon_{2}) \right]
- \pi \int_{\epsilon} \frac{1}{2}\left[ V_{\mathrm{t}}(\epsilon_{1},\epsilon)  -   V_{\mathrm{t}}(\epsilon_{1}, -\epsilon) \right]\frac{{\cal F}_{\epsilon}}{\epsilon}
\Gamma_{\mathrm{t}}(\epsilon,\epsilon_{2}) ,
\label{equationTripletSM}
\end{align}
where recall $\int_{\epsilon} (..) \equiv \int_{-\infty}^{\infty}\frac{d\epsilon}{2\pi} (..)$.
In the figures we defined even and odd in frequencies components,
\begin{align}
&
V_{\mathrm{even}}(\epsilon_{1},\epsilon_{2}) = \frac{1}{2}\left[ V_{\mathrm{s}}(\epsilon_{1},\epsilon_{2})  +   V_{\mathrm{s}}(\epsilon_{1}, -\epsilon_{2}) \right],
\\
&
V_{\mathrm{odd}}(\epsilon_{1},\epsilon_{2}) = \frac{1}{2}\left[ V_{\mathrm{t}}(\epsilon_{1},\epsilon_{2})  -   V_{\mathrm{t}}(\epsilon_{1}, -\epsilon_{2}) \right],
\end{align}
effective in the corresponding channels. It is worth noting that the minus sign in the definition of $V_{\mathrm{odd}}$ is due to $\sigma_{y}{\bm \sigma}^{\mathrm{T}}\sigma_{y} = - {\bm \sigma}$. A good discussion on the structure of the effective interaction depending on the symmetry can be found in Ref. \cite{SamokhinMineev2008}.

\begin{figure}[h] \centerline{\includegraphics[clip, width=0.7  \columnwidth]{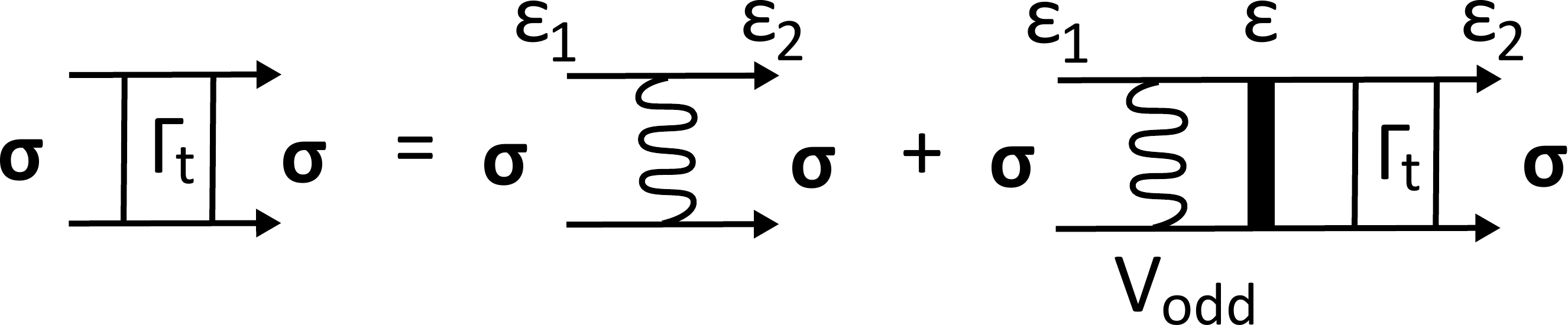}}

\protect\caption{Equation for the interaction in the spin-triplet part of the Cooper channel. Here ${\bm \sigma}$ correspond to the spin Pauli
matrices selecting the spin-triplet part of the Cooper channel.}

\label{fig:EquationTripletSM}

\end{figure}

\section{Effective interaction in the triplet part of the Cooper channel}\label{appendixC}
Here we derive an amplitude $V_{\mathrm{t}}(\epsilon_{1},\epsilon_{2})$ effective for the spin-triplet part of
the Cooper channel.
In the beginning of Sec.~\ref{e-einteractioninodd} we have explained why the amplitudes shown in Fig.~\ref{fig:diagramCDtext}
are the most relevant for this channel. In this Appendix we will calculate these diagrams.
Here, it will be convenient to use their slightly
modified version presented in Fig.~\ref{fig:diagramCDsm}, with frequencies labelled in a way they appear in the course of calculations.

\begin{figure}[h] \centerline{\includegraphics[clip, width=0.5  \columnwidth]{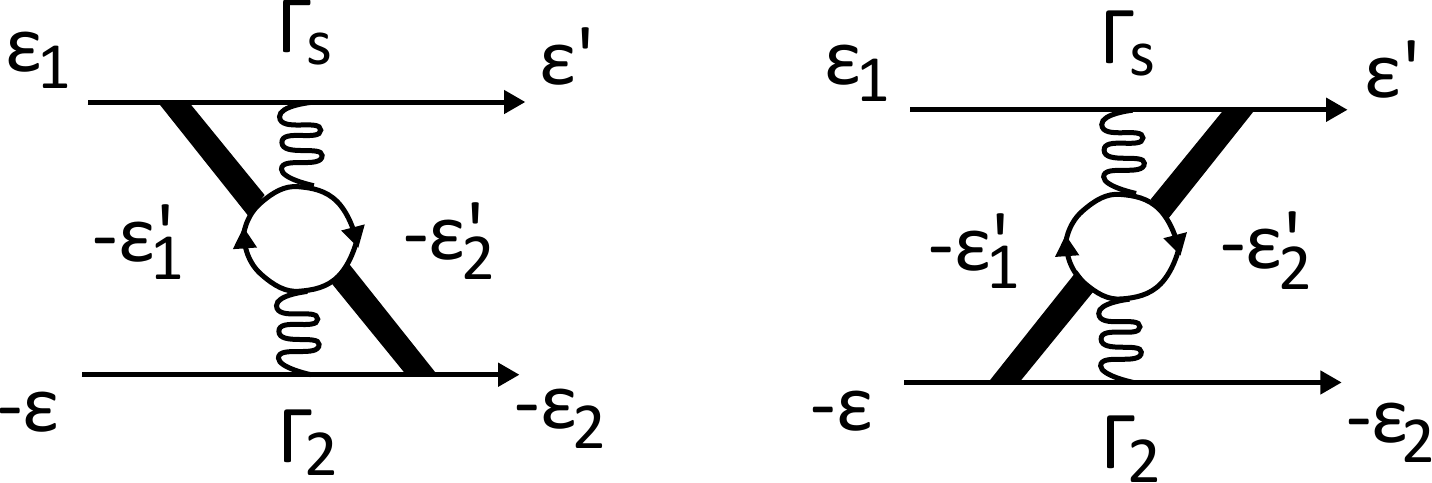}}

\protect\caption{Amplitude efficient for the spin-triplet pairing. The same as in Fig.~\ref{fig:diagramCDtext},
but for a general set of frequencies.}

\label{fig:diagramCDsm}

\end{figure}

\subsection{Scheme of the derivation}\label{sectionmixed1}

In order to evaluate the diagrams in Fig.~\ref{fig:diagramCDsm}, we expand the interaction action, which is expanded to third order in $\hat{W}$ matrices,
\begin{align}
iS_{\mathrm{ee}}^{[3]}
=
-
\frac{\pi^2 \nu^2 }{4}
\sum_{m=\rho,\sigma,\mathrm{c}}
\int_{{\bf r},{\bf r}^\prime}
\mathrm{Tr}\left[ \underline{ \hat{\phi}_{m;\epsilon_{1}\epsilon_{1}^\prime} ({\bf r})} \hat{\sigma}_{3}
\hat{W}_{\epsilon_{1}^\prime \epsilon_{1}}({\bf r})  \right]
\mathrm{Tr}\left[ \underline{ \hat{\phi}_{m;\epsilon_{2}\epsilon_{2}^\prime} ({\bf r}^\prime)} \hat{\sigma}_{3}
\hat{W}_{\epsilon_{2}^\prime \epsilon}({\bf r}^\prime)
\hat{W}_{\epsilon \epsilon_{2}}({\bf r}^\prime)
  \right],
\end{align}
to second order
and then contract certain $\hat{W}$ matrices, and get an action for the interaction in the triplet part of the Cooper channel,
\begin{align}
iS_{\mathrm{mixed}}
= \frac{1}{2} \langle \langle iS_{\mathrm{ee}}^{[3]}   iS_{\mathrm{ee}}^{[3]} \rangle_{\phi} \rangle_{W}
=  \frac{1}{2}\left(\frac{\pi^2 \nu^2}{4}\right)^2
\int_{{\bf r}_{1},{\bf r}_{2},{\bf r}_{3},{\bf r}_{4}}
\langle
{\bf K}_{\mathrm{A}}({\bf r}_{1},{\bf r}_{3})
{\bf K}_{\mathrm{B}}( {\bf r}_{2},{\bf r}_{4})
+
{\bf K}_{\mathrm{B}}({\bf r}_{1},{\bf r}_{3})
{\bf K}_{\mathrm{A}}( {\bf r}_{2},{\bf r}_{4})
\rangle_{\phi},
\end{align}
where vectors (vectors due to $\hat{\bm{\phi}}_{\sigma}$) ${\bf K}_{\mathrm{A}}$ and ${\bf K}_{\mathrm{B}}$ are defined as
\begin{align}\label{KA}
&
{\bf K}_{\mathrm{A}}( {\bf r}_{1},{\bf r}_{3})
 =
\langle
\mathrm{Tr}\left[ \underline{\hat{\phi}_{\mathrm{c};\epsilon_{1}\epsilon_{1}^\prime}({\bf r}_{1})}
\hat{\sigma}_{3} \hat{W}_{\epsilon_{1}^\prime \epsilon_{1}}({\bf r}_{1})\right]
\mathrm{Tr}\left[ \underline{\hat{\bm{\phi}}_{\sigma;\epsilon_{2}\epsilon_{2}^\prime}({\bf r}_{3})}\hat{\sigma}_{3}
\hat{W}_{\epsilon_{2}^\prime\epsilon}({\bf r}_{3})\hat{W}_{\epsilon\epsilon_{2}}({\bf r}_{3})\right]
\rangle_{W}
\\
\rightarrow
-\frac{4}{\pi \nu}
&
\mathrm{Tr}
\left\{
\left[ \underline{\hat{\phi}_{\mathrm{c}}({\bf r}_{1})}\hat{\sigma}_{3}
\right]_{\alpha\beta;\perp\mathrm{K}\perp\mathrm{N};\epsilon_{1}\epsilon_{1}^\prime}
-
\left[ \tau_{1}\sigma_{y} \underline{\hat{\phi}_{\mathrm{c}}({\bf r}_{1})}\hat{\sigma}_{3} \sigma_{y}\tau_{1}\right]^{\mathrm{T}}_{\alpha\beta;\perp\mathrm{K}\perp\mathrm{N};\epsilon_{1}\epsilon_{1}^\prime}
\right\}
\hat{C}_{\mathrm{N}}(\epsilon_{1}+\epsilon_{1}^\prime, {\bf r}_{1}-{\bf r}_{3})
\\
&
\left\{
\left[ \underline{\hat{\bm{\phi}}_{\sigma}({\bf r}_{3})}\hat{\sigma}_{3}\hat{W}({\bf r}_{3}) \right]_{\beta\alpha;\perp\mathrm{K}\perp\mathrm{N};\epsilon_{1}^\prime \epsilon_{1}}
+
\left[ \hat{W}({\bf r}_{3}) \underline{\hat{\bm{\phi}}_{\sigma}({\bf r}_{3})}\hat{\sigma}_{3}\right]_{\beta\alpha;\perp\mathrm{K}\perp\mathrm{N};\epsilon_{1}^\prime \epsilon_{1}}
\right\},
\end{align}
which is the part of the diagram with the Cooperon mode,
and
\begin{align}\label{KB}
&
{\bf K}_{\mathrm{B}}( {\bf r}_{1},{\bf r}_{3})
 = \langle
\mathrm{Tr}\left[ \underline{\hat{{\bm \phi}}_{\sigma}({\bf r}_{1})}\hat{\sigma}_{3} \hat{W}({\bf r}_{1})\right]
\mathrm{Tr}\left[ \underline{\hat{\phi}_{\mathrm{c}}({\bf r}_{3})}\hat{\sigma}_{3} \hat{W}({\bf r}_{3})\hat{W}({\bf r}_{3})\right]
\rangle_{W}
\\
\rightarrow
-\frac{4}{\pi \nu}
&
\mathrm{Tr}
\left\{
\left[ \underline{\hat{\bm{\phi}}_{\sigma}({\bf r}_{1})}\hat{\sigma}_{3} \right]_{\alpha\beta;\perp\mathrm{K}\parallel\mathrm{N};\epsilon_{1}\epsilon_{1}^\prime}
-
\left[ \tau_{1}\sigma_{y} \underline{\hat{\bm{\phi}}_{\sigma}({\bf r}_{1})}\hat{\sigma}_{3} \sigma_{y}\tau_{1}\right]^{\mathrm{T}}_{\alpha\beta;\perp\mathrm{K}\parallel\mathrm{N};\epsilon_{1}\epsilon_{1}^\prime}
\right\}
\hat{\Pi}_{\mathrm{N}}(\epsilon_{1}-\epsilon_{1}^\prime, {\bf r}_{1}-{\bf r}_{3})
\\
&
\left\{
\left[ \underline{\hat{\phi}_{\mathrm{c}}({\bf r}_{3})}\hat{\sigma}_{3}\hat{W}({\bf r}_{3}) \right]_{\beta\alpha;\perp\mathrm{K}\parallel\mathrm{N};\epsilon_{1}^\prime \epsilon_{1}}
+
\left[ \hat{W}({\bf r}_{3}) \underline{\hat{\phi}_{\mathrm{c}}({\bf r}_{3})}\hat{\sigma}_{3}\right]_{\beta\alpha;\perp\mathrm{K}\parallel\mathrm{N};\epsilon_{1}^\prime \epsilon_{1}}
\right\},
\end{align}
which is the part of the diagram with the diffuson mode.
By right arrow we mean selection of terms which will contribute to the interaction in the Cooper channel. Some combinations do not contribute and can be ignored.
In particular, deriving them, we used
\begin{align}
\label{NambuProjectiveIdentities}
&
\tau^{+}\tau_{-} = \tau^{+}, ~~ \tau_{-}\tau^{+} = 0, ~~ \tau^{-}\tau_{-} = 0, ~~ \tau_{-}\tau^{-} = \tau^{-},
\\
&
\tau^{-}\tau_{+} = \tau^{-}, ~~ \tau_{+}\tau^{-} = 0, ~~ \tau^{+}\tau_{+} = 0, ~~ \tau_{+}\tau^{+} = \tau^{+}
\end{align}
identities that selected only certain combinations of the interaction amplitudes.

Having defined ${\bf K}_{\mathrm{A}}( {\bf r}_{1},{\bf r}_{3})$ in Eq. \ref{KA} and ${\bf K}_{\mathrm{B}}( {\bf r}_{1},{\bf r}_{3})$ in Eq. \ref{KB}, we are now in position to contract the $\hat{\phi}$ fields.
For the $\hat{\phi}_{\mathrm{c}}$ fields we will study two cases, $\Gamma_{\mathrm{s}} = \mathrm{const}(\omega,{\bf q}) = \frac{V_{\mathrm{s}}}{1+V_{\mathrm{s}}\ln\left[\frac{\gamma}{\pi T \tau} \right]}$ and $\Gamma_{\mathrm{s}}\rightarrow \Gamma_{\mathrm{s}}(\omega,{\bf q},H,T)$ in magnetic field.

\subsection{Case of $\Gamma_{\mathrm{s}} = \mathrm{const}(\omega,{\bf q})$}\label{sectionmixed2}
Here we suppress all of the frequency dependence of the derived $\Gamma_{\mathrm{s}}$ for the model of constant $V_{\mathrm{s}}$.
Overall, after tedious but straightforward calculations one picks (denoted with the arrow below) the term effective in the triplet part of the Cooper channel,
\begin{align}
&
\langle
{\bf K}_{\mathrm{A}}({\bf r}_{1},{\bf r}_{3})
{\bf K}_{\mathrm{B}}( {\bf r}_{2},{\bf r}_{4})
\rangle_{\phi}
\rightarrow
-
16 \frac{1}{(\pi\nu)^2} \frac{1}{(2\nu)^2}
\Gamma_{\mathrm{s}}
\Gamma_{\mathrm{2}}
\int_{\epsilon_{1},\epsilon_{1}^\prime,\epsilon_{2}, \epsilon_{2}^\prime,\epsilon,\epsilon^\prime}
\delta_{\epsilon_{1}-\epsilon_{1}^\prime, \epsilon^\prime - \epsilon_{2}^\prime}
\delta_{\epsilon_{1}^\prime - \epsilon , \epsilon_{2}^\prime - \epsilon_{2}}
\delta_{{\bf r}_{1},{\bf r}_{4}}
\delta_{{\bf r}_{2},{\bf r}_{3}}
\\
&
\times
\{
{\cal F}_{\epsilon_{1}^\prime }
\underline{C}_{\mathrm{N}}(\epsilon_{1}+\epsilon_{1}^\prime, {\bf r}_{1}-{\bf r}_{3})
\bar{\Pi}_{\mathrm{N}}(\epsilon_{2}-\epsilon_{2}^\prime, {\bf r}_{1}-{\bf r}_{3})
+
{\cal F}_{\epsilon_{2}^\prime }
\bar{C}_{\mathrm{N}}(\epsilon_{1}+\epsilon_{1}^\prime, {\bf r}_{1}-{\bf r}_{3})
\underline{\Pi}_{\mathrm{N}}(\epsilon_{2}-\epsilon_{2}^\prime, {\bf r}_{1}-{\bf r}_{3})
\}
\\
&
\times
\mathrm{Tr}_{\mathrm{KSN}} \left[ {\bm \sigma} \tau^{\pm} \hat{\gamma}^{1/2} \underline{\hat{\sigma_{3}} \hat{W}_{\epsilon \epsilon_{1}} ({\bf r}_{3})}\right]
\mathrm{Tr}_{\mathrm{KSN}} \left[ {\bm \sigma} \tau^{\mp} \hat{\gamma}^{2/1} \underline{\hat{\sigma_{3}} \hat{W}_{\epsilon^\prime \epsilon_{2}} ({\bf r}_{4})}\right]. \label{twotraceblocks1}
\end{align}
Here $\Gamma_{\mathrm{s}} = \frac{V_{\mathrm{s}}}{1+V_{\mathrm{s}}\ln\left[\frac{\gamma}{\pi T \tau} \right]}$ is a constant as a functioin of frequencies. In case of attraction, the bare value is $V_{\mathrm{s}}<0$.
It can be immediately shown that the $\propto {\cal F}_{\epsilon_{1}^\prime }
\underline{C}_{\mathrm{N}}(\epsilon_{1}+\epsilon_{1}^\prime, {\bf r}_{1}-{\bf r}_{3})
\bar{\Pi}_{\mathrm{N}}(\epsilon_{2}-\epsilon_{2}^\prime, {\bf r}_{1}-{\bf r}_{3})$ part does not withstand the procedure of anti-symmetrization in frequency required for the triplet part of the Cooper channel.
This is because its elements do not couple the two trace blocks in frequencies in line (\ref{twotraceblocks1}).
To show it, use $\delta_{\epsilon_{1}^\prime - \epsilon , \epsilon_{2}^\prime - \epsilon_{2}}$ to integrate over $\epsilon_{2}^\prime$ frequency, then
\begin{align}
\int_{\epsilon_{2}^\prime}
\delta_{\epsilon_{1}^\prime - \epsilon , \epsilon_{2}^\prime - \epsilon_{2}}
{\cal F}_{\epsilon_{1}^\prime }
\underline{C}_{\mathrm{N}}(\epsilon_{1}+\epsilon_{1}^\prime, {\bf r}_{1}-{\bf r}_{3})
\bar{\Pi}_{\mathrm{N}}(\epsilon_{2}-\epsilon_{2}^\prime, {\bf r}_{1}-{\bf r}_{3})
=
{\cal F}_{\epsilon_{1}^\prime }
\underline{C}_{\mathrm{N}}(\epsilon_{1}+\epsilon_{1}^\prime, {\bf r}_{1}-{\bf r}_{3})
\bar{\Pi}_{\mathrm{N}}(\epsilon-\epsilon_{1}^\prime, {\bf r}_{1}-{\bf r}_{3}),
\end{align}
and all the frequencies are only from one of the trace blocks, namely $\mathrm{Tr}_{\mathrm{KSN}} \left[ {\bm \sigma} \tau^{\pm} \hat{\gamma}^{1/2} \underline{\hat{\sigma_{3}} \hat{W}_{\epsilon \epsilon_{1}} ({\bf r}_{3})}\right]$.

The second part with imaginary Cooperon survives only due to the dependence of ${\cal F}_{\epsilon_{2}^\prime }$ on $\epsilon_{2}^\prime$, rather than on $\epsilon_{1}^\prime$ as compared to the first vanishing term.
We then will focus on this term in the following. First let us integrate over the frequency $\epsilon_{2}^\prime$,
\begin{align}
\int_{\epsilon_{2}^\prime}
\delta_{\epsilon_{1}^\prime - \epsilon , \epsilon_{2}^\prime - \epsilon_{2}}
{\cal F}_{\epsilon_{2}^\prime }
\bar{C}_{\mathrm{N}}(\epsilon_{1}+\epsilon_{1}^\prime, {\bf r}_{1}-{\bf r}_{3})
\underline{\Pi}_{\mathrm{N}}(\epsilon_{2}-\epsilon_{2}^\prime, {\bf r}_{1}-{\bf r}_{3})
=
{\cal F}_{\epsilon_{1}^\prime - \epsilon +\epsilon_{2} }
\bar{C}_{\mathrm{N}}(\epsilon_{1}+\epsilon_{1}^\prime, {\bf r}_{1}-{\bf r}_{3})
\underline{\Pi}_{\mathrm{N}}(\epsilon-\epsilon_{1}^\prime, {\bf r}_{1}-{\bf r}_{3}).
\end{align}
We then Fourier transform the coordinate-dependent part of the second term
\begin{align}
&
\int_{{\bf r}_{1},{\bf r}_{3}}
\bar{C}_{\mathrm{N}}(\epsilon_{1}+\epsilon_{1}^\prime, {\bf r}_{1} - {\bf r}_{3})
\underline{\Pi}_{\mathrm{N}}(\epsilon-\epsilon_{1}^\prime, {\bf r}_{1} - {\bf r}_{3})
\mathrm{Tr}_{\mathrm{KSN}} \left[ {\bm \sigma} \tau^{\pm} \hat{\gamma}^{1/2} \underline{\hat{\sigma_{3}} \hat{W}_{\epsilon \epsilon_{1}} ({\bf r}_{3})}\right]
\mathrm{Tr}_{\mathrm{KSN}} \left[ {\bm \sigma} \tau^{\mp} \hat{\gamma}^{2/1} \underline{\hat{\sigma_{3}} \hat{W}_{\epsilon^\prime \epsilon_{2}} ({\bf r}_{1})}\right]
\\
&
=
\int_{{\bf q},{\bf q}^\prime}
\bar{C}_{\mathrm{N}}(\epsilon_{1}+\epsilon_{1}^\prime, {\bf q}^\prime - {\bf q} )
\underline{\Pi}_{\mathrm{N}}(\epsilon-\epsilon_{1}^\prime, {\bf q}^\prime )
\mathrm{Tr}_{\mathrm{KSN}} \left[ {\bm \sigma} \tau^{\pm} \hat{\gamma}^{1/2} \underline{\hat{\sigma_{3}} \hat{W}_{\epsilon \epsilon_{1}} ({\bf q})}\right]
\mathrm{Tr}_{\mathrm{KSN}} \left[ {\bm \sigma} \tau^{\mp} \hat{\gamma}^{2/1} \underline{\hat{\sigma_{3}} \hat{W}_{\epsilon^\prime \epsilon_{2}} (-{\bf q})}\right].
\end{align}
If we are interested in long wave-length behavior of the resulting interaction, we can neglect ${\bf q}$ as compared to ${\bf q}^{\prime}$.
This will be the case in the equation for the interaction amplitude (see Cooper ladder) where the frequency of the linking Cooperon is fast, while the momentum is slow.
Integration over ${\bf q}^\prime$ then reads as (we integrate over the angle first),
\begin{align}
\int_{0}^{1/\ell} \frac{q^\prime dq^\prime}{2\pi}
\bar{C}_{\mathrm{N}}(\epsilon_{1}+\epsilon_{1}^\prime, {\bf q}^\prime  )
\underline{\Pi}_{\mathrm{N}}(\epsilon-\epsilon_{1}^\prime, {\bf q}^\prime )
\approx
\frac{i}{8\pi D} \frac{\epsilon_{1}^\prime + \epsilon_{1}}{(\epsilon_{1} + \epsilon)(2\epsilon_{1}^\prime + \epsilon_{1} - \epsilon)}
\ln\left[ \frac{( \epsilon_{1}^\prime - \epsilon)^2}{( \epsilon_{1}^\prime + \epsilon_{1})^2} \right],
\end{align}
where the upper limit dropped out in our approximation.
Overall, the action corresponding to mixed amplitudes is
\begin{align}
iS_{\mathrm{mixed}} =
i\left( \frac{\pi^2 \nu}{8}\right)
\rho
(- \Gamma_{\mathrm{s}})
\Gamma_{2}
\mathrm{Tr}_{\mathrm{F}}
V_{\mathrm{t}}(\epsilon_{1},\epsilon_{2})
\int_{\bf r}
\mathrm{Tr}_{\mathrm{KSN}} \left[ {\bm \sigma} \tau^{\pm} \hat{\gamma}^{1/2} \underline{\hat{\sigma_{3}} \hat{W}_{\epsilon \epsilon_{1}} ({\bf r})}\right]
\mathrm{Tr}_{\mathrm{KSN}} \left[ {\bm \sigma} \tau^{\mp} \hat{\gamma}^{2/1} \underline{\hat{\sigma_{3}} \hat{W}_{\epsilon^\prime \epsilon_{2}} ({\bf r})}\right]
\delta_{\epsilon -\epsilon_{1}, \epsilon_{2} - \epsilon^\prime},
\end{align}
where $V_{\mathrm{s}} = -\vert V_{\mathrm{s}} \vert$ was assumed such that $(- \Gamma_{\mathrm{s}}) = \frac{\vert V_{\mathrm{s}} \vert}{1-\vert V_{\mathrm{s}} \vert\ln\left[\frac{\gamma}{\pi T \tau} \right]}$, $\rho = \frac{1}{(2\pi)^2 \nu_{2\mathrm{d}}D }$ is the thin film's resistance with $\nu_{2\mathrm{d}}$ being the two-dimensional density of states. In two-dimensional metalic systems $\rho \propto \frac{1}{k_{\mathrm{F}}\ell} \ll 1$, where $\ell$ is fermion's mean-free path, is the dimensionless resistance. This small parameter inevitably appears due to integration over the momenta of the diffuson and Cooperon modes.
We have set $\epsilon=\epsilon_{1}$ and $\epsilon^{\prime} = \epsilon_{2}$ (set up relevant for the Cooper ladder when frequencies sum to zero) in deriving the interaction amplitude $V_{\mathrm{t}}(\epsilon_{1},\epsilon_{2}) $,
\begin{align}
V_{\mathrm{t}}(\epsilon_{1},\epsilon_{2}) = -\frac{1}{2} \left[ f(\epsilon_{1},\epsilon_{2}) + f(\epsilon_{2},\epsilon_{1}) \right],
\end{align}
where
\begin{align}
 f(\epsilon_{1}, \epsilon_{2})
=
-2\pi
\int_{\epsilon_{1}^\prime}
{\cal F}_{\epsilon_{1}^\prime + \epsilon_{2} - \epsilon_{1} }
 \frac{\epsilon_{1}^\prime + \epsilon_{1}}{(2\epsilon_{1} )(2\epsilon_{1}^\prime )}
\ln\left[ \frac{( \epsilon_{1}^\prime - \epsilon_{1})^2}{( \epsilon_{1}^\prime + \epsilon_{1})^2} \right]
=
\frac{2\pi}{2\epsilon_{1}}
\int_{\epsilon_{1}^\prime}
{\cal F}_{\epsilon_{1}^\prime + \epsilon_{2} - \epsilon_{1} }
 \frac{\epsilon_{1}^\prime + \epsilon_{1}}{2\epsilon_{1}^\prime }
\ln\left[ \frac{( \epsilon_{1}^\prime + \epsilon_{1})^2}{( \epsilon_{1}^\prime - \epsilon_{1})^2} \right].
\end{align}

We can analytically calculate $f(\epsilon_{1},\epsilon_{2})$ only at $T=0$,
\begin{align}
\label{analyticsCD}
\frac{1}{2\pi}f(\epsilon_{1},\epsilon_{2})
&
=
\frac{1}{2\epsilon_{1}}
\int_{x}
\mathrm{sign}( x + \epsilon_{2} - \epsilon_{1} )
\frac{x + \epsilon_{1}}{2x }
\ln\left[  \frac{(x+\epsilon_{1})^2}{(x-\epsilon_{1})^2}  \right]
\\
&
=
\frac{1}{2\epsilon_{1}}
\int_{\vert \epsilon_{1} - \epsilon_{2} \vert}^{1/\tau}
\frac{dx}{2\pi}
\ln\left[  \frac{(x+\epsilon_{1})^2}{(x-\epsilon_{1})^2}  \right]
-
\frac{1}{4}
\mathrm{sign}(\epsilon_{1} - \epsilon_{2})
\int_{-\vert \epsilon_{1} - \epsilon_{2} \vert}^{\vert \epsilon_{1} - \epsilon_{2} \vert}
\frac{dx}{2\pi}
\frac{1}{x}
\ln\left[  \frac{(x+\epsilon_{1})^2}{(x-\epsilon_{1})^2}  \right]
\\
&
\approx
\frac{1}{2\pi}
\ln\left[
\frac{1}
{\tau^2
\left( \vert \vert \epsilon_{1} - \epsilon_{2}\vert + \epsilon_{1}\vert \right)
\left(   \vert \vert \epsilon_{1} - \epsilon_{2}\vert - \epsilon_{1} \vert  \right) }
\right]
+
\frac{1}{2\pi}
\frac{\vert \epsilon_{1} - \epsilon_{2} \vert }{\epsilon_{1}}
\ln\left[
\frac{ \left( \vert \vert \epsilon_{1} - \epsilon_{2}\vert - \epsilon_{1} \vert \right)  }
{\left( \vert \vert  \epsilon_{1} - \epsilon_{2}\vert + \epsilon_{1} \vert \right)}
\right]
\\
&
-
\frac{1}{2\pi}
\frac{(\epsilon_{1}-\epsilon_{2})}{\epsilon_{1}}
\Theta(\vert  \epsilon_{1}\vert  -  \vert\epsilon_{1}-\epsilon_{2}\vert )
+
\frac{1}{2\pi}
\left[
\frac{2\epsilon_{1}}{\vert\epsilon_{1}-\epsilon_{2}\vert}
-4 \mathrm{sign}(\epsilon_{1})
\mathrm{sign}(\epsilon_{1}-\epsilon_{2})
\right]
\Theta(  \vert\epsilon_{1}-\epsilon_{2}\vert - \vert \epsilon_{1}\vert   ),
\end{align}
where in the second integral $\ln\left[  \frac{(x+\epsilon_{1})^2}{(x-\epsilon_{1})^2}  \right] \approx 4 \frac{\mathrm{min}(\epsilon_{1},x)}{\mathrm{max}(\epsilon_{1},x)}$ approximation was used.
We plot analytical expression for
\begin{align}
V_{\mathrm{odd}}(\epsilon_{1},\epsilon_{2})
 = \frac{1}{2} \left[ V_{\mathrm{t}}(\epsilon_{1},\epsilon_{2}) - V_{\mathrm{t}}(\epsilon_{1}, -\epsilon_{2})   \right],
\end{align}
interaction amplitude entering the equation in the triplet Cooper ladder,
in where
$V_{\mathrm{t}}(\epsilon_{1},\epsilon_{2})$ is based on Eq. (\ref{analyticsCD})
in the right plot of Fig.~\ref{fig:diagramCDfig1} in the MT.
At non-zero temperatures, we plot numerial calculation of the interaction amplitude $V_{\mathrm{odd}}(\epsilon_{1},\epsilon_{2})$ in the left and center of Fig.~\ref{fig:diagramCDfig1} in the MT.
Importantly, the sign of the interaction $V_{\mathrm{odd}}$ is attractive when $V_{\mathrm{s}}<0$.
This will lead to instability in the spin-triplet part of the Cooper channel when $T>T_{\mathrm{c}}$.


\section{Pairing in the presence of magnetic field}\label{appendixD}

\subsection{Zeroth Landau level approximation}\label{sectionmixed3}
When the magnetic field is non-zero, one has to carefully treat integration over the coordinates \cite{GalitskiLarkinPRB2001,MichaeliTikhonovFinkelsteinPRB2012}.
The Cooperon part which enters the expression above reads,
\begin{align}
\bar{C}_{\mathrm{N}}(\epsilon_{1}^\prime +\epsilon_{1}; {\bf r}_{1},{\bf r}_{3})
 =
-
i(\epsilon_{1}^\prime +\epsilon_{1})
 \sum_{N,\alpha}
\frac{\Psi_{N,\alpha}({\bf r}_{1}) \Psi_{N,\alpha}^{\dag}({\bf r}_{3}) }{\left[\omega_{\mathrm{c}} \left(  N +\frac{1}{2} \right) \right]^2 + (\epsilon_{1}^\prime +\epsilon_{1})^2},
\end{align}
here $\Psi_{N,\alpha}({\bf r}_{1}) $ is the Landau wave function with $N$ being main quantum number, while $\alpha$ - angular.
 Identity for the Landau wave functions is
\begin{align}
\sum_{\alpha}
\Psi_{N,\alpha}({\bf r}_{1})
\Psi_{N,\alpha}^{\dag}({\bf r}_{3})
=
\frac{1}{\sqrt{2\pi}\ell_{\mathrm{B}}}
\Psi_{N,0}({\bf r}_{1} - {\bf r}_{3})
\exp\left[ - \frac{i e }{2\hbar c}{\bf B}({\bf r}_{1}\times{\bf r}_{3})  \right],
\end{align}
where
\begin{align}
\Psi_{N,0}({\bf r})
= \frac{1}{\sqrt{2\pi} \ell_{\mathrm{B}}}
 \exp\left( - \frac{r^2}{4\ell_{\mathrm{B}}^2}\right)
L_{N}\left(\frac{r^2}{2\ell_{\mathrm{B}}^2} \right),
\end{align}
where $L_{N}$ is the regular Laguerre polynomial.
For our purposes we can neglect the phase in the identity.
We will only consider a $N = 0$ term in the sum over the Landau levels, then $L_{0} = 1$.
This is the so-called zeroth Landau level approximation.
We then rewrite the identity in the form we will be using in our further calculations,
\begin{align}
\sum_{\alpha}
\Psi_{0,\alpha}({\bf r}_{1})
\Psi_{0,\alpha}^{\dag}({\bf r}_{3})
\rightarrow
\frac{1}{2\pi\ell_{\mathrm{B}}^2}
 \exp\left[ - \frac{({\bf r}_{1} - {\bf r}_{3}) ^2}{4\ell_{\mathrm{B}}^2}\right].
\end{align}
In this approximation the Cooperon becomes,
\begin{align}
\bar{C}_{\mathrm{N}}(\epsilon_{1}^\prime +\epsilon_{1}; {\bf r}_{1},{\bf r}_{3})
\approx - \frac{i(\epsilon_{1}^\prime +\epsilon_{1})}{2\pi\ell_{\mathrm{B}}^2}
\frac{ e^{ - \frac{({\bf r}_{1} - {\bf r}_{3}) ^2}{4\ell_{\mathrm{B}}^2}} }{\left( \frac{\omega_{\mathrm{c}} }{2} \right)^2 + (\epsilon_{1}^\prime +\epsilon_{1})^2},
\end{align}
which is now dependent on the coordinate difference ${\bf r}_{1} - {\bf r}_{3}$.
To find Fourier image of the approximated Cooperon in magnetic field, we use
\begin{align}
\int dx~e^{iq_{x}x} e^{-\frac{x^2}{4\ell^{2}_{\mathrm{B}}}} = \sqrt{4\pi \ell_{\mathrm{B}}^2} e^{-q_{x}^2 \ell_{\mathrm{B}}^2},
\end{align}
and, therefore, we have
\begin{align}
\int dx~e^{iq_{x}x} e^{-\frac{x^2}{4\ell^{2}_{\mathrm{B}}}}
\int dy~e^{iq_{y}y} e^{-\frac{y^2}{4\ell^{2}_{\mathrm{B}}}}
 = 4\pi \ell_{\mathrm{B}}^2 e^{-q^2 \ell_{\mathrm{B}}^2}.
\end{align}
Then
\begin{align}
\bar{C}_{\mathrm{N}}(\epsilon_{1}+\epsilon_{1}^\prime, {\bf q} ) =-
2e^{-q^2 \ell_{\mathrm{B}}^2}
\frac{ i(\epsilon_{1}^\prime +\epsilon_{1}) }{\left( \frac{\omega_{\mathrm{c}} }{2} \right)^2 + (\epsilon_{1}^\prime +\epsilon_{1})^2}.
\end{align}
Diffuson mode stays the same as in the case of zero magnetic field, namely
\begin{align}
\underline{\Pi}_{\mathrm{N}}(\epsilon_{1}^\prime - \epsilon_{1}, {\bf q})  =
 \frac{ Dq^2}{(Dq^2)^2 + (\epsilon_{1}^\prime - \epsilon_{1})^2}.
\end{align}

\subsection{Effective interaction at non-zero magnetic field. Case of $\Gamma_{\mathrm{s}}(\omega, {\bf q},H,T)$.}\label{sectionmixed4}
We now study the magnetic field dependence of the diagrams presented in Fig.~\ref{fig:diagramCDsm}.
In case of magnetic field $\Gamma_{\mathrm{s}}(\omega, {\bf q},H,T)$ is now a function of the magnetic field, temperature, frequency and momentum (we define below).
Then same steps as above apply, and we get for the relevant expression the following form,
\begin{align}
\langle
{\bf K}_{\mathrm{A}}({\bf r}_{1},{\bf r}_{3})
{\bf K}_{\mathrm{B}}( {\bf r}_{2},{\bf r}_{4})
\rangle_{\phi}
\rightarrow
&
-
16 \frac{1}{(\pi\nu)^2} \frac{1}{(2\nu)^2}
\Gamma_{2}
\int_{\epsilon_{1},\epsilon_{1}^\prime,\epsilon_{2}, \epsilon_{2}^\prime,\epsilon,\epsilon^\prime}
\delta_{\epsilon_{1}-\epsilon_{1}^\prime, \epsilon^\prime - \epsilon_{2}^\prime}
\delta_{\epsilon_{1}^\prime - \epsilon , \epsilon_{2}^\prime - \epsilon_{2}}
\delta_{{\bf r}_{1},{\bf r}_{4}}
\delta_{{\bf r}_{2},{\bf r}_{3}}
\\
&
\times
{\cal F}_{\epsilon_{2}^\prime }
\Gamma_{\mathrm{s}}(\epsilon_{1}^\prime-\epsilon_{1}, {\bf r}_{4} , {\bf r}_{1},H,T)
\bar{C}_{\mathrm{N}}(\epsilon_{1}+\epsilon_{1}^\prime, {\bf r}_{1}-{\bf r}_{3})
\underline{\Pi}_{\mathrm{N}}(\epsilon_{2}-\epsilon_{2}^\prime, {\bf r}_{1}-{\bf r}_{3})
\\
&
\times
\mathrm{Tr}_{\mathrm{KSN}} \left[ {\bm \sigma} \tau^{\pm} \hat{\gamma}^{1/2} \underline{\hat{\sigma_{3}} \hat{W}_{\epsilon \epsilon_{1}} ({\bf r}_{3})}\right]
\mathrm{Tr}_{\mathrm{KSN}} \left[ {\bm \sigma} \tau^{\mp} \hat{\gamma}^{2/1} \underline{\hat{\sigma_{3}} \hat{W}_{\epsilon^\prime \epsilon_{2}} ({\bf r}_{4})}\right],
\end{align}
where we have already discarded the term with imaginary part of the diffuson mode.
After the spin-singlet Cooper channel propagator $\Gamma_{\mathrm{s}}$ is used and zeroth Landau level approximation is made, we get for the effective interaction
\begin{align}
&
\int_{{\bf r}_{1},{\bf r}_{2},{\bf r}_{3},{\bf r}_{4}}
\Gamma_{\mathrm{s}}(\epsilon_{1}^\prime-\epsilon_{1}, {\bf r}_{4} - {\bf r}_{1},H,T)
\bar{C}_{\mathrm{N}}(\epsilon_{1}+\epsilon_{1}^\prime, {\bf r}_{1} - {\bf r}_{3})
\underline{\Pi}_{\mathrm{N}}(\epsilon-\epsilon_{1}^\prime, {\bf r}_{2} - {\bf r}_{4})
\\
&
~~~~~~~~~~~~
\times
\mathrm{Tr}_{\mathrm{KSN}} \left[ {\bm \sigma} \tau^{\pm} \hat{\gamma}^{1/2} \underline{\hat{\sigma_{3}} \hat{W}_{\epsilon \epsilon_{1}} ({\bf r}_{3})}\right]
\mathrm{Tr}_{\mathrm{KSN}} \left[ {\bm \sigma} \tau^{\mp} \hat{\gamma}^{2/1} \underline{\hat{\sigma_{3}} \hat{W}_{\epsilon^\prime \epsilon_{2}} ({\bf r}_{4})}\right]
\delta_{{\bf r}_{1},{\bf r}_{4}}
\delta_{{\bf r}_{2},{\bf r}_{3}}
\\
&
=
\int_{{\bf q},{\bf q}^\prime}
\Gamma_{\mathrm{s}}(\epsilon_{1}^\prime-\epsilon_{1}, {\bf q}^\prime,H,T)
\bar{C}_{\mathrm{N}}(\epsilon_{1}+\epsilon_{1}^\prime, {\bf q}^\prime )
\underline{\Pi}_{\mathrm{N}}(\epsilon-\epsilon_{1}^\prime, {\bf q}^\prime -{\bf q})
\mathrm{Tr}_{\mathrm{KSN}} \left[ {\bm \sigma} \tau^{\pm} \hat{\gamma}^{1/2} \underline{\hat{\sigma_{3}} \hat{W}_{\epsilon \epsilon_{1}} ({\bf q})}\right]
\mathrm{Tr}_{\mathrm{KSN}} \left[ {\bm \sigma} \tau^{\mp} \hat{\gamma}^{2/1} \underline{\hat{\sigma_{3}} \hat{W}_{\epsilon^\prime \epsilon_{2}} (-{\bf q})}\right].
\end{align}
Here we again neglect ${\bf q}$ as compared to ${\bf q}^\prime$ in the expressions for diffuson.
The zeroth Landau level assumption is
\begin{align}
\Gamma_{\mathrm{s}}(\epsilon_{1}^{\prime} - \epsilon_{1}, {\bf q},H,T)
 \approx
2e^{-q^2 \ell_{\mathrm{B}}^2} \frac{V_{\mathrm{s}}}{1+ V_{\mathrm{s}}M(\epsilon_{1}^\prime-\epsilon_{1}, \vert {\bf q} \vert=\sqrt{\frac{\omega_{c}}{2D}},H,T)}.
\end{align}
This propagator besides here also enters in the equation for the s-wave spin-singlet Cooper channel interaction amplitude.
When $\Gamma_{\mathrm{s}}(\epsilon_{1}^{\prime} - \epsilon_{1}, {\bf q},H,T)$ diverges, there is a phase transition. In the metallic regime, sign of the propagtor is negative corresponding to attraction in the s-wave spin-singlet Cooper channel.
Let us now set $T = 0$ and define critical magnetic field of the transition from the s-wave superconducting state to metallic phase.
\begin{align}
M(\epsilon_{1}^{\prime} - \epsilon_{1},{\bf q},H,T=0) \approx
\frac{1}{4}
\ln\left\{ \frac{\Lambda^4 \left(\frac{2\gamma}{\pi}\right)^4}{ \left[ \frac{\left(\frac{\omega_{c}}{2}\right)^2 + (\epsilon_{1}^{\prime} - \epsilon_{1})^2}{4}\right]^2  } \right\}.
\end{align}
Then
\begin{align}\label{CooperPropagatorMF1}
\Gamma_{\mathrm{s}}(\epsilon_{1}^{\prime} - \epsilon_{1}, {\bf q},H,T=0) = -  2e^{-q^2 \ell_{\mathrm{B}}^2} \frac{1}{\ln\left[ \frac{\omega_{\mathrm{c}}}{\omega_{\mathrm{c}2}} \right] + \ln\sqrt{1 + \frac{(\epsilon_{1}^{\prime} - \epsilon_{1})^2}{\left(\frac{\omega_{c}}{2}\right)^2}}},
\end{align}
where $\omega_{\mathrm{c}2} = \frac{8\gamma \Lambda}{\pi} e^{-\frac{1}{\vert V_{\mathrm{s}}\vert}}$ corresponds to the critical magnetic field at $T=0$, i.e. $H_{\mathrm{c}2}$.
For magnetic fields $\omega_{c}>\omega_{\mathrm{c}2}$, i.e. when the system is in the metallic regime, the sign of the Cooper propagator does not change, it is negative. At the transition the sign of Cooper propagator changes signalling a phase transition in to superconducting state. Dynamics of the Cooper propagator can make it to be sign changing as a function of frequency as we are approaching the phase transition in magnetic field from below, namely $\omega_{c}\lesssim \omega_{\mathrm{c}2}$, but we wish to study the system in the metallic regime away from the transition. We question the stability of the metallic phase in the vicinity to a transition to the regular s-wave spin-triplet superconductivity

Both $\Gamma_{\mathrm{s}}(\epsilon_{1}^{\prime} - \epsilon_{1}, {\bf q}^\prime,H,T) $ and $\bar{C}_{\mathrm{N}}(\epsilon_{1}+\epsilon_{1}^\prime, {\bf q}^\prime  )$ have the same momentum, therefore integration over ${\bf q}^\prime$ is similar to the  one above for $\Gamma_{\mathrm{s}} = \mathrm{const}(\omega,{\bf q})$ case. It is, however, important to keep the upper limit, which would now be equal to $1/\ell_{\mathrm{B}}$ due to the $e^{-q^2 \ell_{\mathrm{B}}^2}$ factor, in the integration over ${\bf q}^{\prime}$.

Next, when $T \neq 0$ but $H \gtrsim H_{\mathrm{c}2}$
\begin{align}
M(\epsilon_{1}^{\prime} - \epsilon_{1},{\bf q},H,T) \approx
\frac{1}{4}
\ln\left\{ \frac{\Lambda^4 \left(\frac{2\gamma}{\pi}\right)^4}{ \left[ T^2 + \frac{1}{4}\left( \frac{\omega_{\mathrm{c}}}{2}\right)^2 \right]^2  + \frac{(\epsilon_{1}^{\prime} - \epsilon_{1})^2}{4}
\left[ \frac{1}{2}\left( \frac{\omega_{\mathrm{c}}}{2}\right)^2  - T^2 \right]  + \left[ \frac{(\epsilon_{1}^{\prime} - \epsilon_{1})^2}{4}\right]^2 } \right\}.
\end{align}
Then
\begin{align}\label{CooperPropagatorMF2}
\Gamma_{\mathrm{s}}(\epsilon_{1}^{\prime} - \epsilon_{1}, {\bf q},H,T)
&
= -  2e^{-q^2 \ell_{\mathrm{B}}^2}
\frac{1}{\ln\left\{ \frac{ 4\left[ \left[ T^2 + \frac{1}{4}\left( \frac{\omega_{\mathrm{c}}}{2}\right)^2 \right]^2  + \frac{(\epsilon_{1}^{\prime} - \epsilon_{1})^2}{4}
\left[ \frac{1}{2}\left( \frac{\omega_{\mathrm{c}}}{2}\right)^2  - T^2 \right]  + \left[ \frac{(\epsilon_{1}^{\prime} - \epsilon_{1})^2}{4}\right]^2 \right]^{1/4}}{\omega_{\mathrm{c}2}} \right\}}
\\
&
\equiv
-  2e^{-q^2 \ell_{\mathrm{B}}^2} \tilde{\Gamma}_{\mathrm{s}}\left(\epsilon_{1}^{\prime} - \epsilon_{1}, \vert {\bf q} \vert=\sqrt{\frac{\omega_{c}}{2D}},H,T\right)
\end{align}
where first term under the logarithm measures how far the system is from the phase transition, while the other one describes the dynamics.

Overall, the action for the effective interaction in the triplet part of the Cooper channel is
\begin{align}
iS_{\mathrm{mixed}}
=
i\left( \frac{\pi^2 \nu}{8}\right)
4
\rho
\Gamma_{2}
\mathrm{Tr}_{\mathrm{F}}
V_{\mathrm{t}}(\epsilon_{1},\epsilon_{2})
\int_{{\bf r}_{3}}
\mathrm{Tr}_{\mathrm{KSN}} \left[ {\bm \sigma} \tau^{\pm} \hat{\gamma}^{1/2} \underline{\hat{\sigma_{3}} \hat{W}_{\epsilon \epsilon_{1}} ({\bf r}_{3})}\right]
\mathrm{Tr}_{\mathrm{KSN}} \left[ {\bm \sigma} \tau^{\mp} \hat{\gamma}^{2/1} \underline{\hat{\sigma_{3}} \hat{W}_{\epsilon^\prime \epsilon_{2}} ({\bf r}_{3})}\right]
\delta_{\epsilon -\epsilon_{1}, \epsilon_{2} - \epsilon^\prime},
\end{align}
where the effective interaction $V_{\mathrm{t}}$ is
\begin{align}
&
V_{\mathrm{t}}(\epsilon_{1},\epsilon_{2})
= -\frac{1}{2} \left[ f_{\mathrm{MF}}(\epsilon_{1},\epsilon_{2})  +  f_{\mathrm{MF}}(\epsilon_{2},\epsilon_{1}) \right],
\\
&
\frac{1}{2\pi} f_{\mathrm{MF}}(\epsilon_{1},\epsilon_{2})
= \int_{x}
{\cal F}_{x-(\epsilon_{1}-\epsilon_{2})}
\tilde{\Gamma}_{\mathrm{s}}\left(x - \epsilon_{1}, \vert {\bf q} \vert=\sqrt{\frac{\omega_{c}}{2D}},H,T\right)
\frac{x+\epsilon_{1}}{\left(\frac{\omega_{\mathrm{c}}}{2} \right)^2 + (x+\epsilon_{1})^2}
\ln\left[  \frac{\left(D\ell_{\mathrm{B}}^{-2} \right)^2+(x -\epsilon_{1})^2}{(x -\epsilon_{1})^2 }  \right],
\label{effectiveMFsm}
\end{align}
where $\Lambda$ is a high frequency cut-off. We can only study this interaction numerically.

In Fig.~\ref{fig:diagramCDfig2} of the MT a plot of the interaction amplitude $V_{\mathrm{odd}}(\epsilon_{1},\epsilon_{2}) = \frac{1}{2}\left[ V_{\mathrm{t}}(\epsilon_{1},\epsilon_{2}) - V_{\mathrm{t}}(\epsilon_{1},-\epsilon_{2}) \right]$ corresponding to Eq. (\ref{effectiveMFsm}) at different magnetic fields and frequencies is shown. In Fig.~\ref{fig:MatLabCD4} a dependence of the minimum eigenvalue of the corresponding Eq. (\ref{equationTripletSM}) on the $\Gamma_{\mathrm{t}}$ is plot.
There is a phase transition at magnetic field $H_{\mathrm{odd}}$ when $\lambda_{\mathrm{min}}$ crosses $-1$.
Let us picture this instability.
If we are increasing the magnetic field from zero, then at $\omega_{\mathrm{c}} \leq \omega_{\mathrm{c}2}$ the system is the s-wave spin-singlet superconductor, when the magnetic field reaches $\omega_{\mathrm{c}} = \omega_{\mathrm{c}2}$ a phase transition occurs to the spin-triplet odd-frequency paired state, then at $\omega_{\mathrm{c}} \approx 2.5\omega_{\mathrm{c}2}$, finally, the system becomes a metal.

\end{widetext}

\bibliographystyle{apsrev}
\bibliography{oddbib}

\end{document}